\documentclass[12pt]{iopart}
\usepackage{iopams}
\usepackage{combelow}

\usepackage{graphicx}
\usepackage{bm}
\usepackage{braket}
\usepackage{combelow}
\usepackage{slashed}
\usepackage{color}

\usepackage{array}   
\newcolumntype{C}{>{$}c<{$}} 

\eqnobysec

\def\eqref#1{(\ref{#1})}
\renewenvironment{pmatrix}{\left(\begin{tabular}{CC}}{\end{tabular}\right)}

\def\feq{f^{(\mathrm{eq})}}

\def\betat{{\widetilde{\beta}}}

\def\psibar{{\overline{\psi}}}
\def\tr{{{\rm tr}}}

\def\hatt{{\hat{0}}}
\def\hati{{\hat{\imath}}}
\def\hatj{{\hat{\jmath}}}
\def\hatk{{\hat{k}}}

\def\halpha{{\hat{\alpha}}}
\def\hbeta{{\hat{\rho}}}
\def\hgamma{{\hat{\sigma}}}

\def\hsigma{{\hat{\sigma}}}

\def\vx{\bm{x}}
\def\vgamma{\bm{\gamma}}
\def\vsigma{\bm{\sigma}}
\def\vn{\bm{n}}

\begin{document}

\title[Thermal expectation values of fermions on adS]
{Thermal expectation values of fermions on anti-de Sitter space-time}

\author{Victor E. Ambru\cb{s}}
\address{Department of Physics, West University of Timi\cb{s}oara,
Bd.~Vasile P\^arvan 4, Timi\cb{s}oara 300223, Romania}
\ead{Victor.Ambrus@e-uvt.ro}

\author{Elizabeth Winstanley}
\address{Consortium for Fundamental Physics, School of Mathematics and Statistics, University of Sheffield,
Hicks Building, Hounsfield Road, Sheffield. S3 7RH \\United Kingdom}
\ead{E.Winstanley@sheffield.ac.uk}

\begin{abstract}
Making use of the symmetries of anti-de Sitter space-time,
we derive an analytic expression for the
bispinor of parallel transport, from which we construct in closed form the vacuum Feynman Green's function of
the Dirac field on this background. Using the
imaginary time anti-periodicity property of the thermal Feynman Green's function,
we calculate the thermal expectation values
of the fermion condensate and stress-energy tensor and highlight the effect of quantum corrections as
compared to relativistic kinetic theory results.
\end{abstract}



%
%
%


\section{Introduction}

The formulation of the adS/CFT (anti-de Sitter space-time/conformal field theory) correspondence  \cite{Maldacena:1997re}
sparked an explosion of interest in the behaviour of classical and quantum fields on asymptotically adS space-times (see, for example \cite{art:aharony00} for a review).
In an appropriate limit, it is anticipated that results derived in a full theory of quantum gravity should reduce to those obtained within the semi-classical
framework of quantum field theory (QFT) in curved space-time.
In this setting, the space-time geometry is fixed and purely classical, with a quantum field propagating on this space-time background.
Since the simplest asymptotically adS space-time is pure adS itself, a first step in understanding the behaviour of quantum fields on asymptotically adS geometries is to study quantum fields on pure adS space-time.

The fact that adS space-time is a maximally symmetric space-time simplifies many aspects of QFT on this classical background.
At the same time, adS space-time is not globally hyperbolic, since it possesses closed time-like curves.
While the closed time-like curves can be removed by considering the covering space of adS space-time, this is still not globally hyperbolic,
which complicates QFT on this background.
As a result, boundary conditions have to be applied on the time-like boundary \cite{Avis:1977yn} in order for the resulting QFT to be well-defined.

Quantum fields on four-dimensional adS space-time were first studied in \cite{Avis:1977yn}, where a scalar field with arbitrary mass and general coupling to the space-time curvature is considered and closed-form expressions for the scalar field modes and the Feynman Green's function are derived.
This work was extended to $n$-dimensional adS space-time in \cite{Burgess:1984ti}.
When the quantum scalar field is in the global adS vacuum state,
the Feynman Green's function can be used to derive the renormalized vacuum expectation values (v.e.v.s) of the square of the scalar field and the stress-energy tensor (SET) using Hadamard renormalization \cite{Kent:2014nya}
(these quantities can also be calculated using zeta-function regularization \cite{Camporesi:1992wn,Caldarelli:1998wk}).
Thermal expectation values (t.e.v.s) for a massless, conformally coupled scalar field on four-dimensional adS space-time have also been computed \cite{Allen:1986ty}.

The maximal symmetry of adS space-time also simplifies the study of higher-spin fields, again allowing closed-form expressions for the Feynman Green's function to be derived in both the bosonic \cite{art:allen86a,art:belo}
and fermionic \cite{art:allen86b,art:muck00} cases.
The field modes for a massive fermion field can also be found in closed form using a suitable choice of tetrad basis vectors \cite{art:cotaescu07}.
For a massive fermion field on four-dimensional adS space-time, renormalized v.e.v.s of the fermion condensate (FC) and SET have been computed using various regularization techniques: Pauli-Villars, zeta-function \cite{Camporesi:1992wn} and Schwinger-de Witt, Hadamard \cite{art:ambrus15plb}, when the quantum field is in the global adS vacuum state.
On the time-like space-time boundary, regular boundary conditions are imposed in \cite{art:ambrus15plb}, which correspond to the reflective boundary conditions in \cite{Avis:1977yn} for the scalar field case.
The v.e.v.s for both scalar \cite{Kent:2014nya,Camporesi:1992wn,Caldarelli:1998wk} and fermion \cite{Camporesi:1992wn,art:ambrus15plb} fields on adS space-time have recently found application in a one-loop test of gauge-gravity duality \cite{Buhl-Mortensen:2016jqo}.

In this paper we calculate t.e.v.s for a massive quantum fermion field on a four-dimensional adS space-time background, extending the work of \cite{Allen:1986ty} to a higher-spin field.
As in \cite{Allen:1986ty}, the regular boundary conditions imposed on the fermion field allow adS space-time to act like a perfectly reflecting box, and to be filled with thermal radiation.
Due to the curvature of the space-time, the local temperature \cite{tolman30} of the radiation is not constant, and this breaks some of the space-time symmetries.
The authors of \cite{Allen:1986ty} consider only a massless, conformally coupled scalar field, which means that the short-distance singularity structure of the vacuum Feynman Green's function possesses only poles, and no logarithmic singularities \cite{art:decanini08}.
This enables them to use an elegant method to compute the t.e.v.s, by writing the thermal Feynman Green's function (which is doubly-periodic on pure adS space-time with closed time-like curves) in terms of an elliptic function.
For a massive fermion field, in general the vacuum Feynman Green's function has logarithmic short-distance singularities \cite{art:ambrus15plb} and therefore the method in \cite{Allen:1986ty} cannot be extended to our situation.
Nonetheless, we are able to derive closed-form expressions for the t.e.v.s of the FC and SET on adS space-time.
We consider the differences in expectation values between the thermal state and the vacuum state, which do not require renormalization.
The v.e.v.s derived in \cite{art:ambrus15plb} could be added to the differences computed here to find the renormalized t.e.v.s.
Remarkably, the quantum SET we compute here for the Dirac field takes the form of an ideal fluid, while the quantum SET for the Klein-Gordon field does not \cite{Allen:1986ty}.

Recently, the energy density and pressure of a relativistic thermal gas of particles on adS space-time have been derived using kinetic theory, by solving the relativistic Boltzmann equation \cite{ambrus16cota}.
Using Maxwell-J\"uttner statistics (the results can be extended to Bose-Einstein and Fermi-Dirac statistics following \cite{ambrus15wut,florkowski15}), it is found that the energy density and pressure depend only on the local temperature of the gas.
For a massless, conformally coupled scalar field, this property is not shared by the quantum t.e.v.s \cite{Allen:1986ty}.
We therefore compare our fermion quantum t.e.v.s with the results of classical relativistic kinetic theory, to examine the effect of quantum corrections.

The structure of the paper is as follows.
In section~\ref{sec:dirac}, we review the Dirac equation on curved space-times and introduce our notation before describing, in section~\ref{sec:geometry}, those geometric properties of adS space-time
relevant to our work.
Section~\ref{sec:parallel} contains a discussion of parallel transport on adS space-time using
the bivector (for tensors) and bispinor (for spinors) of parallel transport, for which we present analytic expressions.
In section~\ref{sec:feynman}, we review the
construction of the vacuum Feynman Green's function for a Dirac field on adS space-time following \cite{art:muck00,art:ambrus15plb}.
The computation of QFT t.e.v.s is performed in section~\ref{sec:thermal}, while the corresponding expressions in relativistic kinetic theory are obtained in section \ref{sec:kinetic}.
In section~\ref{sec:comparison}, we compare the QFT t.e.v.s with the results from relativistic kinetic theory.
Finally, our conclusions are presented in section~\ref{sec:conc}.

\section{Dirac equation on curved space-times}
\label{sec:dirac}

In this section we briefly review the formalism for Dirac spinors on a general curved space-time (see, for example, \cite{Brill:1957fx} for more details).

The Dirac equation for fermions of mass $m$ which is covariant with respect to general
coordinate transformations takes the form:
\begin{equation}
 (i \gamma^\halpha D_\halpha - m) \psi(x) = 0,
 \label{eq:dirac}
\end{equation}
where $\psi(x)$ is a spinor with four components.
Throughout this paper, we use the mostly $+$ convention for the metric signature and
Planck units such that $G=c=\hbar=k_B=1$.

In \eqref{eq:dirac}, hatted indices denote tensor components with respect to an orthonormal tetrad,
comprising the frame vectors $e_\halpha = e_\halpha^\mu \partial_\mu$ and coframe one-forms
$\omega^\halpha = \omega^\halpha_\mu dx^\mu$, chosen such that:
\begin{equation}
 g_{\mu\nu} = \eta_{\halpha\hbeta} \omega^\halpha_\mu \omega^\hbeta_\nu, \qquad
 \braket{\omega^\halpha, e_\hbeta} \equiv \omega^\halpha_\mu e^\mu_\hbeta = \delta^\halpha{}_\hbeta,
\end{equation}
where $g_{\mu\nu}$ are the metric components of the background geometry, while
$\eta_{\halpha\hbeta} = {\rm diag}(-1, 1,1, 1)$ is the Minkowski metric.
We employ the Dirac representation for the gamma matrices $\gamma^\halpha$, such that:
\begin{equation}
 \gamma^\hatt =
 \begin{pmatrix}
  1 & 0 \\
  0 & -1
 \end{pmatrix},\qquad
 \gamma^\hati =
 \begin{pmatrix}
  0 & \sigma_i \\
  -\sigma_i & 0
 \end{pmatrix},
 \label{eq:gamma}
\end{equation}
where $\sigma_i$ are the Pauli matrices:
\begin{equation}
 \sigma_1 =
 \begin{pmatrix}
  0 & 1 \\ 1 & 0
 \end{pmatrix},\qquad
\sigma_2 =
\begin{pmatrix}
 0 & -i \\ i & 0
\end{pmatrix},\qquad
\sigma_3 =
\begin{pmatrix}
 1 & 0 \\
 0 & -1
\end{pmatrix}.
\end{equation}
The gamma matrices satisfy the canonical anticommutation relations:
\begin{equation}
 \{\gamma^\halpha, \gamma^\hbeta\} \equiv \gamma^\halpha \gamma^\hbeta + \gamma^\hbeta \gamma^\halpha = -2\eta^{\halpha\hbeta}.
\end{equation}

The spinor covariant derivative
\begin{equation}
D_\halpha \psi = e_\halpha^\mu \partial_\mu \psi - \Gamma_\halpha \psi
\end{equation}
involves the spinor connection $\Gamma_\halpha$, which has the following expression:
\begin{equation}
 \Gamma_\halpha = \frac{1}{2} \Gamma_{\hbeta\hgamma\halpha} \Sigma^{\hbeta\hgamma}.
 \label{eq:spincon}
\end{equation}
In \eqref{eq:spincon}, $\Sigma^{\hbeta\hgamma}$ are the anti-Hermitian generators of Lorentz transformations:
\begin{equation}
 \Sigma^{\hbeta\hgamma} = \frac{1}{4} [\gamma^\hbeta, \gamma^\hgamma],
\end{equation}
while the connection coefficients $\Gamma_{\hbeta\hgamma\halpha}$ can be obtained using the following formula:
\begin{equation}
 \Gamma_{\hbeta\hgamma\halpha} = \frac{1}{2}(c_{\hbeta\hgamma\halpha} + c_{\hbeta\halpha\hgamma} - c_{\hgamma\halpha\hbeta}).
 \label{eq:conncoef}
\end{equation}
The expression \eqref{eq:conncoef} is written in terms of the Cartan coefficients $c_{\halpha\hbeta}{}^\hgamma$, which can be obtained
from the expression for the commutator of the tetrad vectors:
\begin{equation}
 c_{\halpha\hbeta}{}^\hgamma = \braket{\omega^\hgamma, [e_\halpha, e_\hbeta]} \equiv
 \omega^\hgamma_\mu (e_\halpha^\nu \partial_\nu e_\hbeta^\mu - e_\hbeta^\nu \partial_\nu e_\halpha^\mu).
 \label{eq:Cartancoeff}
\end{equation}

The classical fermion charge current (CC) has tetrad components
\begin{equation}
J^{\halpha } = {\overline {\psi }} \gamma ^{\halpha } \psi ,
\end{equation}
where the conjugate spinor ${\overline {\psi }}$ is defined by ${\overline {\psi }}=\psi ^{\dagger } \gamma ^{\hatt} $.
The classical SET for the spinor field has tetrad components \cite{Brill:1957fx}
\begin{equation}
T_{\halpha \hbeta } = \frac{i}{2} \left\{ {\overline {\psi }} \gamma _{(\halpha } D_{\hbeta )} \psi - \left[ D_{(\halpha } {\overline {\psi }}\, \right]  \gamma _{\hbeta )} \psi
 \right\} ,
\end{equation}
where the covariant derivative of the conjugate spinor is
\begin{equation}
D_{\halpha }{\overline {\psi }} = e_\halpha^\mu \partial_\mu {\overline {\psi }} +{\overline {\psi }} \Gamma_\halpha .
\end{equation}

\section{Geometry of anti-de Sitter space-time}
\label{sec:geometry}

Our focus in this paper is Dirac fermions on adS space-time. In this section
we introduce the adS space-time
metric, tetrad and connection coefficients, as well as
the geodetic interval and its corresponding tangent vectors.

\subsection{Coordinate system and metric}
\label{sec:metric}

The adS space-time metric is given by:
\begin{equation}
ds^{2} = \frac {1}{(\cos \omega r)^2 } \left[ -dt^{2} + dr^{2}
+ \left(\frac {\sin \omega r}{\omega}\right)^2 \left( d\theta ^{2} + \sin ^{2} \theta \, d\varphi ^{2} \right) \right],
\label{eq:metric}
\end{equation}
where $\omega$ is the inverse radius of curvature,
$x^i = \{x, y, z\}$ and we use the standard spherical polar coordinates $\{ r, \theta ,\varphi \}$ with $r = \sqrt{x^2 + y^2 + z^2}$.
While pure adS space-time is periodic in time $t$, in this paper we consider the covering space
of adS, where $t \in (-\infty, \infty)$. The radial coordinate $r$ is restricted to take values between
$r = 0$ and $r = \pi / 2\omega$, where the boundary of the space-time lies.
The Ricci scalar for the metric (\ref{eq:metric}) is $R = -12\omega^2$.

The metric \eqref{eq:metric} admits the following Cartesian-gauge tetrad \cite{art:cotaescu07}:
\begin{eqnarray}
 e_{\hatt} = \cos \omega r \, \partial_t, \qquad
 e_{\hat{i}} = \cos \omega r \left[ \frac{\omega r}{\sin \omega r} \left(
 \delta_{ij} - \frac{x^ix^j}{r^2}\right) + \frac{x^ix^j}{r^2}\right]
 \label{eq:frame}
 \partial_j,\\
 \omega^\hatt = \frac{dt}{\cos\omega r},\qquad
 \omega^{\hat{i}} = \frac{1}{\cos\omega r}\left[\frac{\sin \omega r}{\omega r}
 \left(\delta_{ij} - \frac{x^ix^j}{r^2}\right) + \frac{x^ix^j}{r^2}\right]dx^j.
 \label{eq:coframe}
\end{eqnarray}
The nonvanishing Cartan coefficients (\ref{eq:Cartancoeff}) corresponding to the tetrad (\ref{eq:frame}--\ref{eq:coframe}) are:
\begin{equation}\label{eq:cartan}
 c_{\hatt\hat{i}}{}^\hatt = \omega (\sin \omega r) \frac{x^i}{r},\quad
 c_{\hat{i}\hat{j}}{}^{\hat{k}} =
 -\omega \tan\frac{\omega r}{2}
 \left(\frac{x^i}{r} \delta_{jk} - \frac{x^j}{r} \delta_{ik}\right),
\end{equation}
while the ensuing nonvanishing connnection coefficients (\ref{eq:conncoef}) are:
\begin{equation}\label{eq:connection}
 \Gamma^\hatt{}_{\hati\hatt} = \omega (\sin \omega r) \frac{x^i}{r}, \quad
 \Gamma^\hati{}_{\hatj\hatk} = -\omega \tan\frac{\omega r}{2}
 \left(\frac{x^i}{r} \delta_{jk} - \frac{x^j}{r} \delta_{ik}\right).
\end{equation}
Finally, the spin connection coefficients $\Gamma_\halpha$ (\ref{eq:spincon}) can be computed:
\begin{eqnarray}
 \Gamma_\hatt = -\frac{\omega}{2} (\sin \omega r) \gamma^{\hatt}
 \left(\frac{\vx \cdot \vgamma}{r}\right), \nonumber\\
 \Gamma_{\hat{k}} = \frac{\omega}{2} \tan\frac{\omega r}{2}
 \left[\frac{x^k}{r} + \gamma^\hatk \left(\frac{\vx\cdot\vgamma}{r}\right)\right],
 \label{eq:Gamma}
\end{eqnarray}
while their contraction with the $\gamma$ matrices (\ref{eq:gamma}) reads:
\begin{equation}
 \slashed{\Gamma} \equiv \gamma^\halpha \Gamma_\halpha =
 -\omega \tan\frac{\omega r}{2}  \left(1 + \cos^2\frac{\omega r}{2}\right)
 \left(\frac{\vx \cdot \vgamma}{r}\right).
 \label{eq:Gammaslash}
\end{equation}

\subsection{Geodesic structure}
\label{sec:geodesics}

On adS, the geodetic interval $s(x,x')$ between two points $x=(t,\vx)$ and $x'=(t',\vx')$ is given
by the following relation \cite{Allen:1986ty}:
\begin{equation}
 \cos(\omega s) = \frac{\cos\omega \Delta t}{\cos\omega r \cos\omega r'} - \cos\gamma \tan\omega r \tan\omega r',
 \label{eq:geodetic}
\end{equation}
where $\Delta t = t - t'$ and $\cos\gamma = \vx \cdot \vx' / rr'$ represents the cosine of the
angle between $\vx$ and $\vx'$.

The tangent $n_{\mu}(x, x') = \nabla_{\mu} s(x,x')$ at $x$ to the
geodesic connecting the points $x$ and $x'$ has the following tetrad components
\begin{eqnarray}
 n_{\hatt} = \frac{\sin\omega \Delta t}{\sin\omega s \cos\omega r'},\nonumber\\
 n_{\hati} = -\frac{x^i}{r} \frac{\cos\omega \Delta t \sin\omega r - \cos\gamma \sin\omega r'(1-\cos\omega r)}
 {\sin\omega s \cos\omega r \cos\omega r'}
 + \frac{\tan\omega r'}{\sin\omega s} \frac{x'{}^i}{r'},
 \label{eq:ntetrad}
\end{eqnarray}
while the components of the tangent $n_{\mu'} (x, x') = \nabla_{\mu'} s$ at $x'$
can be obtained from (\ref{eq:ntetrad}) by swapping the coordinates $x^\mu$ and $x'{}^\mu$.
On maximally symmetric space-times,
the tangents $n_{\mu}$ and $n_{\mu'}$ obey the following relations (which are not identical in form to those in \cite{art:allen86a} as we are using different conventions):
\begin{eqnarray}
 \nabla_{\halpha} n_{\hbeta} = -A (\eta _{\halpha\hbeta} + n_{\halpha} n_{\hbeta}), \label{eq:nn}\\
 \nabla_{\halpha'} n_{\hbeta} = -C (g_{\halpha'\hbeta} - n_{\halpha'} n_{\hbeta}), \label{eq:npn}\\
 \nabla_{\hgamma} g_{\halpha\hbeta'} = -(A + C) (\eta_{\hgamma\halpha} n_{\hbeta'} +
 g_{\hgamma \hbeta'} n_{\halpha}),
 \label{eq:nbivec}
\end{eqnarray}
where $A$ and $C$ are given on adS by \cite{art:allen86a}:
\begin{equation}
 A = \omega \cot\omega s, \qquad C = -\frac{\omega}{\sin\omega s},
 \label{eq:AC}
\end{equation}
and $g_{\halpha\hbeta'}$ is the bivector of parallel transport,
which will be introduced in the next section.

\section{Parallel transport on adS space-time}
\label{sec:parallel}

In order to compute t.e.v.s in section \ref{sec:thermal}, we  first require the vacuum Feynman Green's function for the Dirac field $S_{F}(x,x')$, which we construct in section \ref{sec:feynman}.
The form of $S_{F}(x,x')$ derived in \cite{art:muck00} involves the bispinor of parallel transport $\Lambda (x,x')$ (see (\ref{eq:sf_muck})).
The t.e.v.~of the SET (\ref{eq:ev_set}) also requires the bivector of parallel transport $g_{\mu \nu'}$.
In this section we therefore derive analytic expressions for the bivector and bispinor of parallel transport on
adS space-time.

\subsection{Bivector of parallel transport} \label{sec:bivector}

The bivector of parallel transport $g_{\mu\nu'}(x,x')$ is defined such that it performs the
parallel transport of a tensor index $\nu'$ from $x'$ to $x$.
For example, for a vector field $V^{\mu'}(x')$, the following
relation holds \cite{Christensen:1976vb}:
\begin{equation}
 V^\mu_{||}(x) = g^\mu{}_{\nu'} V^{\nu'}(x'),
\end{equation}
where $V^\mu_{||}(x)$ represents the vector $V^{\nu'}(x')$ evaluated at $x'$, which is then  parallel-transported to $x$ along the
geodesic connecting these two points. Thus $g_{\mu \nu'}$ satisfies the parallel transport equation \cite{poisson}:
\begin{equation}
 n^\lambda \nabla_\lambda g_{\mu\nu'} = 0,\qquad
 n^{\lambda'}\nabla_{\lambda'} g_{\mu\nu'} = 0,
\end{equation}
where the last equation is a statement that $g_{\mu\nu'}$ performs parallel transport both ways (from $x$ to
$x'$, as well as from $x'$ to $x$).
The bivector of parallel transport also satisfies the conditions:
\begin{equation}
 g^\mu{}_{\nu'} g^{\nu'}{}_\lambda = \delta^\mu{}_\lambda, \qquad
 g^{\mu'}{}_\lambda g^\lambda{}_{\nu'} = \delta^{\mu'}{}_{\nu'}.
\end{equation}

On adS space-time, \eqref{eq:npn} can be used to obtain an expression for the tetrad
components $g_{\halpha\hbeta'}$ of the bivector of parallel transport:
\begin{equation}
 g_{\halpha'\hbeta} = n_{\halpha'} n_\hbeta + \frac{\sin\omega s}{\omega} \nabla_{\halpha'} n_\hbeta.
\end{equation}
Taking into account that all tetrad components \eqref{eq:ntetrad}
of $n_\hbeta$ have a factor of $\sin\omega s$ in their denominator, it is convenient to
introduce the reduced bivector of parallel transport $\widetilde{g}_{\halpha'\hbeta}$, defined by:
\begin{equation}
 \widetilde{g}_{\halpha'\hbeta} = g_{\halpha'\hbeta} - n_{\halpha'}n_\hbeta (1 - \cos\omega s)
 = \frac {1}{\omega }\nabla_{\halpha'} (n_\hbeta \sin\omega s).
 \label{eq:gtilde}
\end{equation}
Since $n_\hbeta(x,x')$ is a vector at $x$ and a scalar at $x'$, the covariant
derivative $\nabla_{\halpha'}$ acts as a simple differential operator:
\begin{equation}
 \widetilde{g}_{\halpha'\hbeta} = \frac{1}{\omega } e_{\halpha'}^{\mu'} \partial_{\mu'} [n_\hbeta(x,x') \sin\omega s].
\end{equation}
The derivative with respect to $x'{}^\mu$ only acts on the primed coordinates in $n_\hbeta \sin\omega s$.
Working through the algebra, we find:
\begin{eqnarray}
\fl \widetilde{g}_{\hatt'\hatt} &=& -\cos\omega \Delta t, \nonumber \\
\fl \widetilde{g}_{\hatt'\hati} &=& -\frac{x^i}{r} \sin\omega\Delta t \tan\omega r,\nonumber \\
\fl \widetilde{g}_{\hati'\hatt} &=& \frac{x'{}^i}{r'} \sin\omega\Delta t \tan\omega r',\nonumber \\
\fl \widetilde{g}_{\hati'\hatj} &=& \delta_{ij} + \frac{1-\cos\omega r}{\cos\omega r} \frac{x^i x^j}{r^2} +
 \frac{1 -\cos\omega r'}{\cos\omega r'} \frac{x'{}^i x'{}^j}{(r')^2} \nonumber\\
\fl & & - \frac{x'{}^i x^j}{r' r} \left(\cos\omega\Delta t \tan \omega r \tan\omega r' -
 \cos\gamma \frac{1- \cos\omega r}{\cos\omega r} \frac{1 - \cos\omega r'}{\cos\omega r'}\right),
\end{eqnarray}
from which it is straightforward to calculate $g_{\halpha '\hbeta }$ using (\ref{eq:gtilde}).

\subsection{Bispinor of parallel transport}
\label{sec:bispinor}

In analogy to the role played by the bivector of parallel transport, the bispinor of parallel transport
$\Lambda(x,x')$ performs the parallel transport of spinors, such that \cite{art:muck00,art:christensen78,art:groves02}:
\begin{equation}\label{eq:Lambda_def}
 \psi_{||}(x) = \Lambda(x,x') \psi(x'),
\end{equation}
where $\psi_{||}(x)$ represents the spinor $\psi(x')$ evaluated at $x'$, parallel-transported to $x$ along the
geodesic connecting these two points. The bispinor $\Lambda(x,x')$
must satisfy the parallel transport equations for spinors \cite{art:muck00}:
\begin{equation}
\label{eq:lambda_pt}
 n^\halpha D_\halpha \Lambda(x,x') = 0, \qquad n^{\halpha'} D_{\halpha'} \Lambda(x,x') = 0,
\end{equation}
where
\begin{equation}
D_{\halpha'} \Lambda(x,x') \equiv
e_{\halpha'}^{\mu'} \partial_{\mu'} \Lambda(x,x')
+ \Lambda(x,x') \Gamma_{\halpha'}(x')
\end{equation}
is the spinor covariant derivative of $\Lambda (x,x')$ with respect to the coordinate $x'$.
The initial conditions for (\ref{eq:lambda_pt}) are \cite{art:muck00}:
\begin{equation}\label{eq:lambda_ic}
 \Lambda(x,x) = 1, \qquad \Lambda^{-1}(x,x') = \overline{\Lambda}(x,x') = \Lambda(x',x),
\end{equation}
where the first equation implies that parallel transport from $x$ to the same point $x$ is the identity
operation, while the second equation ensures that no parallel transport is performed on scalars of the form
$\overline{\chi} \psi$.
The parallel transport of the $\gamma$ matrices is given by:
\begin{equation}
\label{eq:gamma_pt}
 \Lambda(x,x')\gamma^{\mu'} = g^{\mu'}{}_\nu \gamma^\nu \Lambda(x,x').
\end{equation}

On adS space-time, equations  (\ref{eq:lambda_pt}, \ref{eq:gamma_pt}) can be used to
obtain the following equations \cite{art:muck00}:
\begin{eqnarray}
 D_{\halpha} \Lambda(x,x') = -\frac{A+C}{2}(n_\halpha + \gamma_\halpha \slashed{n}) \Lambda(x,x'),\nonumber \\
 D_{\halpha'} \Lambda(x,x') = -\frac{A+C}{2} \Lambda(x,x') (n_{\halpha'} + \slashed{n}' \gamma_{\halpha'}),
 \label{eq:dLambda}
\end{eqnarray}
where $\slashed{n} = \gamma^\halpha n_\halpha$.
Using the expressions given in \eqref{eq:AC} for $A$ and $C$,
(\ref{eq:dLambda}) can be written as:
\begin{eqnarray}
 D_{\halpha} \Lambda(x,x') = \frac{\omega}{2} \tan\left( \frac{\omega s}{2} \right) (n_\halpha + \gamma_\halpha \slashed{n}) \Lambda(x,x'),
 \nonumber\\
 D_{\halpha'} \Lambda(x,x') = \frac{\omega}{2} \tan \left( \frac{\omega s}{2} \right) \Lambda(x,x')
 (n_{\halpha'} + \slashed{n}' \gamma_{\halpha'}).
 \label{eq:DLambda}
\end{eqnarray}
Since constructing the solution $\Lambda (x,x')$ of (\ref{eq:DLambda}) is a rather lengthy task, we
simply present the result here and refer the reader to the Appendix for details of the derivation:
\begin{eqnarray}
\label{eq:Lambda}
 \fl \Lambda(x,x') = \frac{\sec\frac{\omega s}{2}}{\sqrt{\cos\omega r \cos\omega r'}} \Bigg[
 \cos\frac{\omega \Delta t}{2} \left(
 \cos\frac{\omega r}{2} \cos\frac{\omega r'}{2} +
 \sin\frac{\omega r}{2} \sin\frac{\omega r'}{2}
 \frac{\vx \cdot \vgamma}{r} \frac{\vx' \cdot \vgamma}{r'}\right)\nonumber\\
 + \sin\frac{\omega \Delta t}{2} \left(
 \sin\frac{\omega r}{2} \cos\frac{\omega r'}{2} \frac{\vx \cdot\vgamma}{r} \gamma^\hatt
 +\sin\frac{\omega r'}{2} \cos\frac{\omega r}{2} \frac{\vx' \cdot \vgamma}{r'} \gamma^\hatt\right)\Bigg].
\end{eqnarray}
In computing the t.e.v.s in section \ref{sec:thermal}, we will require the quantity $\slashed{n} \Lambda$, which is also derived in the Appendix and given by:
\begin{eqnarray}
\label{eq:Lambdan}
 \fl \slashed{n} \Lambda(x,x')
 \nonumber \\
\fl \qquad = \frac{ {\rm cosec} \, \frac{\omega s}{2}}{\sqrt{\cos\omega r \cos\omega r'}} \Bigg[
 \sin\frac{\omega \Delta t}{2} \left(
 \cos\frac{\omega r}{2} \cos\frac{\omega r'}{2}\gamma^\hatt  -
 \sin\frac{\omega r}{2} \sin\frac{\omega r'}{2}
 \frac{\vx \cdot \vgamma}{r} \frac{\vx' \cdot \vgamma}{r'}\gamma^\hatt \right)\nonumber\\
 - \cos\frac{\omega \Delta t}{2} \left(
 \sin\frac{\omega r}{2} \cos\frac{\omega r'}{2} \frac{\vx \cdot \vgamma}{r} -
 \cos\frac{\omega r}{2} \sin\frac{\omega r'}{2} \frac{\vx' \cdot \vgamma}{r'}\right)\Bigg].
\end{eqnarray}

\section{Feynman Green's function for the global maximally-symmetric vacuum}\label{sec:feynman}

In this section, we review
the construction of the Feynman Green's function $S_F(x,x')$ for
the global vacuum state of the Dirac field on adS space-time \cite{art:muck00,art:ambrus15plb}.

The Feynman Green's function $S_F(x,x')$ satisfies the inhomogeneous Dirac equation:
\begin{equation}
 \left( i \slashed{D} - m \right) S_F(x,x') = \frac{1}{\sqrt{-g}} \delta^4(x - x').
 \label{eq:sf_eq}
\end{equation}
Due to the maximal symmetry of adS space-time, the fermion Feynman Green's function $S_F(x,x')$ for the global adS vacuum state can be written as \cite{art:muck00}:
\begin{equation}
i S_F(x,x') = ({\mathcal {A}}_F + {\mathcal {B}}_F \slashed{n}) \Lambda(x,x'),
\label{eq:sf_muck}
\end{equation}
where ${\mathcal {A}}_F$ and ${\mathcal {B}}_F$ are scalar functions of the geodetic interval $s$.
Substituting the ansatz \eqref{eq:sf_muck} into \eqref{eq:sf_eq} and using the properties
 (\ref{eq:nn}, \ref{eq:DLambda})  of the derivatives of
$n_\mu$ and $\Lambda (x,x')$, we can reduce \eqref{eq:sf_eq} to two coupled equations:
\begin{eqnarray}
 i\omega \frac{\partial {\mathcal {A}}_F}{\partial \left( \omega s \right) } - \frac{3i\omega}{2} {\mathcal {A}}_{F} \tan \left( \frac{\omega s}{2} \right)
 - m {\mathcal {B}}_F = 0,
 \label{eq:beta_from_alpha} \\
 i \omega \frac{\partial {\mathcal {B}}_F}{\partial \left( \omega s \right)} + \frac{3i\omega}{2} {\mathcal {B}}_{F} \cot \left( \frac{\omega s}{2} \right)
 - m {\mathcal {A}}_F = \frac{i}{\sqrt{-g}} \delta(x,x').
 \label{eq:alpha_from_beta}
\end{eqnarray}
The solution of the system (\ref{eq:beta_from_alpha}--\ref{eq:alpha_from_beta}) can be written as \cite{art:ambrus15plb}:
\begin{eqnarray}
\fl {\mathcal {A}}_{F} = \frac {\omega ^{3}\Gamma \left( 2 + k \right)}{16 \pi ^{\frac {3}{2}} 4^{k} \Gamma \left( \frac {1}{2} + k \right) }
\cos \left( \frac {\omega s}{2} \right) \left[
-\sin^{2} \left( \frac {\omega s}{2} \right) \right] ^{-2-k}
\nonumber\\
\times
{}_{2}F_{1}\left( 1+k, 2+k; 1+2k ; {\rm cosec} ^{2} \left( \frac {\omega s}{2} \right) \right),
\label{eq:a_z}
\end{eqnarray}
where $k$ is given in terms of the fermion mass $m$ and inverse radius of curvature $\omega $ by
\begin{equation}
k = \frac {m}{\omega }
\end{equation}
and ${}_{2}F_{1}(a,b;c;z)$ is a hypergeometric function.
Inserting the expression (\ref{eq:a_z}) for ${\mathcal {A}}_F$ in \eqref{eq:beta_from_alpha} gives:
\begin{eqnarray}
 \fl {\mathcal {B}}_F = \frac {i\omega ^{3}\Gamma \left( 2 + k \right)}{16 \pi ^{\frac {3}{2}} 4^{k} \Gamma \left( \frac {1}{2} + k \right) }
\sin \left( \frac {\omega s}{2} \right) \left[
-\sin^{2} \left( \frac {\omega s}{2} \right) \right] ^{-2-k}\nonumber\\
\times {}_{2}F_{1}\left(k, 2+k; 1+2k ; {\rm cosec} ^{2} \left( \frac {\omega s}{2} \right) \right).
 \label{eq:b_z}
\end{eqnarray}
The expressions (\ref{eq:a_z}, \ref{eq:b_z}) simplify considerably in the massless case ($k = 0$):
\begin{equation}
 {\mathcal {A}}_F\rfloor_{k = 0} = \frac{\omega^3}{16\pi^2} \left(\cos\frac{\omega s}{2}\right)^{-3}, \qquad
 {\mathcal {B}}_F\rfloor_{k = 0} = \frac{i\omega^3}{16\pi^2} \left(\sin\frac{\omega s}{2}\right)^{-3}.
\end{equation}

\section{Thermal expectation values}\label{sec:thermal}

We now have all the ingredients necessary for a computation of the t.e.v.s of the FC $\braket{\psibar \psi}$, CC $\braket{J^{\halpha}}$ and SET $\braket{T_{\halpha\hbeta}}$ for a massive fermion field at inverse temperature $\beta $.
The required quantities are: the exact expression \eqref{eq:Lambda} for the bispinor of parallel transport and
the functions ${\mathcal {A}}_F$ \eqref{eq:a_z} and ${\mathcal {B}}_F$ \eqref{eq:b_z} necessary to construct the vacuum Feynman Green's
function \eqref{eq:sf_muck}.

\subsection{Finite temperature Feynman Green's function}\label{sec:beta}

The expectation values of the FC $\braket{\psibar \psi}$, CC $\braket{J^\halpha}$
and SET $\braket{T_{\halpha\hbeta}}$ for a given state can be calculated from the Feynman Green's function $S_{F}(x,x')$
corresponding to that state using the following expressions \cite{art:ambrus15plb,art:groves02}:
\begin{eqnarray}
\fl \braket{\psibar \psi} = -\lim_{x'\rightarrow x} \tr \left[ i S_F(x,x')
 \Lambda(x',x)\right],
 \label{eq:ev_ppsi}\\
\fl \braket{J^\halpha} = -\lim_{x'\rightarrow x} \tr\left[\gamma^\halpha i S_F(x, x')
 \Lambda(x',x)\right],
 \label{eq:ev_j}\\
\fl \braket{T_{\halpha\hbeta}} = \frac{i}{2} \lim_{x'\rightarrow x} \tr\left\{
 \left[\gamma_{(\halpha} D_{\hbeta)} iS_F(x, x') -
D_{\hbeta'} [iS_F(x,x')]
 \gamma_{\halpha'} g^{\halpha'}{}_{(\halpha}
 g^{\hbeta'}{}_{\hbeta)}\right] \Lambda(x',x)\right\}.
 \label{eq:ev_set}
\end{eqnarray}
Taking the trace of (\ref{eq:ev_set}), since the fermion Feynman Green's function satisfies the inhomogeneous Dirac equation (\ref{eq:sf_eq}), we find the following relation between the unrenormalized FC and the SET trace
\begin{equation}
\braket{T_{\halpha}^{\halpha}} = -m\braket{\psibar \psi}.
\label{eq:trace}
\end{equation}
This relationship is not preserved during the renormalization process.
In particular, (\ref{eq:trace}) does not hold for the v.e.v.s
of the FC and SET calculated in \cite{art:ambrus15plb}, regardless of the method of renormalization.

To compute the expectation values (\ref{eq:ev_ppsi}--\ref{eq:ev_set}) at finite temperature, the thermal Feynman Green's function can be constructed
as follows \cite{birrell82}:
\begin{equation}
 S_F^\beta(x, x') = \sum_{j} (-1)^j S_F(t + ij\beta, \vx; t', \vx'),
\end{equation}
where $j$ runs over all integers, so that $S_{F}^{\beta }(x,x')$ is
anti-periodic in imaginary time, with period $\beta$.
It is convenient to introduce the following notation:
\begin{equation}
 S_F^\beta(x,x') = \sum_j (-1)^j[{\mathcal {A}}_F(s_j) + {\mathcal {B}}_F(s_j) \slashed{n}_j] \Lambda_j(x_{j},x'),
 \label{eq:sf_beta}
\end{equation}
where the quantity $s_{j}$ is the geodetic interval between the points $x_{j}=(t+ij\beta , \vx)$ and $x'=(t',\vx')$, the vector $n_{j}$ is the corresponding tangent vector and $\Lambda _{j}(x_{j},x')$ is the bispinor of parallel transport between these two points.

The v.e.v.s of the FC, CC and SET when the fermion field is in the global adS vacuum were calculated in \cite{art:ambrus15plb}. In this section we compute the difference between the
t.e.v.s at finite inverse temperature $\beta $ and the v.e.v.s.  Since the short-distance singularity structure of the Feynman Green's function is independent of the state of the quantum field (see, for example, \cite{art:ambrus15plb}), these differences do not require renormalization and the relationship (\ref{eq:trace}) between the FC and the trace of the SET will hold.

The $j=0$ term in (\ref{eq:sf_beta}) is the vacuum Feynman Green's function.
Therefore, to find the differences between the t.e.v.s and the v.e.v.s, we simply subtract the $j=0$ term from (\ref{eq:sf_beta}), and substitute into (\ref{eq:ev_ppsi}--\ref{eq:ev_set}) as applicable.
The bispinor of parallel transport written explicitly in (\ref{eq:ev_ppsi}--\ref{eq:ev_set}) is between the points $x=(t,\vx )$ and $x'=(t',\vx ')$ as the shift $s\rightarrow s_{j}$ occurs only in the thermal Feynman Green's function (\ref{eq:sf_beta}).
Since the quantities resulting from (\ref{eq:ev_ppsi}--\ref{eq:ev_set}) do not require renormalization, we can then straightforwardly bring the space-time points together by setting $\vx ' =\vx $ and $t'=t$.
However, in doing so $s_{j}$ will remain nonzero because it will be the geodetic interval between the points $x_{j}=(t+ij\beta ,\vx )$ and $x=(t,\vx )$.
Similarly, $\Lambda _{j}(x_{j},x)$ will be nontrivial.

We use the notation $\braket{:\psibar \psi:}_\beta $,   $\braket{:J^\halpha:}_\beta$ and $\braket{:T_{\halpha\hbeta}:}_\beta$ to denote these differences in expectation values for the FC, CC and SET respectively.
The full t.e.v.s for each quantity can easily be found from the results in this section by simply adding the v.e.v.s of the corresponding quantity from \cite{art:ambrus15plb}\footnote{The renormalized expectation value of the FC reported in equations (38a, 51a) of \cite{art:ambrus15plb} must be multiplied by $-1$.}.

\subsection{Fermion condensate}\label{sec:ppsi}

Subtracting the $j=0$ term from \eqref{eq:sf_beta} and inserting into \eqref{eq:ev_ppsi},
the following expression is obtained for the difference between the t.e.v.~and the v.e.v.~of the FC:
\begin{eqnarray}
 FC_{\beta } =\braket{:\psibar \psi:}_\beta = -\sum_{j \neq 0} (-1)^j \lim_{\vx'\rightarrow \vx\atop \Delta t \rightarrow ij\beta} {\mathcal {A}}_F(s_j)
 \tr[\Lambda_j(x_{j},x')],
 \label{eq:tev_ppsi_ab}
\end{eqnarray}
where we have set $\Lambda (x,x)=1$ in \eqref{eq:ev_ppsi}.
When $\vx' = \vx$, we find from \eqref{eq:Lambda} that:
\begin{equation}
\fl \Lambda_{j}(x_{j},x')\rfloor_{\vx' = \vx} = \frac{\sec\frac{\omega s_{j}}{2}}{\cos\omega r}
 \left(\cos\frac{\omega \Delta t_j}{2} \cos\omega r +
 \sin\frac{\omega \Delta t_j}{2} \sin\omega r \frac{\vx \cdot \vgamma}{r} \gamma^{\hatt}\right),
 \label{eq:xx_Lambda}
\end{equation}
where $\Delta t_j = t-t'+ij\beta $.
The difference between the t.e.v.~and the v.e.v.~of the FC then becomes, using (\ref{eq:a_z}):
\begin{eqnarray}
\fl FC_{\beta } = \braket{:\psibar \psi:}_\beta = -\sum_{j \neq 0} (-1)^j
 \frac {\omega ^{3} \cos\left( \frac{\omega \Delta t_j}{2} \right) \Gamma \left( 2 + k \right)}
 {\pi ^{3/2} 4^{1+k}
 \Gamma \left( \frac {1}{2} + k \right) }
 \left[-\sin^{2} \left( \frac {\omega s_{j}}{2} \right) \right] ^{-2-k}\nonumber\\
 \times {}_{2}F_{1}\left( 1+k, 2+k; 1+2k ; {\rm cosec} ^{2} \left( \frac {\omega s_{j}}{2} \right) \right).
\label{eq:tev_ppsi_aux}
\end{eqnarray}
To further simplify the above expression, the limit $\vx' = \vx$ of the geodetic interval (\ref{eq:geodetic}) gives:
\begin{equation}
 \cos\omega s_{j}\rfloor_{\vx' = \vx} = 1 - \frac{2\sin^2 \frac{\omega \Delta t_j}{2}}{\cos^2\omega r},\qquad
 \left.\sin^2\frac{\omega s_{j}}{2}\right\rfloor_{\vx' = \vx} = \frac{\sin^2\frac{\omega \Delta t_j}{2}}{\cos ^{2} \omega r}.
 \label{eq:xx_s}
\end{equation}
Setting $t'=t$, we then have  $\Delta t_j = ij\beta$ and \eqref{eq:tev_ppsi_aux} reduces to:
\begin{eqnarray}
\fl FC_{\beta } = \braket{:\psibar\psi:}_\beta = -\frac{2\omega^3 \Gamma(2+k) (\cos\omega r)^{4 + 2k}}{\pi^{3/2}4^{1+k}\Gamma(\frac{1}{2} + k)}
 \sum_{j = 1}^\infty (-1)^j
 \frac{\cosh\frac{\omega j \beta}{2}}{(\sinh\frac{\omega j \beta}{2})^{4+2k}}\nonumber\\
\times {}_2F_1\left(1+k, 2+k; 1+2k; -\frac{\cos^2\omega r}{\sinh^2 \frac{\omega j \beta}{2}}\right) .
\label{eq:ppsi}
\end{eqnarray}
In the massless case ($k=0$), this simplifies even further to the nonzero expression:
\begin{equation}
 FC_{\beta } = \braket{:\psibar\psi:}_\beta = -\frac{\omega^3 (\cos\omega r)^4}{2\pi^2}
 \sum_{j  = 1}^{\infty} (-1)^j \frac{\cosh\frac{\omega j \beta}{2}}{(\sinh^2\frac{\omega j \beta}{2} + \cos^2\omega r)^{2}}.
 \label{eq:qft_fc_mo}
\end{equation}

\subsection{Charge current}
\label{sec:j}

The difference between the t.e.v.~and the v.e.v.~for the CC has the following expression:
\begin{equation}
 \braket{:J^\halpha:}_\beta = -\sum_{j \neq 0} \lim_{\vx'\rightarrow \vx \atop \Delta t \rightarrow ij\beta}
 {\mathcal {B}}_F(s_j) \tr[\gamma^\halpha \slashed{n}_j \Lambda_j(x_{j},x')],
 \label{eq:tev_j_ab}
\end{equation}
where again we have used the fact that $\Lambda(x,x) = 1$.
To evaluate the above trace, we note that $\slashed{n}_j \Lambda_j(x_{j},x')$ (\ref{eq:Lambdan}) reduces in the
limit $\vx' \rightarrow \vx$ to:
\begin{equation}
 \slashed{n}_j \Lambda_j(x_{j},x')\rfloor_{\vx' = \vx} =
 \frac{\sin(\omega \Delta t_j/2)}{\cos\omega r \sin(\omega s_j / 2)}
 \gamma^{\hatt}.
 \label{eq:xx_nLambda}
\end{equation}
Inserting the result \eqref{eq:xx_nLambda} into \eqref{eq:tev_j_ab} and using \eqref{eq:xx_s} to eliminate the
geodetic interval gives:
\begin{eqnarray}
 \fl \braket{:J^\halpha:}_\beta =
 -\eta^{\halpha\hatt} \frac{\omega^3 \Gamma(2+k)(\cos\omega r)^{3 + 2k}}
 {\pi^{3/2} 4^{1+k} \Gamma(\frac{1}{2} + k)}
 \sum_{j \neq 0} \frac{(-1)^j \sinh \frac{\omega j \beta}{2}}{\left( \sinh ^{2}\frac{\omega j \beta}{2}\right) ^{2+k}}\nonumber\\
 \times {}_2F_1\left(k, 2+k; 1+2k; -\frac{\cos^2\omega r}{\sinh^2 \frac{\omega j \beta}{2}}\right),
\end{eqnarray}
where we have also used (\ref{eq:b_z}).
It can be seen that $\braket{:J^\halpha:}_\beta$ vanishes, since the summand above is odd with respect
to $j\rightarrow -j$.

This result is not unexpected. In \cite{art:ambrus15plb} we found that the v.e.v.~of the CC vanished when the fermion field is in the global adS vacuum.
In a thermal state, we would anticipate that particle and anti-particle configurations would be equally populated, resulting in a net vanishing expectation value for the CC.

\subsection{Stress-energy tensor}
\label{sec:set}

Before attempting to compute the difference between the t.e.v.~and the v.e.v.~of the SET, we use \eqref{eq:DLambda} to show that the covariant derivative of the vacuum Feynman Green's function (\ref{eq:sf_muck}) takes the form
\begin{eqnarray}
\fl D_\hbeta \left[ i
 S_F(x,x') \right] = \omega\Bigg[
 \left(\frac{\partial {\mathcal {A}}_{F}}{\partial \left(\omega s \right) } + \frac{{\mathcal {A}}_{F}}{2} \tan\frac{\omega s}{2}\right) n_\hbeta +
 \frac{{\mathcal {A}}_{F}}{2}\tan\left(\frac{\omega s}{2}\right) \gamma_\hbeta \slashed{n} \nonumber\\
 + \left(\frac{\partial {\mathcal {B}}_{F}}{\partial \left(\omega s \right) } - \frac{{\mathcal {B}}_{F}}{2} \cot\frac{\omega s}{2}\right) n_\hbeta \slashed{n}
 - \frac{{\mathcal {B}}_{F}}{2}\cot\left(\frac{\omega s}{2}\right) \gamma_\hbeta\Bigg] \Lambda(x,x').
 \label{eq:DSFx}
\end{eqnarray}
Similarly, the derivative of $S_F(x,x')$ with respect to $x'$ can
be written as:
\begin{eqnarray}
\fl D_{\hbeta'} \left[ i
 S_F(x,x') \right] = \omega \Lambda(x,x')\Bigg[
 \left(\frac{\partial {\mathcal {A}}_{F}}{\partial \left(\omega s \right)} + \frac{{\mathcal {A}}_{F}}{2} \tan\frac{\omega s}{2}\right) n_{\hbeta'} +
 \frac{{\mathcal {A}}_{F}}{2}\tan\left(\frac{\omega s}{2}\right) \slashed{n}' \gamma_{\hbeta'} \nonumber\\
 - \left(\frac{\partial {\mathcal {B}}_{F}}{\partial \left(\omega s \right)} - \frac{{\mathcal {B}}_{F}}{2} \cot\frac{\omega s}{2} \right) \slashed{n}' n_{\hbeta'}
 + \frac{{\mathcal {B}}_{F}}{2}\cot\left(\frac{\omega s}{2}\right) \gamma_\hbeta\Bigg].
 \label{eq:DSFxp}
\end{eqnarray}
The results (\ref{eq:DSFx}--\ref{eq:DSFxp}) also hold for the case when $S_F(x,x')$ is replaced by $S_F(t+ij\beta , \vx ; t', \vx' )$ and $s$ is
replaced by $s_j$.
Using the following expressions:
\begin{eqnarray}
\fl
 \tr[\gamma_\halpha \slashed{n} \Lambda(x,x')]_{\vx' = \vx} =
 -\tr[\Lambda(x, x') \slashed{n}' \gamma_{\halpha'}]_{\vx' = \vx} =
 -\frac{4 \sin(\omega\Delta t/2)}{\cos\omega r \sin(\omega s / 2)} \delta^\hatt{}_{\halpha},
\end{eqnarray}
the difference between the t.e.v.~and the v.e.v.~of the SET can be written as:
\begin{eqnarray}
\fl \braket{:T_{\halpha\hbeta}:}_\beta = 2 i\omega \sum_{j \neq 0} (-1)^j
 \left[
 \eta_{\halpha\hbeta} \frac{\cos(\omega \Delta t_j/2)}{\sin(\omega s_j/2)} {\mathcal {B}}_{F} \right.\nonumber\\
 \left. -
 \frac{\sin(\omega \Delta t / 2)}{\cos\omega r \sin(\omega s/2)}
 \left(\frac{\partial {\mathcal {B}}_{F}}{\partial \left(\omega s_j\right) } - \frac{{\mathcal {B}}_{F}}{2}\cot\frac{\omega s_j}{2}\right)
 \delta^\hatt{}_{(\halpha}(n_{j\hbeta)} - n_{j\hbeta')})\right] ,
 \label{eq:tev_set_aux}
\end{eqnarray}
where $n_{j\hbeta }$ is the $\hbeta $ tetrad component of the vector $n_{j}$.
The primed indices in (\ref{eq:tev_set_aux}) were not parallel-transported to $x$, since the bivectors of parallel
transport in \eqref{eq:ev_set} reduce to Kronecker deltas when $x' = x$.
The coincidence limits of $n_{j\halpha}$ and $n_{j\halpha'}$ are:
\begin{eqnarray}
 n_{j\hatt}\rfloor_{\vx'=\vx} = -n_{j\hatt'}\rfloor_{\vx' = \vx} =
 \frac{\cos(\omega\Delta t_j/2)}{\cos(\omega s_{j}/2)}, \nonumber\\
 n_{j\hati}\rfloor_{\vx' = \vx} = n_{j\hati'}\rfloor_{\vx' = \vx} =
 \frac{x^i}{r} \frac{\sin(\omega \Delta t_j/2)}{\cos(\omega s_{j}/2)}\tan\omega r,
\end{eqnarray}
and hence the second term in \eqref{eq:tev_set_aux} contributes
only when $\halpha = \hbeta = t$, so that $\braket{:T_{\hatt \hati}:}_\beta = 0$.

Thus, the non-vanishing components of $\braket{:T_{\halpha\hbeta}:}_\beta$ are:
\begin{eqnarray}
 \braket{:T_{\hatt\hatt}:}_\beta = -2 i\omega \sum_{j \neq 0} (-1)^j
 \frac{\cos(\omega \Delta t_j/2)}{\cos(\omega s_j/2)}
 \left.\frac{\partial {\mathcal {B}}_{F}(s)}{\partial (\omega s/2)}\right\rfloor_{s = s_j},\\
 \braket{:T_{\hati\hat{\ell}}:}_\beta = 2 i\omega \delta_{i\ell} \sum_{j \neq 0} (-1)^j
 \frac{\cos(\omega \Delta t_j/2)}{\sin(\omega s_j/2)} {\mathcal {B}}_{F}(s_j).
\end{eqnarray}
The above results indicate that the difference between the t.e.v.~and the v.e.v.~of the SET corresponds to that of an ideal fluid \cite{rezzolla13}:
\begin{equation}
 \braket{:T^{\halpha\hbeta}:}_\beta = (E_\beta + P_\beta) u^\halpha u^\hbeta + \eta^{\halpha\hbeta} P_\beta,
 \label{eq:SETfluid}
\end{equation}
where $E_\beta = \braket{:T_{\hatt\hatt}:}_\beta$ and
$P_\beta = \frac{1}{3} \delta^{ij} \braket{:T_{\hati\hatj}:}_\beta$ are, respectively,
the energy density and isotropic pressure at inverse temperature $\beta$, while the macroscopic
velocity $u^\halpha=(1,0,0,0)^T$ corresponds to that of a fluid at rest.
Using (\ref{eq:b_z}), we find the following expressions for the energy density $E_{\beta }$ and pressure $P_{\beta }$:
\begin{eqnarray}
\fl E_\beta + P_\beta = -\frac{2\omega^4 \Gamma(3+k) (\cos\omega r)^{4+2k}}{\pi^{3/2}4^{1+k} \Gamma(\frac{1}{2} + k)}
\sum_{j = 1}^\infty (-1)^j \frac{\cosh\frac{\omega j \beta}{2}}{(\sinh\frac{\omega j \beta}{2})^{4 + 2k}}\nonumber\\
\times {}_2F_1\left[k, 3+k; 1+2k; -\frac{\cos^2\omega r}{\sinh^2\frac{\omega j \beta}{2}}\right],\label{eq:qft_m_E}\\
\fl P_\beta = -\frac{\omega^4 \Gamma(2+k) (\cos\omega r)^{4+2k}}{\pi^{3/2}4^{1+k} \Gamma(\frac{1}{2} + k)}
\sum_{j = 1}^\infty (-1)^j \frac{\cosh\frac{\omega j \beta}{2}}{(\sinh\frac{\omega j \beta}{2})^{4 + 2k}}\nonumber\\
\times {}_2F_1\left[k, 2+k; 1+2k; -\frac{\cos^2\omega r}{\sinh^2\frac{\omega j \beta}{2}}\right].
\label{eq:qft_m_P}
\end{eqnarray}
In the massless limit ($k=0$), the energy density reduces to
\begin{eqnarray}
 E_\beta = -\frac{3\omega ^4}{4\pi^2} (\cos\omega r)^4 \sum_{j = 1}^\infty (-1)^j
 \frac{\cosh\frac{\omega j \beta}{2}}{(\sinh\frac{\omega j \beta}{2})^4},
 \label{eq:qft_m0}
\end{eqnarray}
while $P_{\beta } = E_{\beta } / 3$.
Therefore, in the massless limit the trace $\braket{:T^{\halpha}_{\halpha }:}_\beta$ vanishes, as expected from (\ref{eq:trace}), although the FC (\ref{eq:qft_fc_mo}) is nonzero when the fermion is massless.

\section{Kinetic theory results}
\label{sec:kinetic}

In the kinetic theory approach, the dynamics of a gas of particles of momentum $p^\halpha$ and mass $m$ in general
relativity can be described using the Boltzmann equation written with respect to tetrad components \cite{ambrus16cota,ambrus15wut,mezzacappa13}:
\begin{equation}
p^{\halpha } e^{\mu }_{\halpha} \frac{\partial f}{\partial x^{\mu }}
- \Gamma ^{\hati }_{\halpha \hbeta } p^{\halpha } p^{\hbeta } \frac {\partial f}{\partial p^{\hati }}
= C[f]
 \label{eq:boltz_cons}
\end{equation}
where $f$ is the particle distribution function and
$C[f]$ represents the collision operator which drives the system towards equilibrium.
The time component $p^\hatt$ of the particle momentum four-vector is determined from
the mass-shell condition $p^\hatt = \sqrt{m^2 + \bm{p}^2}$.
States in thermal equilibrium are described by the equilibrium distribution function:
\begin{equation}
 \feq_\epsilon (\beta ) = \frac{Z/\left( 2\pi \right) ^{3}}{e^{-\betat\mu - \betat p_\halpha u^\halpha} - \epsilon},
 \label{eq:feq}
\end{equation}
where $Z$ represents the number of degrees of freedom per particle and $\epsilon$ takes the
values $-1$, $1$ and $0$ for Fermi-Dirac, Bose-Einstein and Maxwell-J\"uttner statistics respectively.
In \eqref{eq:feq}, $\betat$ represents the local inverse temperature of the state, $\mu$ is the chemical
potential and $u^\halpha$ is the fluid velocity four-vector.
In order for the distribution function \eqref{eq:feq} to satisfy the Boltzmann equation \eqref{eq:boltz_cons},
$\betat \mu$ must be constant and $\betat u^\halpha$ must satisfy the Killing equation \cite{cercignani02}:
\begin{equation}
 \nabla_\halpha (\betat \mu) = 0, \qquad
 \nabla_\halpha (\betat u_\hsigma) + \nabla_{\hsigma} (\betat u_\halpha) = 0.
\end{equation}

In this paper, we are interested in states with vanishing chemical potential $\mu = 0$,
where the fluid is at rest (that is,~$u^\halpha = (1, 0,0, 0)^T$). The Killing equation
is satisfied with the above choice for $u^\halpha$ if the inverse
temperature $\betat$ takes the form \cite{tolman30,ambrus16cota}:
\begin{equation}
 \betat = \beta\sqrt{-g_{tt}} = \frac{\beta}{\cos\omega r},
 \label{eq:tolman}
\end{equation}
where $\beta \equiv \betat(r = 0)$ represents the local inverse temperature at the coordinate origin.
The kinetic theory formulation gives the macroscopic SET $T^{\halpha\hbeta}$
as the second order moment of the distribution function $f$:
\begin{equation}
 T^{\halpha\hsigma} = \int \frac{d^3p}{p^\hatt} f\, p^\halpha p^\hsigma,
\end{equation}
where the integration measure $d^3p / p^\hatt$ is Lorentz-invariant.

The SET corresponding to the distribution \eqref{eq:feq} can be found by first performing a series expansion of the equilibrium distribution function (\ref{eq:feq}) in powers of $\epsilon $ \cite{ambrus16cota,ambrus15wut,florkowski15}:
\begin{equation}
 \feq_\epsilon (\beta )= \sum_{j = 0}^\infty \epsilon^j
 \feq_{0}(\beta[j + 1]),
 \label{eq:fseries}
\end{equation}
where the Maxwell-J\"uttner equilibrium distribution function (\ref{eq:feq}) takes the simple form \cite{ambrus16cota}
\begin{equation}
\feq_{0}(\beta)= \frac {Z}{\left( 2\pi \right) ^{3}} e^{-\beta p^{\hatt}}.
\label{eq:MJf}
\end{equation}
The energy density and pressure arising from the Maxwell-J\"uttner distribution function (\ref{eq:MJf}) can be found in closed form \cite{ambrus16cota,ambrus15wut,cercignani02}:
\begin{eqnarray}
 E_{0}(\betat) - 3P_{0}(\betat) = \frac{Zm^3 \cos\omega r}{2\pi^2 \beta}
 K_1\left(\frac{m \beta}{\cos\omega r}\right),\\
 P_{0}(\betat ) = \frac{Z m^2}{2\pi^2 \beta^2} (\cos\omega r)^2
 K_2\left(\frac{m \beta}{\cos\omega r}\right),
\end{eqnarray}
where
$K_n(z)$ is a modified Bessel function of the third kind and $m$ is the particle mass.
The SET arising from the distribution (\ref{eq:fseries})  with general statistics can then
be written as \cite{rezzolla13,cercignani02}:
\begin{equation}
 T^{\halpha\hbeta}_{\epsilon}(\betat) = \left[ E_{\epsilon}(\betat ) + P_{\epsilon}(\betat)\right] u^\halpha u^\hbeta +
 \eta^{\halpha\hbeta} P_{\epsilon}(\betat ),
\end{equation}
where \cite{ambrus16cota,ambrus15wut,florkowski15}
\begin{equation}
 E_{\epsilon}(\betat ) = \sum_{j = 0}^\infty \epsilon^j E_{0}([j + 1]\betat), \qquad
 P_{\epsilon}(\betat ) = \sum_{j = 0}^\infty \epsilon^j P_{0}([j + 1]\betat).
 \label{eq:EPgen}
\end{equation}

For thermal states of fermions ($\epsilon = -1$),  the energy density and pressure (\ref{eq:EPgen}) take the form \cite{ambrus15wut,florkowski15}:
\begin{eqnarray}
 E_{-1}(\betat )- 3P_{-1}(\betat ) = -\frac{2m^3 \cos\omega r}{\pi^2 \beta}
 \sum_{j = 1}^\infty \frac{(-1)^j}{j} K_1\left(\frac{m j\beta}{\cos\omega r}\right),
 \label{eq:kinetic_m_E}\\
 P_{-1}(\betat ) = -\frac{2 m^2}{\pi^2 \beta^2} (\cos\omega r)^2
 \sum_{j = 1}^\infty \frac{(-1)^j}{j^2} K_2\left(\frac{m j\beta}{\cos\omega r}\right),
 \label{eq:kinetic_m_P}
\end{eqnarray}
where $Z = 4$ was taken to account for the spin and charge degrees of freedom
of the Dirac fermions.
The FC in kinetic theory is defined analogously to (\ref{eq:trace}):
\begin{equation}
\fl FC_{-1}(\betat ) =  \frac {1}{m} \left[ E_{-1}(\betat ) - 3P_{-1} (\betat ) \right] =
-\frac{2m^2 \cos\omega r}{\pi^2 \beta}
 \sum_{j = 1}^\infty \frac{(-1)^j}{j} K_1\left(\frac{m j\beta}{\cos\omega r}\right) .
 \label{eq:kineticFC}
\end{equation}

If we set $\omega =0$ (so that the adS radius of curvature is infinite) in (\ref{eq:kinetic_m_E}--\ref{eq:kineticFC}), we recover the Minkowski space-time kinetic theory quantities
\begin{eqnarray}
 \left. E_{-1}(\betat )- 3P_{-1}(\betat )\right\rfloor_{\omega = 0} = -\frac{2m^3}{\pi^2 \beta}
 \sum_{j = 1}^\infty \frac{(-1)^j}{j} K_1\left(m j\beta\right),
\nonumber \\
 \left. P_{-1}(\betat )\right\rfloor_{\omega = 0} = -\frac{2 m^2}{\pi^2 \beta^2}
 \sum_{j = 1}^\infty \frac{(-1)^j}{j^2} K_2\left(m j\beta\right),
 \nonumber \\
 \left. FC_{-1}(\betat )\right\rfloor_{\omega = 0} =  -\frac{2m^2 }{\pi^2 \beta}
 \sum_{j = 1}^\infty \frac{(-1)^j}{j} K_1\left(m j\beta\right),
 \label{eq:minkowski}
\end{eqnarray}
for a thermal distribution of fermions of mass $m$.
On Minkowski space-time, the kinetic theory results (\ref{eq:minkowski}) are identical to the t.e.v.s obtained using QFT (calculated in, for example, \cite{ambrus14phd}).
On adS space-time, the kinetic theory results corresponding to
(\ref{eq:EPgen}) for massless bosons ($\epsilon = 1$)
do not agree with the difference between the QFT t.e.v.s and v.e.v.s computed in
\cite{Allen:1986ty} for a massless, conformally coupled scalar field.
In particular, the kinetic theory results can only depend on the inverse temperature $\beta $ at the origin and the radial coordinate $r$ as functions of the local temperature $\betat$ (\ref{eq:tolman}).
However, the QFT results in \cite{Allen:1986ty} have a much more complicated dependence on $\beta $ and the radial coordinate $r$.
How the kinetic theory results (\ref{eq:kinetic_m_E}--\ref{eq:kineticFC}) on adS space-time compare with the QFT t.e.v.s calculated in section \ref{sec:thermal} will be the focus of section \ref{sec:comparison}.

In the massless limit ($m=0$), using the asymptotic behaviour of the Bessel functions for fixed order $\nu >0$ as the argument $z$ tends to zero \cite{book:olver10}
\begin{equation}
K_{\nu }(z) \simeq \frac {1}{2} \Gamma (\nu )\left( \frac{1}{2} z \right) ^{-\nu }, \qquad z \rightarrow 0,
\end{equation}
 the kinetic theory energy density (\ref{eq:kinetic_m_E}) reduces to:
\begin{eqnarray}
 E_{ -1}(\betat)\rfloor_{m = 0} &= \frac{12}{\pi^2 \beta^4} (\cos \omega r)^4 \sum_{j = 1}^\infty \frac{(-1)^{j+1}}{j^4}
 = \frac{7\pi^2}{60\beta^4} (\cos \omega r)^4,
 \label{eq:kinetic_m0}
\end{eqnarray}
while $P_{-1}(\betat)\rfloor_{m = 0} = \frac{1}{3} E_{-1}(\betat)\rfloor_{m = 0}$.
We note that the kinetic theory FC (\ref{eq:kineticFC}) vanishes in the massless limit, $FC_{-1}(\betat)=0$ when $m=0$, in contrast to the nonzero expression (\ref{eq:qft_fc_mo}) for the QFT $FC_{\beta }$ for massless fermions.
This is our first indication that quantum corrections are significant for t.e.v.s for fermions on adS.

\section{Comparing the QFT and kinetic theory results}
\label{sec:comparison}

We now compare our QFT results obtained in section~\ref{sec:thermal} with the kinetic theory results from section~\ref{sec:kinetic}, both analytically and numerically.
We begin in subsection~\ref{sec:comparison:gen} by comparing the profiles of the FC, energy density, pressure and equation of state
$w = P/E$. Next, subsection~\ref{sec:comparison:boundary} discusses their behaviour
in the vicinity of the adS boundary and subsection~\ref{sec:comparison:origin}
analyses
these quantities at the origin.
We focus on the massless case in subsection~\ref{sec:comparison:m0}.

\subsection{Profiles of the FC, energy density, pressure and equation of state}
\label{sec:comparison:gen}

\begin{figure}
\begin{center}
\begin{tabular}{cc}
\includegraphics[width=0.4\linewidth]{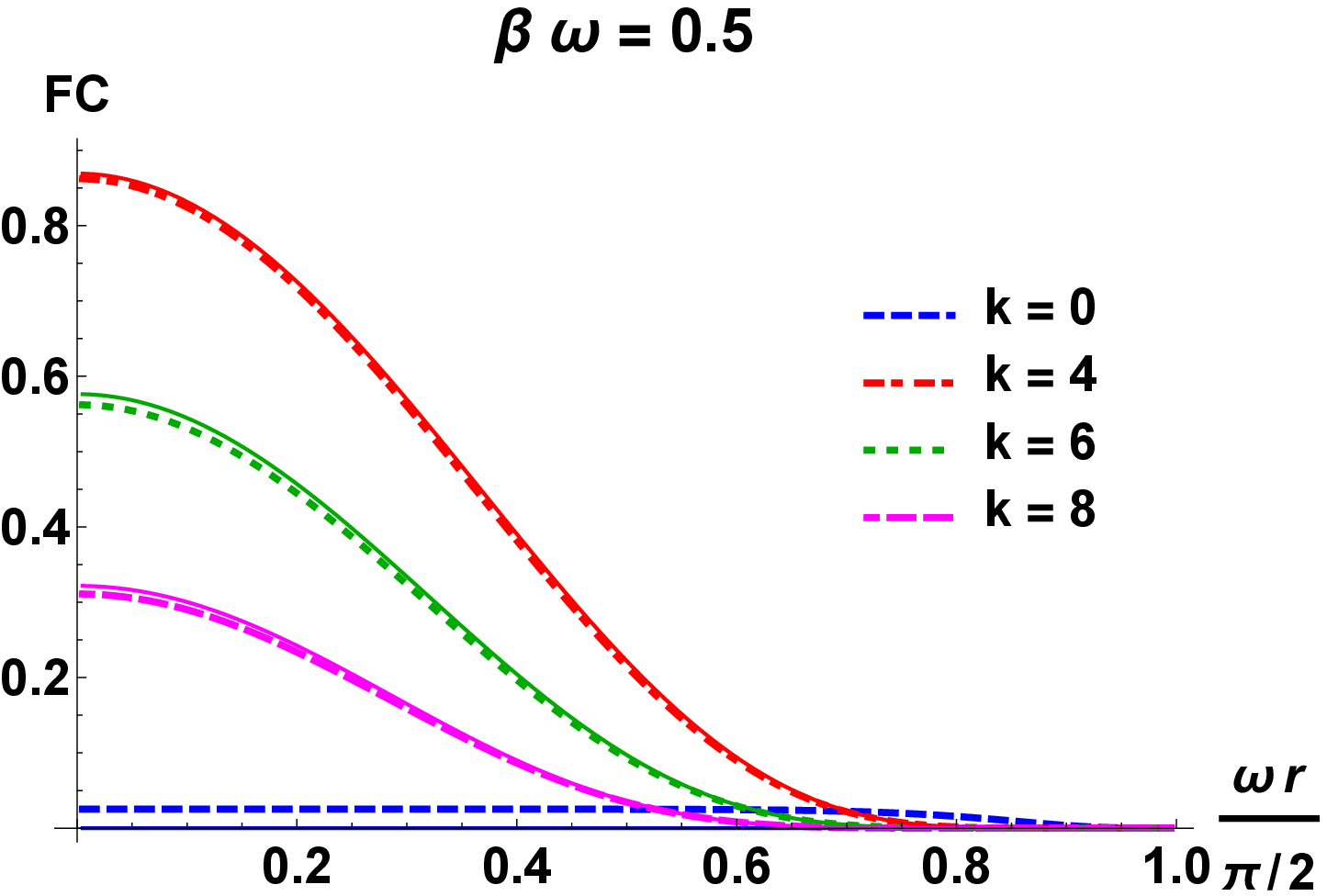} &
\includegraphics[width=0.4\linewidth]{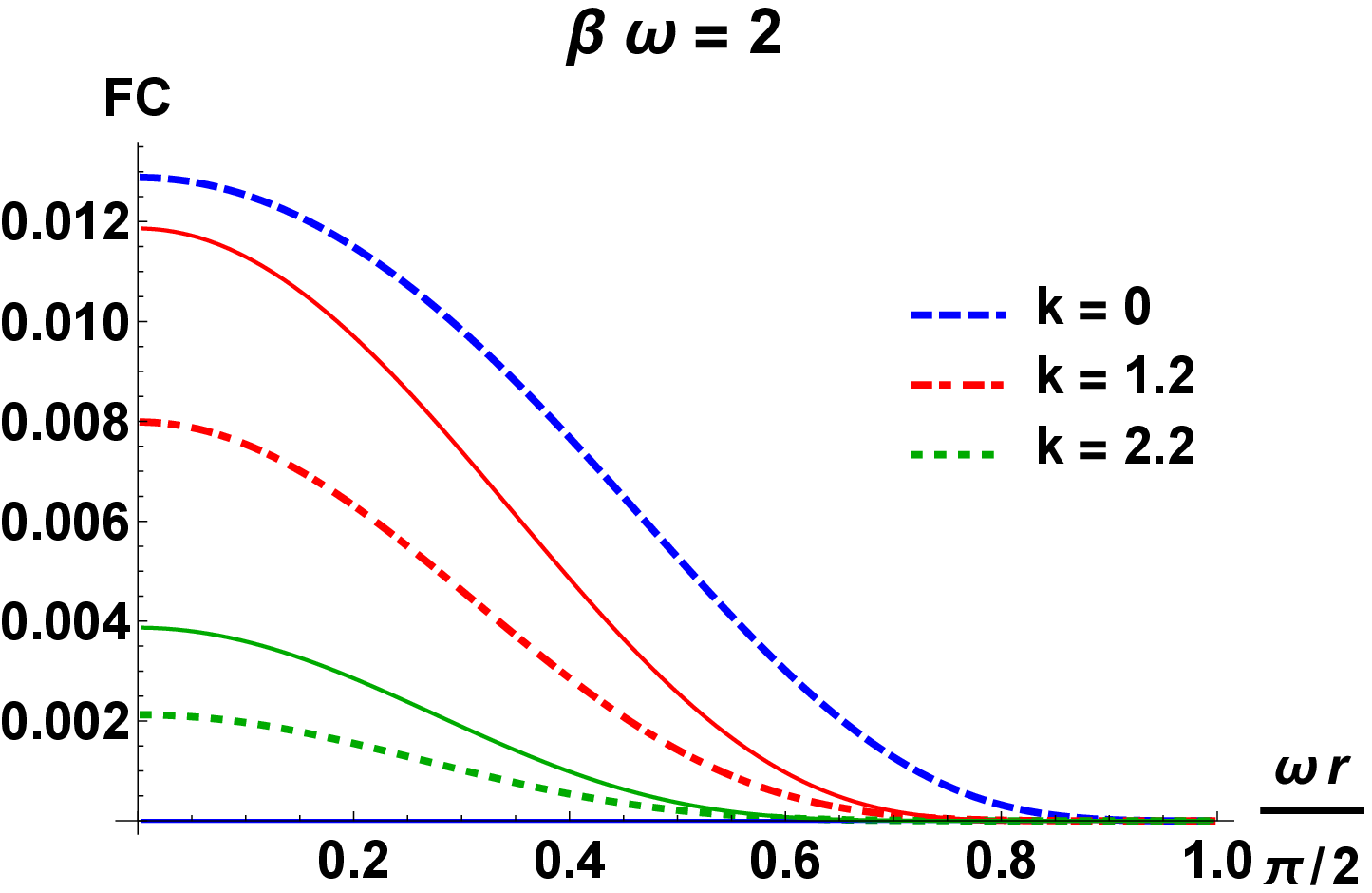}
\\
 (a) & (b)\\
 \includegraphics[width=0.4\linewidth]{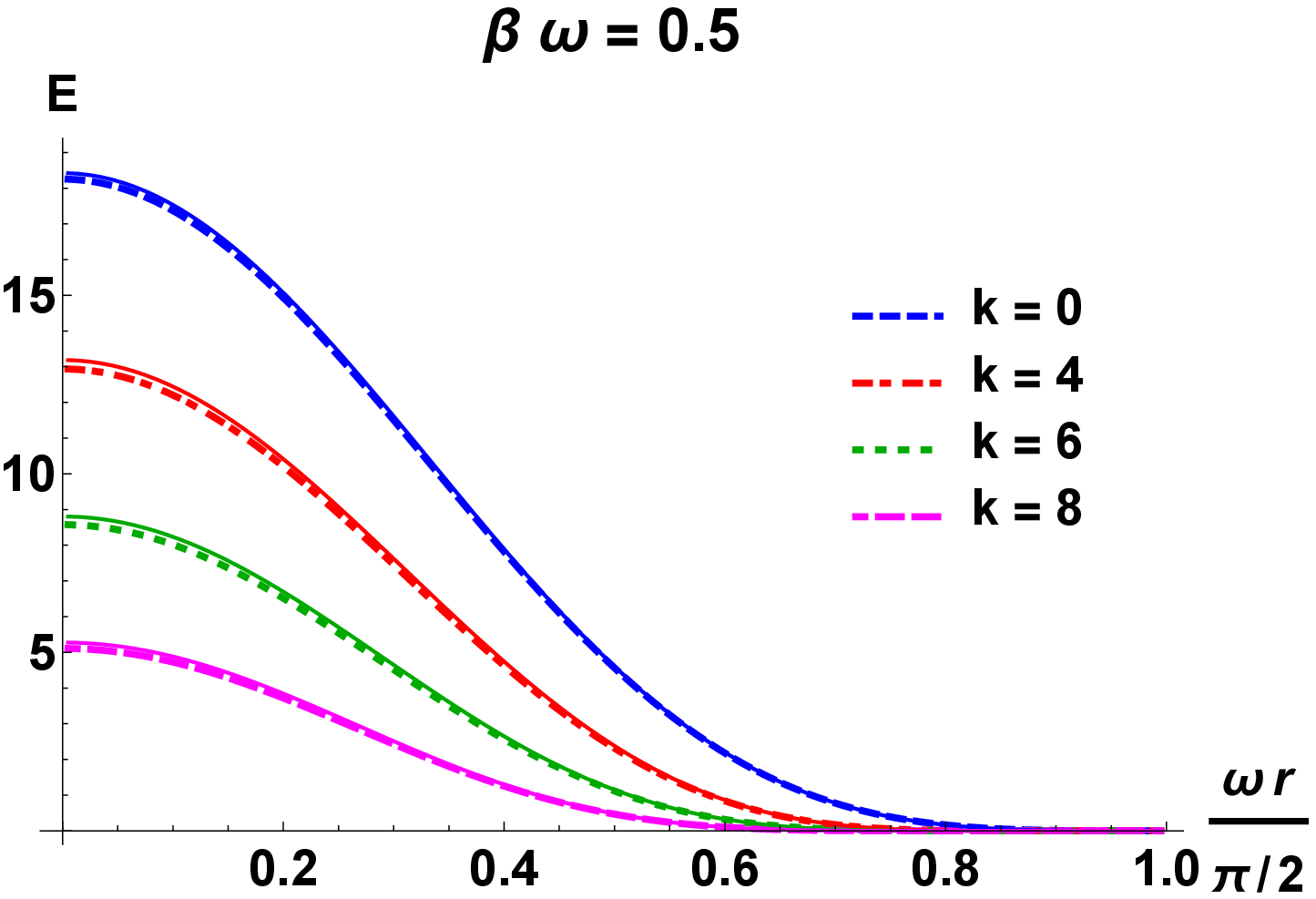} &
 \includegraphics[width=0.4\linewidth]{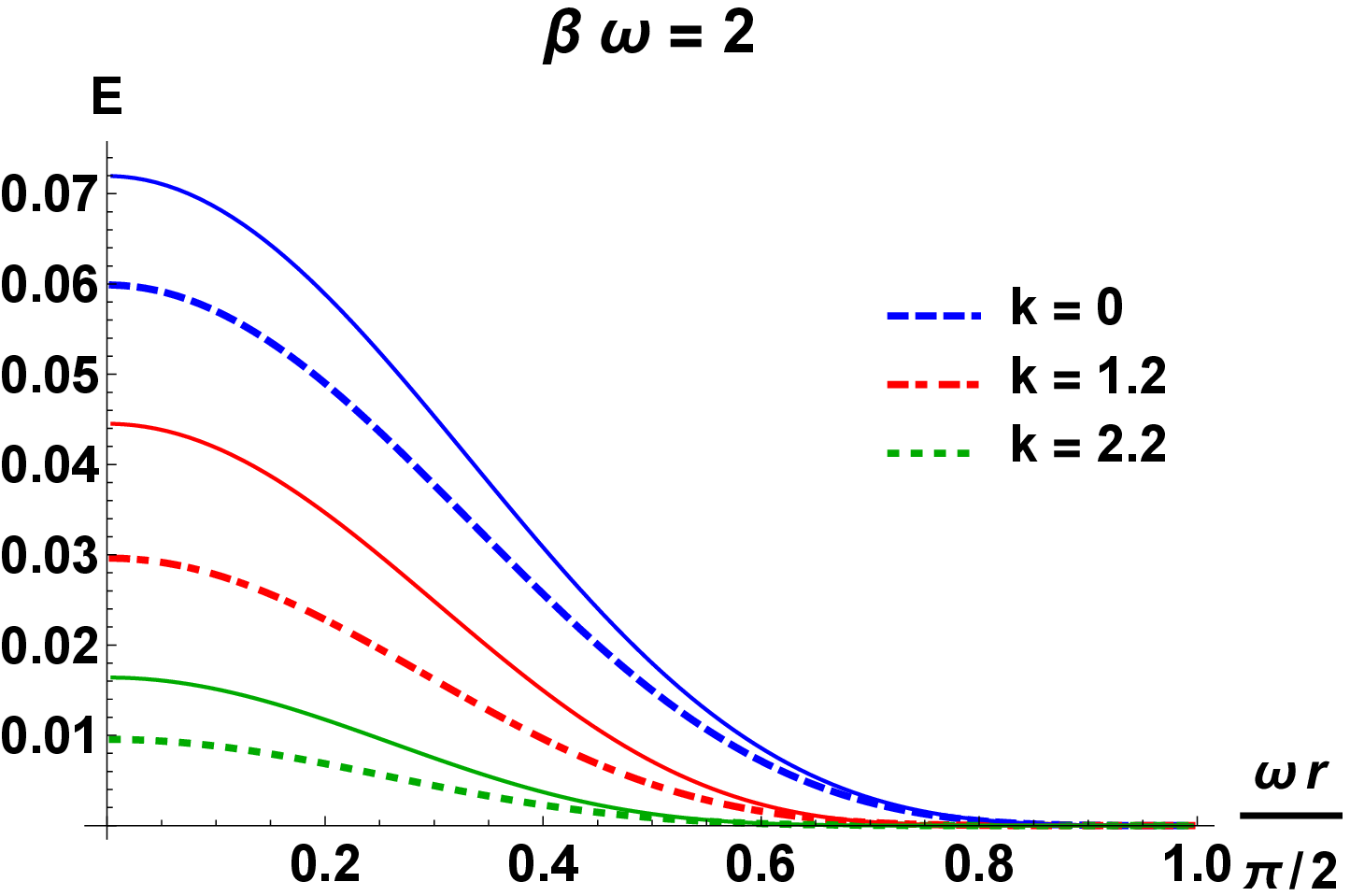} \\
 (c) & (d)\\
 \includegraphics[width=0.4\linewidth]{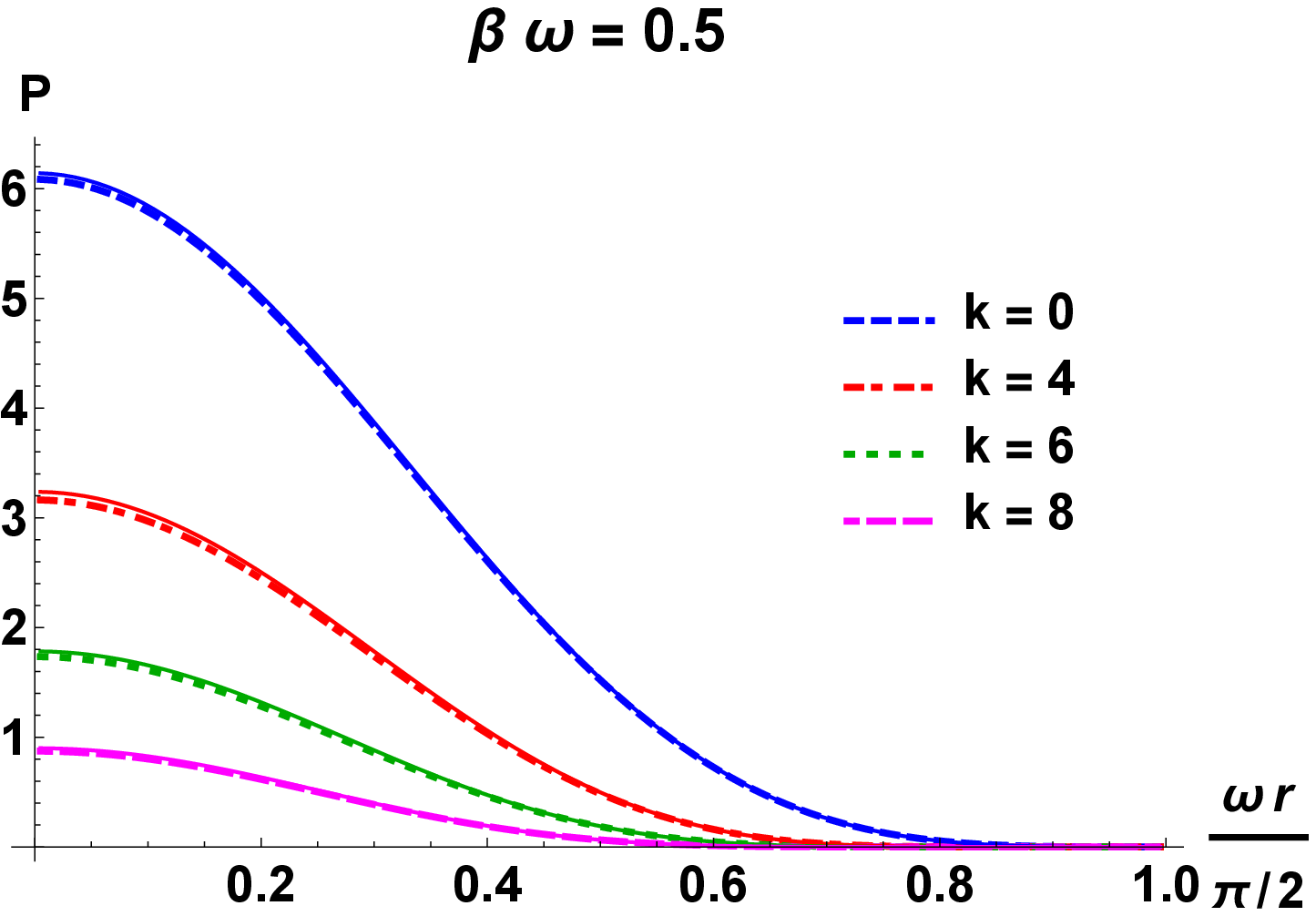} &
 \includegraphics[width=0.4\linewidth]{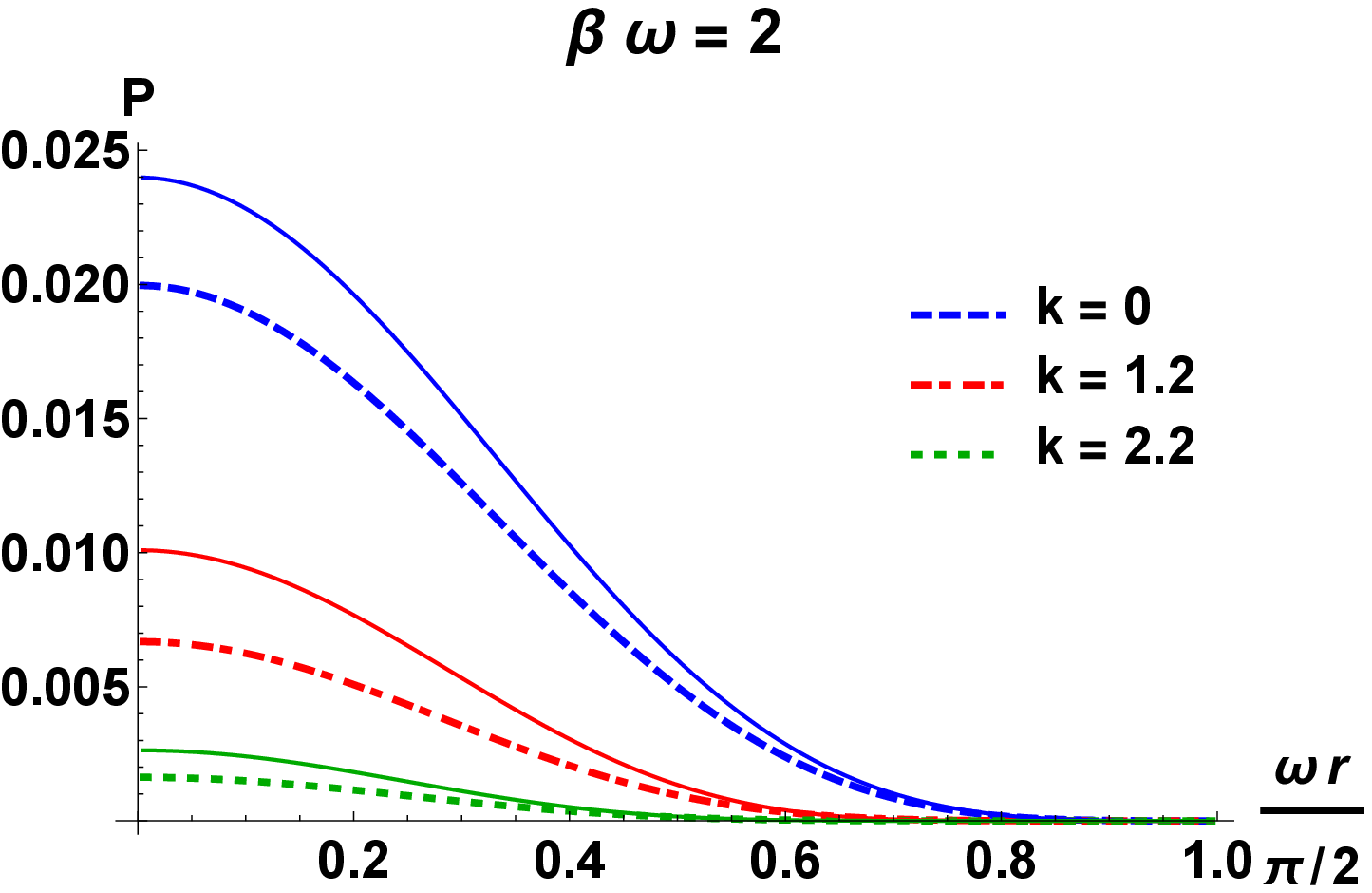} \\
 (e) & (f) \\
 \includegraphics[width=0.4\linewidth]{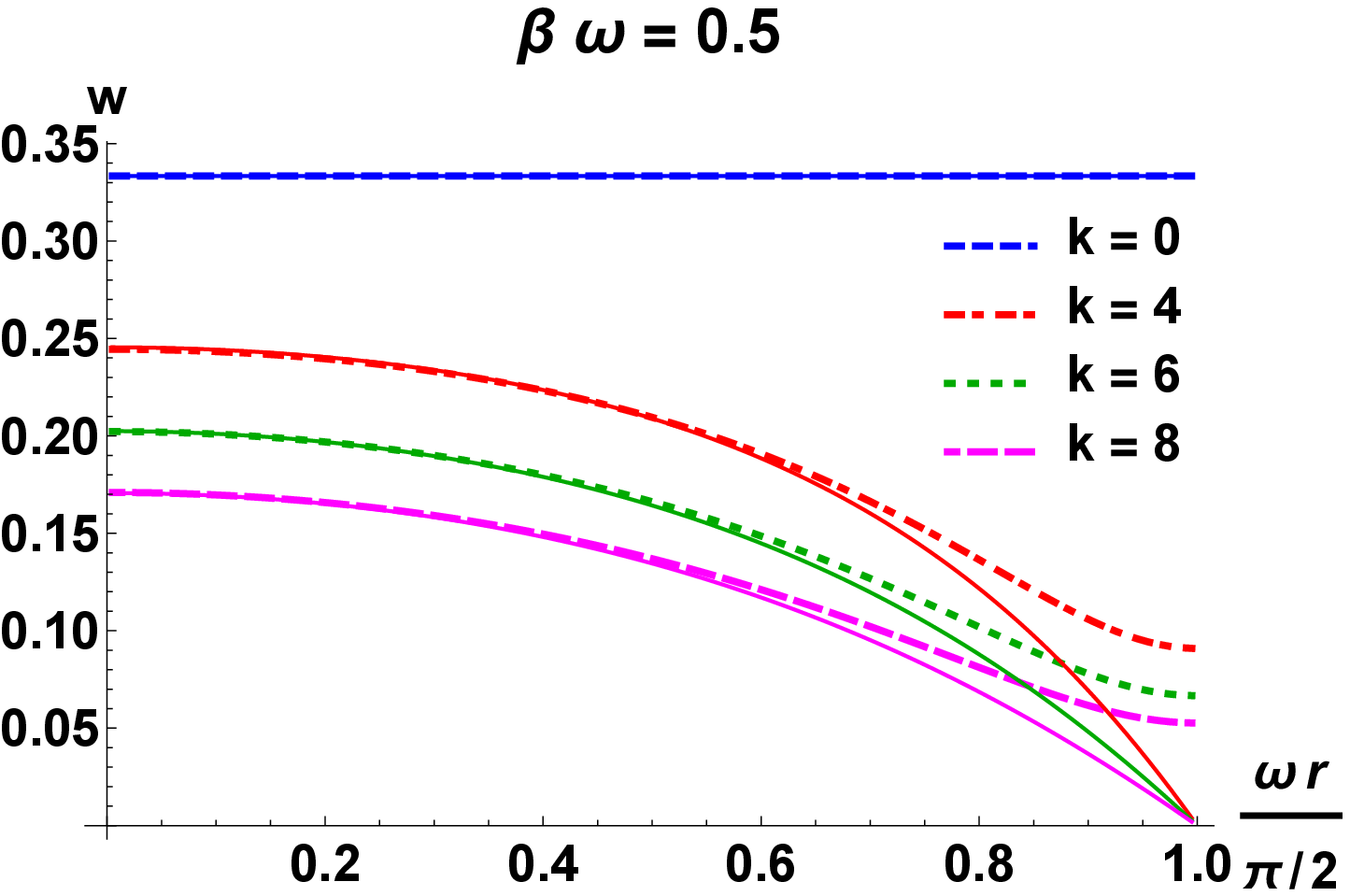} &
 \includegraphics[width=0.4\linewidth]{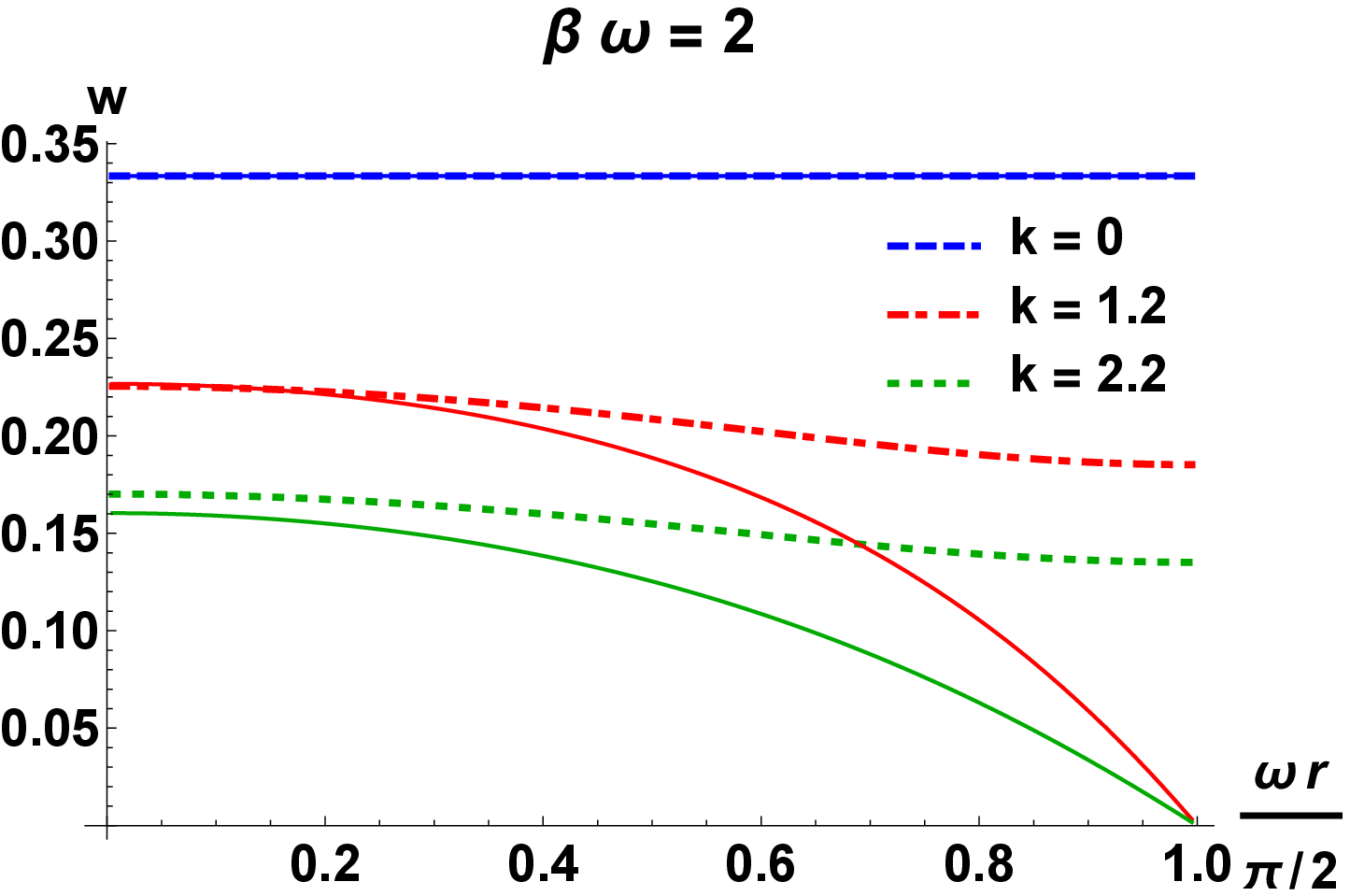} \\
 (g) & (h)
\end{tabular}
\end{center}
\caption{Comparison between the QFT (dashed lines) and kinetic theory (continuous lines)
results for the profiles of the FC (top line), energy density $E$ (second line), pressure $P$ (third line) and
equation of state $w=P/E$ (bottom line), for $\beta \omega = 0.5$ (left) and $\beta\omega = 2.0$ (right).
The curves correspond to various values of $k = m/\omega$.}
\label{fig:gen}
\end{figure}

Figure \ref{fig:gen} shows the profiles of the FC, energy density $E$, pressure $P$ and equation of state $w=P/E$ as derived in QFT (dashed lines) and kinetic theory (solid lines) as functions of the radial coordinate $\omega r$, for two different values of the inverse temperature: $\beta \omega =0.5$ (left) and $\beta \omega =2.0$ (right) and various values of the fermion mass $m=k\omega $.

For the FC, we compare the QFT result (\ref{eq:ppsi}) with that arising from kinetic theory (\ref{eq:kineticFC}) in figure \ref{fig:gen}(a--b).
In both formulations, the profiles of the FC have similar qualitative features for all values of the inverse temperature and fermion mass studied in figure \ref{fig:gen}.
In particular, there is a maximum at the origin and both the energy density and pressure tend to zero as $\omega r \rightarrow \pi /2$ and the space-time boundary is approached.
This arises from the overall
powers of $\cos \omega r$ in (\ref{eq:ppsi}, \ref{eq:kineticFC}).
Thermal radiation is concentrated near the origin, away from the space-time boundary.  This can also be understood from the Tolman relation (\ref{eq:tolman}) for the local temperature $\betat^{-1}$
which has a maximum at the origin and vanishes on the boundary.
It is also clear from figure \ref{fig:gen} that considering a thermal state has broken the space-time symmetries and the t.e.v.s are not constant throughout the space-time, unlike the v.e.v.s computed in \cite{art:ambrus15plb}.
The kinetic theory FC (\ref{eq:kineticFC}) vanishes when $k=0$ and the fermion field is massless.  We see from  figure \ref{fig:gen}(a--b) that the QFT result (\ref{eq:qft_fc_mo}), while nonzero, is very small, even when $\beta \omega =0.5$ and the temperature is large.

Looking at figure \ref{fig:gen}(c--f), the profiles of energy density and pressure in both QFT and kinetic theory have similar qualitative features to the profiles of the FC for all values of the inverse temperature and fermion mass studied.
The overall factors of powers of $\cos  \left( \omega r \right)$ appearing in (\ref{eq:qft_m_E}, \ref{eq:qft_m_P}, \ref{eq:kinetic_m_E}, \ref{eq:kinetic_m_P}) mean that the profiles have a maximum at the origin and tend to zero on the space-time boundary.

In general, quantum effects (that is, deviations from classical kinetic theory) will occur in the presence
of strong gravitational fields, where the coupling between the space-time curvature and the
quantum fields becomes important. On adS, the space-time curvature is proportional to the inverse radius of curvature $\omega$.
We anticipate that at low values of $\beta \omega$ (corresponding to large temperatures) the QFT results derived in section \ref{sec:thermal}
should approach the kinetic theory results from section \ref{sec:kinetic}.
As $\beta\omega$ increases, and the temperature decreases, we expect to see deviations from kinetic theory in the QFT results.
These expectations are confirmed in figure \ref{fig:gen}(a--f),
where it can be seen that when $\beta \omega = 0.5$, the profiles obtained using QFT and
kinetic theory are nearly indistinguishable, and visible deviations between these two approaches
appear when $\beta\omega = 2.0$.

Figure \ref{fig:gen}(g--h) show the profiles of the equation of state $w=P/E$ as a function of $\omega r$.
As expected from sections \ref{sec:thermal} and \ref{sec:kinetic}, the equation of state equals one third for all $\omega r$ when the fermion field is massless ($k=0$).
However, for a massive fermion field, figure \ref{fig:gen} reveals fundamental differences between QFT and kinetic theory
in the properties of the equation of state.
In the framework of kinetic theory, the equation of state $w_{-1}(\betat) = P_{-1}(\betat) / E_{-1}(\betat)$ always approaches $0$ as $\omega r \rightarrow \pi / 2$ as long as $k = m/\omega>0$. This is in contrast with the
QFT results, which show a temperature-independent
nonzero value of the equation of state $w_{\beta }=P_{\beta }/E_{\beta }$ on the adS boundary.
To understand this behaviour in more detail, in the following subsection we examine more closely
the FC, energy density, pressure and equation of state in the vicinity of the boundary.

\subsection{Behaviour near the boundary}
\label{sec:comparison:boundary}

Near the boundary, it is useful to change the radial variable from $\omega r $ to $\delta $, where
\begin{equation}
 \delta = \frac{\pi}{2} - \omega r,
\end{equation}
with $\cos\omega r = \sin \delta$.
For small values of the argument, the hypergeometric
functions appearing in (\ref{eq:ppsi}, \ref{eq:qft_m_E}, \ref{eq:qft_m_P}) have the series expansion \cite{book:olver10}:
\begin{equation}
 {}_2F_1(a, b; c; z) = 1 + \frac{ab}{c} z + O(z^2).
\end{equation}
Thus, the QFT t.e.v.s of the FC, energy density and pressure (\ref{eq:ppsi}, \ref{eq:qft_m_E}, \ref{eq:qft_m_P}) can be approximated by the following expressions:
\begin{equation}
\fl FC_\beta \simeq  \frac{2\omega^3 \Gamma(2+k) (\sin\delta)^{4 + 2k}}{\pi^{3/2}4^{1+k}\Gamma(\frac{1}{2} + k)}
 \left[S_{k} - \frac{(1+k)(2+k)}{1+2k} S_{1+k} \sin^2\delta\right],
 \label{eq:qft_bas_ppsi}
\end{equation}
\begin{eqnarray}
\fl E_\beta &\simeq& \frac{\omega^4 \Gamma(2+k) (\sin\delta)^{4+2k}}{\pi^{3/2}4^{1+k} \Gamma(\frac{1}{2} + k)}
\left[(3+2k) S_k - \frac{k(2+k)}{1+2k} (5 + 2k) S_{1+k} \sin^2\delta\right],
\label{eq:qft_bas_E}
\\
\fl P_\beta &\simeq& \frac{\omega^4 \Gamma(2+k) (\sin\delta)^{4+2k}}{\pi^{3/2}4^{1+k} \Gamma(\frac{1}{2} + k)}
\left[S_k - \frac{k(2+k)}{1+2k} S_{1+k} \sin^2\delta\right],
\label{eq:qft_bas_P}
\end{eqnarray}
where we have retained terms up to and including corrections of second order in $\sin \delta $ and
\begin{equation}
 S_{\nu} = -\sum_{j = 1}^\infty (-1)^j
 \frac{\cosh\frac{\omega j \beta}{2}}{(\sinh\frac{\omega j \beta}{2})^{4+2\nu}} .
\end{equation}
It can be seen that the QFT t.e.v.s of $FC_{\beta }$, the energy density $E_\beta$ and pressure $P_\beta$ all approach $0$ as
$(\sin\delta)^{4+2k}$ for $\delta \rightarrow 0$. For small $\delta $, the equation of state from QFT
can be approximated by the expression:
\begin{equation}
 w_\beta =\frac {P_{\beta }}{E_{\beta }}\simeq \left[3 + 2k - \frac{2k(2+k)}{1+2k}\frac{S_{k+1}}{S_k} \sin^2\delta \right]^{-1}.
 \label{eq:qft_bas_w}
\end{equation}
As $\delta \rightarrow 0$, the equation of state $w_{\beta}\rightarrow \left( 3 + 2k \right) ^{-1}$, which is confirmed by the numerical results presented in
figure \ref{fig:gen}(g--h).

In order to find the behaviour near the boundary of the energy density \eqref{eq:kinetic_m_E},
pressure \eqref{eq:kinetic_m_P} and FC (\ref{eq:kineticFC}) obtained using kinetic theory, the following asymptotic expression
for the modified Bessel functions $K_\nu(z)$  when the argument is large can be employed \cite{book:olver10}:
\begin{equation}
 K_\nu(z) = \sqrt{\frac{\pi}{2z}} e^{-z} \left[1 + \frac{4\nu^2 - 1}{8z} + O(z^{-2})\right] , \qquad z\rightarrow \infty .
\end{equation}
In the following we assume that the fermion mass $m>0$.
We first consider the expansion for $P_{-1}(\betat)$ \eqref{eq:kinetic_m_P}:
\begin{equation}
\fl P_{-1}(\betat ) \simeq -\frac{1}{2} \left(\frac{2m}{\pi}\right)^{3/2}
 \left(\frac{\sin\delta}{\beta}\right)^{5/2} \sum_{j = 1}^\infty\frac{(-1)^j}{j^{5/2}} e^{-mj\beta/\sin\delta}
 \left[1 + \frac{15 \sin\delta}{8mj\beta} + O(\delta^2)\right].
\end{equation}
Due to the exponential decrease of the summand, a suitable approximation can be obtained by truncating the sum over $j$ after the first term,
yielding:
\begin{equation}
 P_{-1}(\betat ) \simeq \frac{1}{2} \left(\frac{2m}{\pi}\right)^{3/2}
 \left(\frac{\sin\delta}{\beta}\right)^{5/2} e^{-m\beta/\sin\delta}
 \left(1 + \frac{15}{8m\beta} \sin\delta\right).
 \label{eq:kinetic_bas_P}
\end{equation}
The difference $E_{-1}(\betat )-3P_{-1}(\betat)$
can be similarly approximated from \eqref{eq:kinetic_m_E}:
\begin{equation}
\fl E_{-1}(\betat) - 3P_{-1}(\betat) \simeq \frac{m}{2} \left(\frac{2m}{\pi}\right)^{3/2}
 \left(\frac{\sin\delta}{\beta}\right)^{3/2} e^{-m\beta/\sin\delta}
 \left(1 + \frac{3}{8m\beta} \sin\delta\right).
 \label{eq:kinetic_bas_E}
\end{equation}
We then also have an approximation for the FC in kinetic theory from (\ref{eq:kineticFC}):
\begin{equation}
FC_{-1}(\betat) \simeq \frac{1}{2} \left(\frac{2m}{\pi}\right)^{3/2}
 \left(\frac{\sin\delta}{\beta}\right)^{3/2} e^{-m\beta/\sin\delta}
 \left(1 + \frac{3}{8m\beta} \sin\delta\right).
 \label{eq:kinetic_bas_FC}
\end{equation}
Since the power of $\sin\delta$ in \eqref{eq:kinetic_bas_E} is smaller than that in
\eqref{eq:kinetic_bas_P}, the energy density dominates over the
pressure as the boundary is approached. This can be seen by considering the behaviour
of the equation of state:
\begin{equation}
 w_{-1}(\betat ) = \frac {P_{-1}(\betat)}{E_{-1}(\betat)} \simeq
 \sin\delta \left( m\beta + \frac{3}{2} \sin\delta\right) ^{-1}.
 \label{eq:kinetic_bas_w}
\end{equation}
As $\delta \rightarrow 0$, the equation of state in kinetic theory tends to $0$ as the boundary is approached, providing the fermion mass $m>0$, as is confirmed in
figure \ref{fig:gen}(g--h).
Our approximations near the boundary for the kinetic theory results are not valid if the fermion is massless $m=0$, when the energy density takes the simple form (\ref{eq:kinetic_m0}) and the equation of state equals one third everywhere.

\begin{figure}
\begin{center}
\begin{tabular}{cc}
\includegraphics[width=0.4\linewidth]{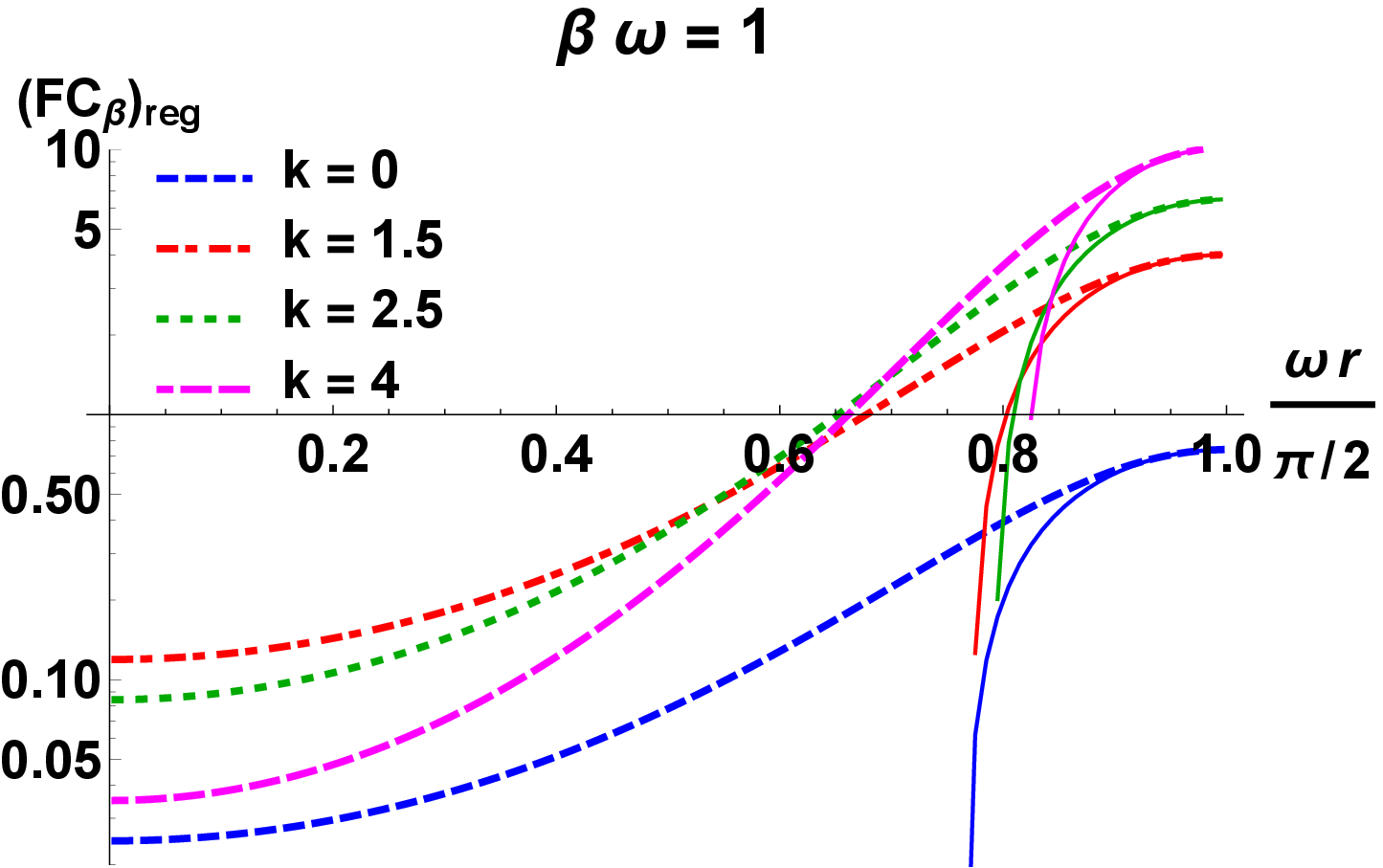} &
\includegraphics[width=0.4\linewidth]{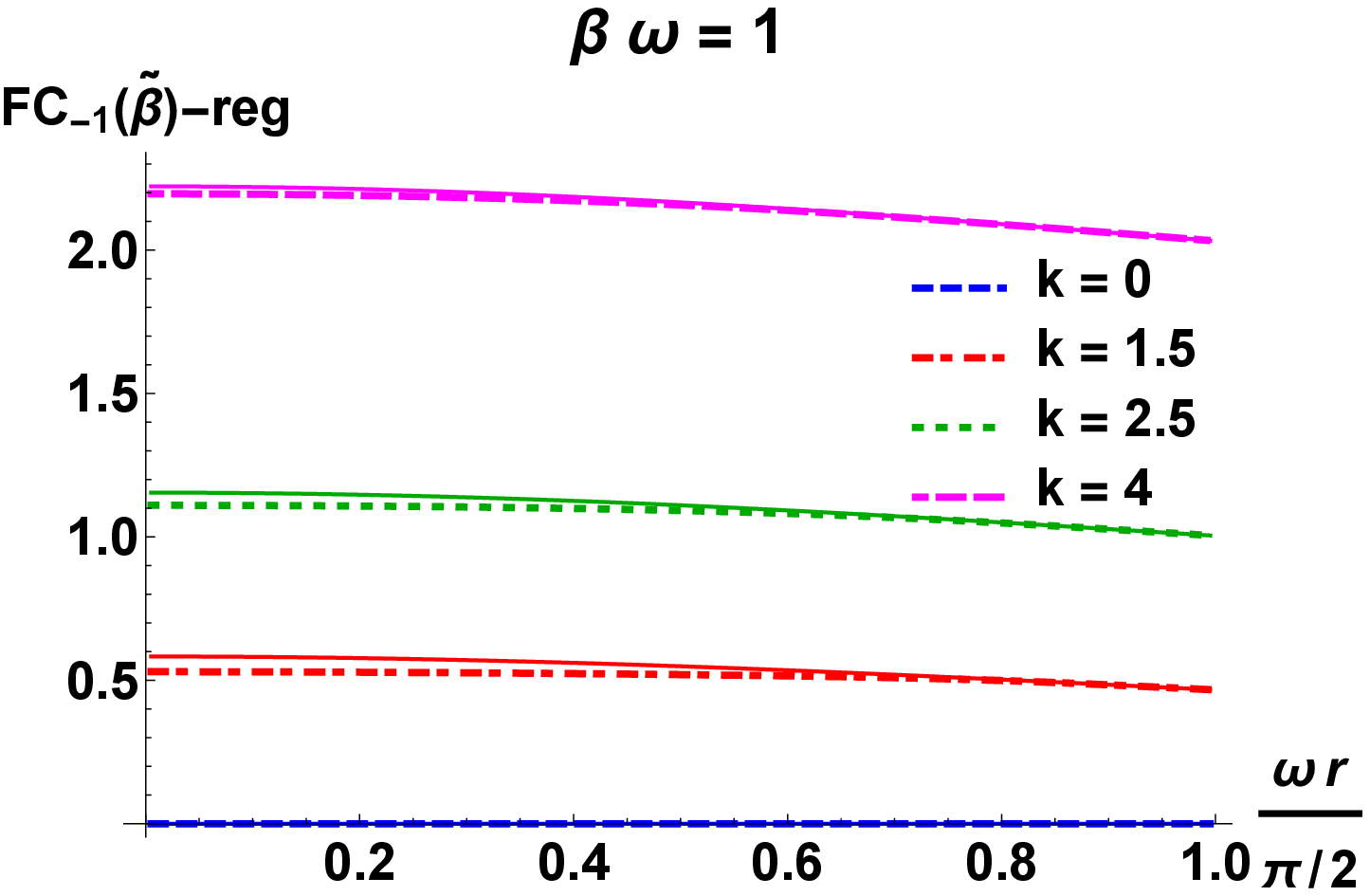}
\\
(a) & (b) \\
 \includegraphics[width=0.4\linewidth]{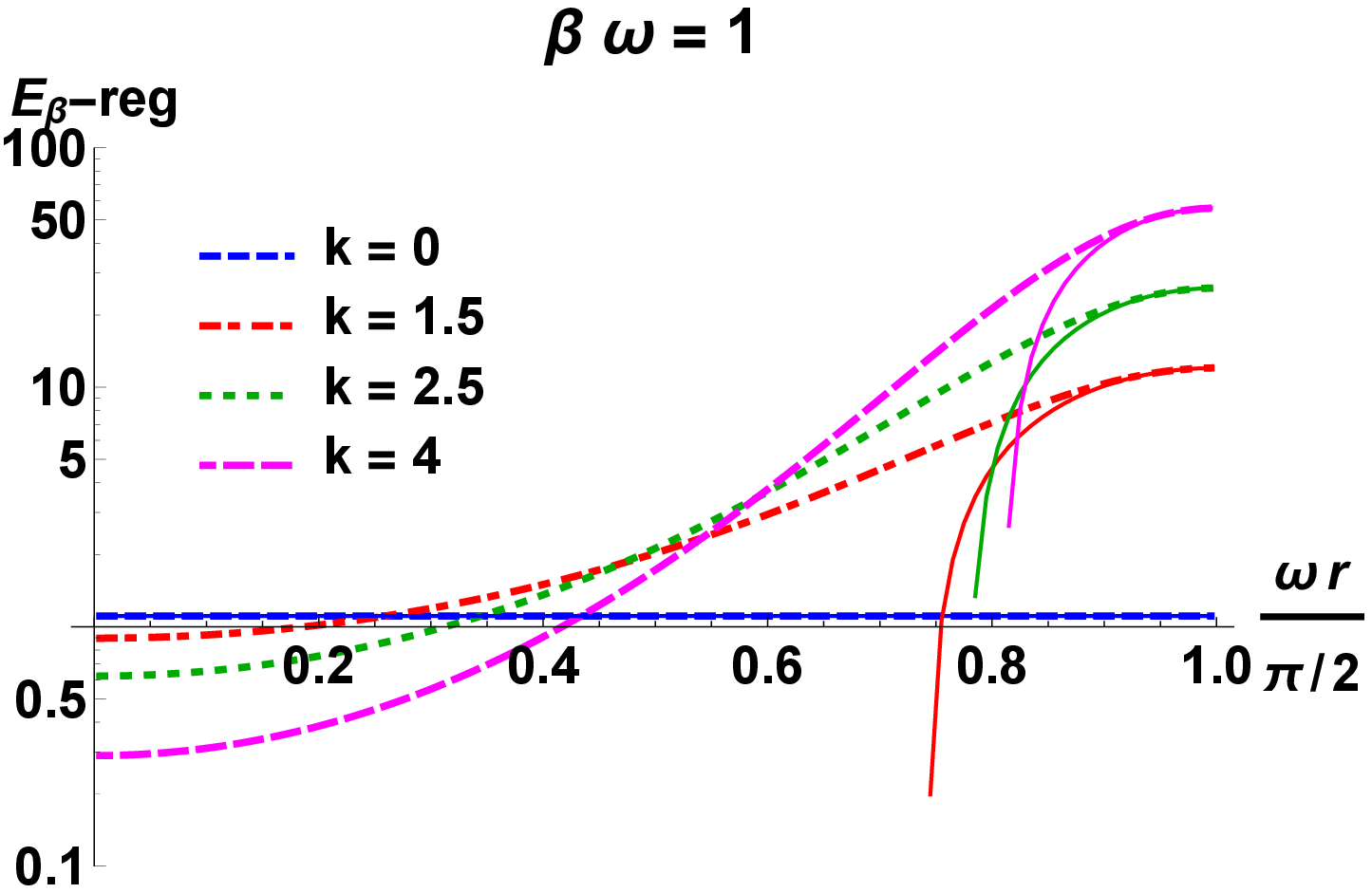} &
 \includegraphics[width=0.4\linewidth]{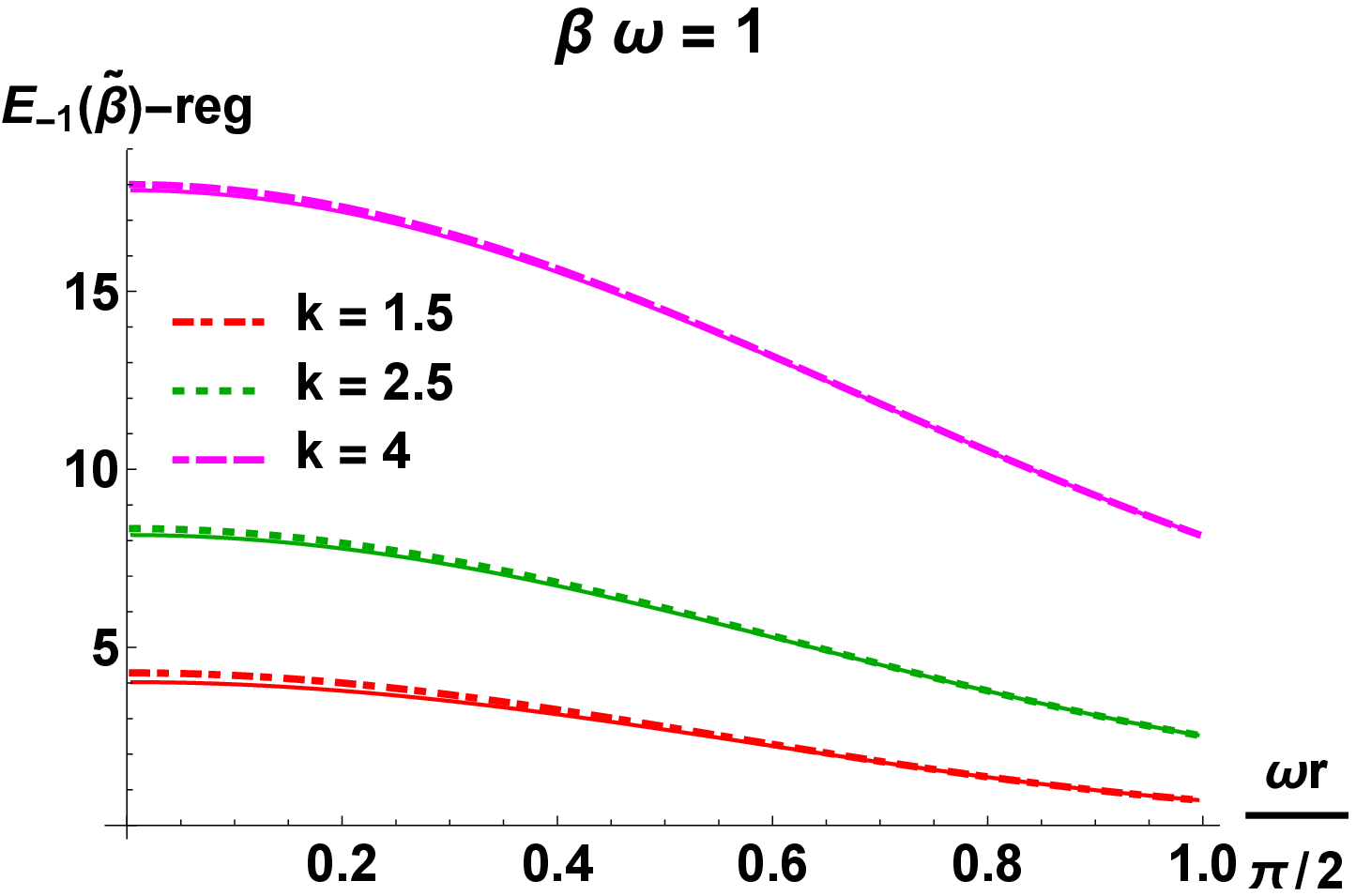} \\
 (c) & (d)\\
 \includegraphics[width=0.4\linewidth]{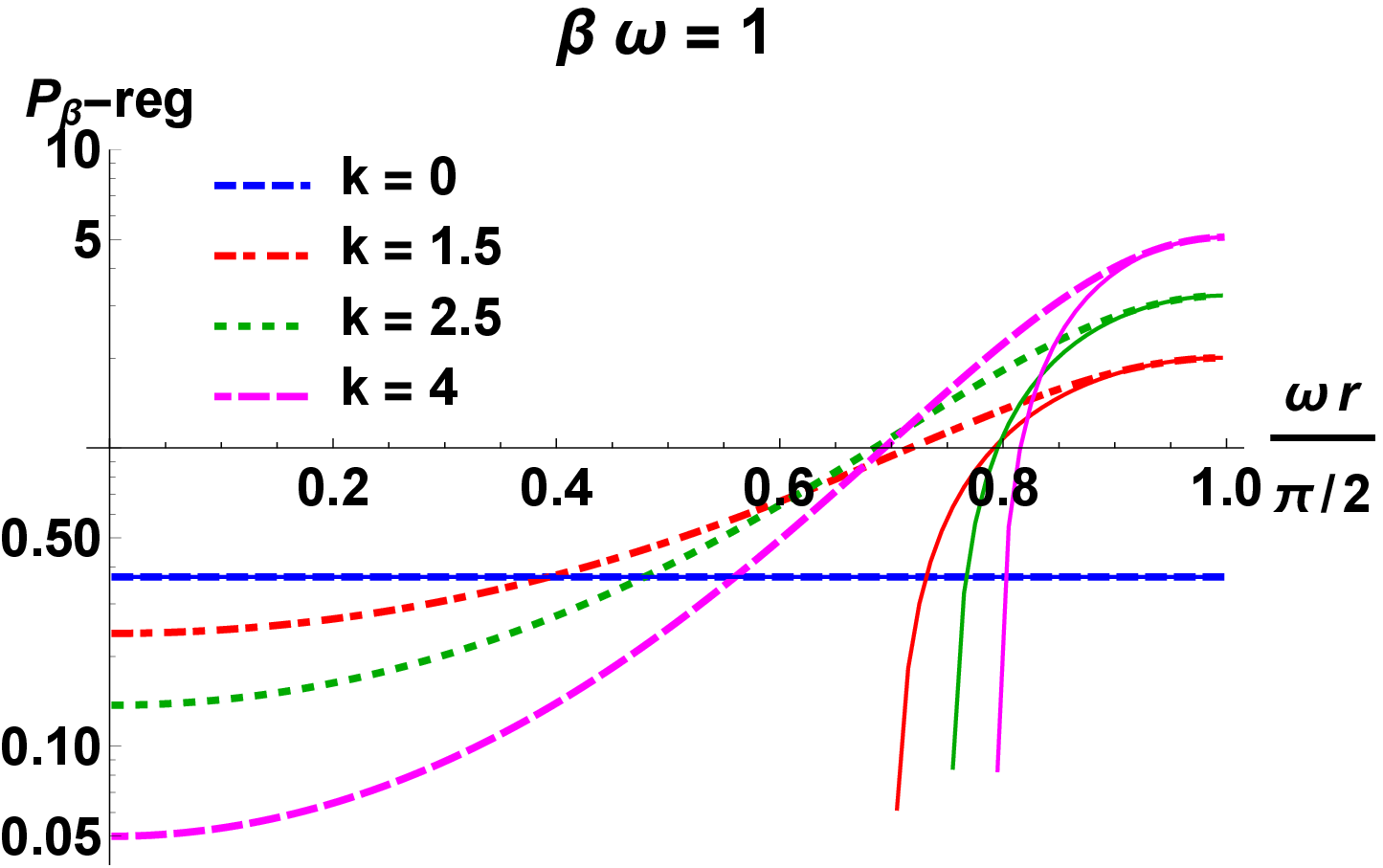} &
 \includegraphics[width=0.4\linewidth]{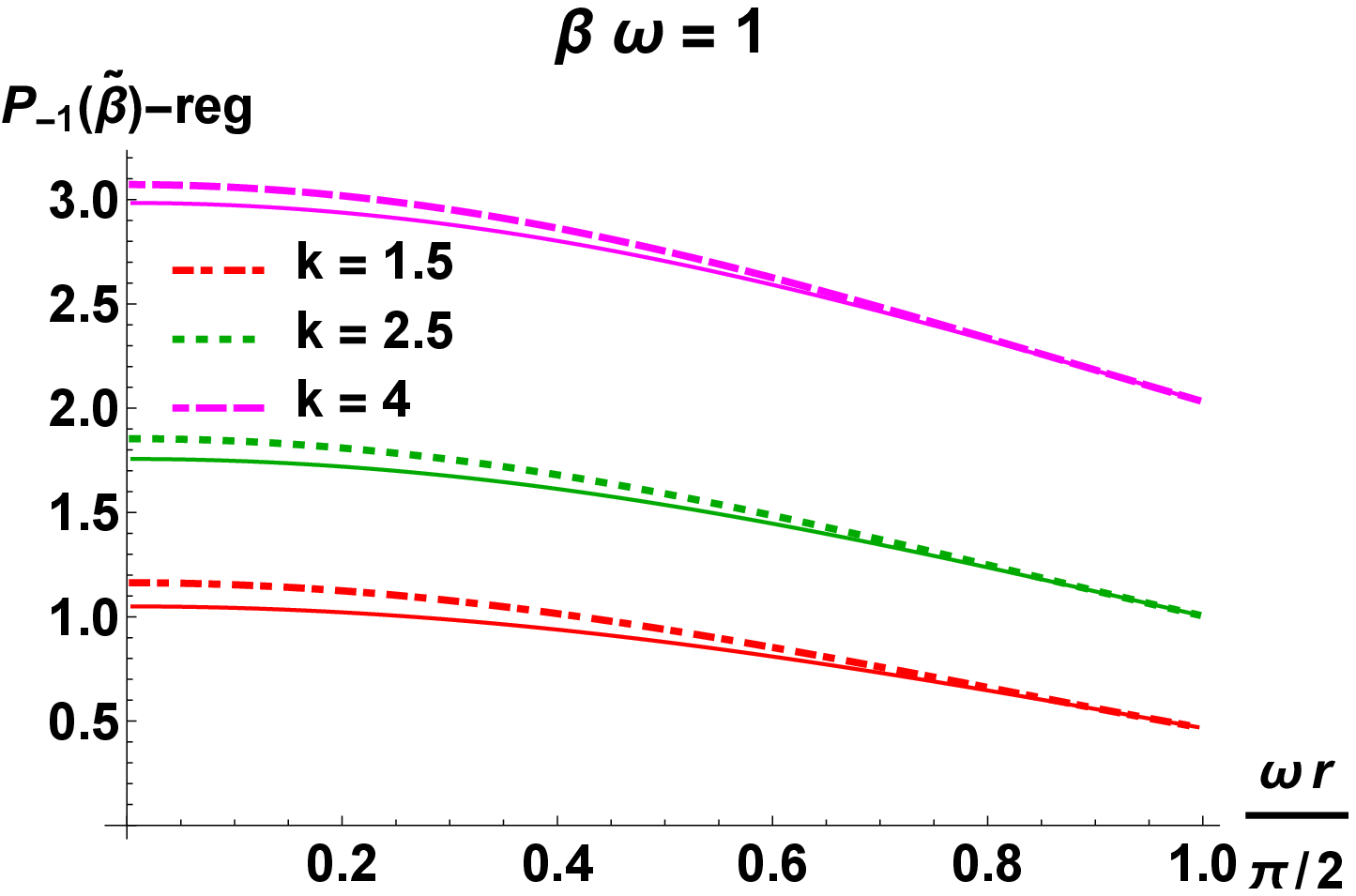} \\
 (e) & (f) \\
 \includegraphics[width=0.4\linewidth]{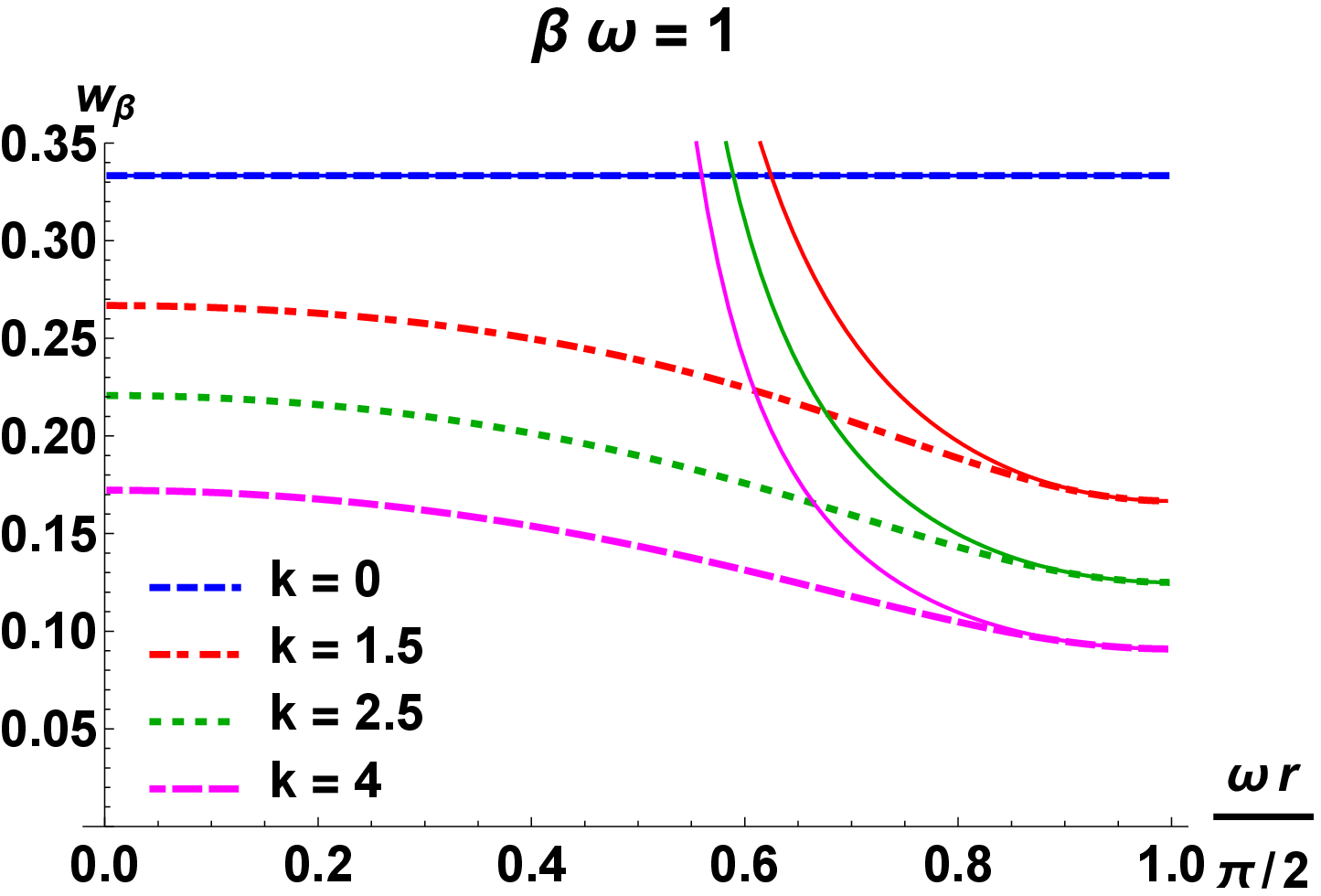} &
 \includegraphics[width=0.4\linewidth]{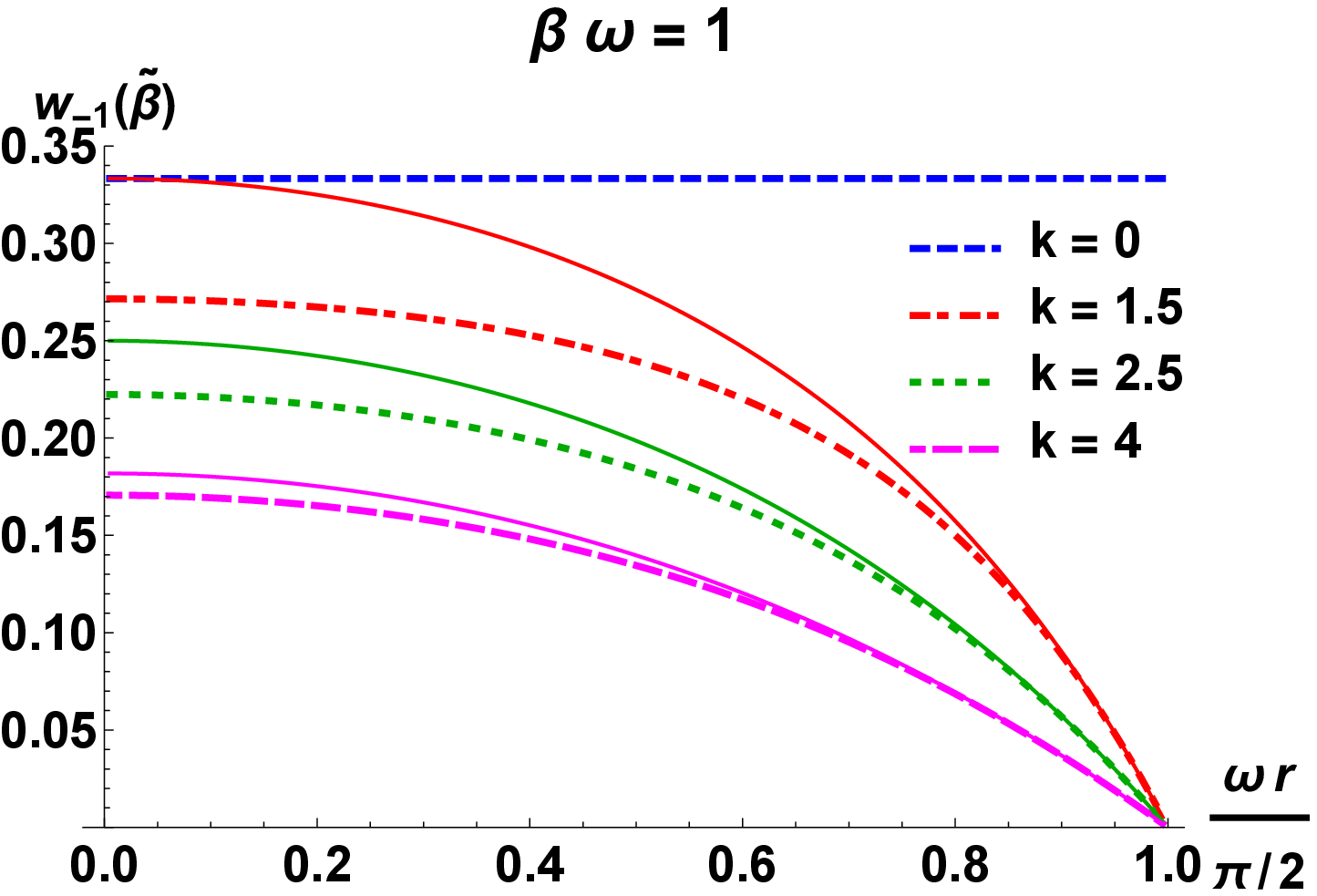} \\
 (g) & (h)
\end{tabular}
\end{center}
\caption{Comparison between the exact numerical profiles of the FC (top line), energy density $E$ (second line),
pressure  $P$ (third line) and equation of state  $w=P/E$ (bottom line) and the asymptotic formulae
presented in section \ref{sec:comparison:boundary}. The QFT results are shown in the left column,
while the kinetic theory results are presented in the right column. In each case, the continuous lines correspond to the asymptotic approximations, while the dashed lines are the exact QFT/kinetic theory results.}
\label{fig:bas}
\end{figure}

In figure \ref{fig:bas} we compare our analytic approximations near the space-time boundary with the full numerical t.e.v.s derived in both QFT (left) and kinetic theory (right), for inverse temperature $\beta \omega =1$.
In order to maximise the visibility of the domain of applicability of our results, the ratio between
the FC, energy density and pressure and their leading order asymptotic behaviour (\ref{eq:qft_bas_ppsi}--\ref{eq:qft_bas_P}, \ref{eq:kinetic_bas_P}--\ref{eq:kinetic_bas_FC}) is taken, in other words we plot the following quantities:
\begin{eqnarray}
 FC_{\beta}^{\rm reg} = \frac{FC_{\beta }}{ (\cos \omega r )^{4+2k}}, \quad
 E_\beta^{\rm reg} = \frac{E_\beta}{(\cos\omega r)^{4+2k}}, \quad
P_\beta^{\rm reg} = \frac{P_\beta}{(\cos\omega r)^{4+2k}}, \nonumber\\
FC_{-1}^{\rm {reg}}(\betat) = \frac{FC_{-1}(\betat)}{e^{-m\beta/\cos\omega r} (\cos \omega r)^{3/2}}, \quad
 E_{-1}^{\rm reg}(\betat) = \frac{E_{-1}(\betat)}{e^{-m\beta/\cos\omega r} (\cos\omega r)^{3/2}}, \nonumber \\
P_{-1}^{\rm reg}(\betat) = \frac{P_{-1}(\betat)}{e^{-m\beta/\cos\omega r} (\cos\omega r)^{5/2}},
 \label{eq:approxplots}
\end{eqnarray}
where the QFT quantities $FC_{\beta }$, the energy density $E_{\beta }$ and pressure $P_{\beta }$ are given by (\ref{eq:ppsi}, \ref{eq:qft_m_E}, \ref{eq:qft_m_P}) and the kinetic theory energy density
$E_{-1}(\betat)$, pressure $P_{-1}(\betat)$ and $FC_{-1}(\betat)$ by (\ref{eq:kinetic_m_E}, \ref{eq:kinetic_m_P}, \ref{eq:kineticFC}).
We also plot the equations of state for QFT and kinetic theory, given respectively by
\begin{equation}
w_{\beta} = \frac {P_{\beta }}{E_{\beta }} = \frac {P_{\beta }^{\rm reg}}{E_{\beta }^{\rm reg}},
\qquad
w_{-1}(\betat) = \frac{P_{-1}(\betat)}{E_{-1}(\betat)} =
\frac{P_{-1}^{\rm reg}(\betat) \cos \omega r}{E_{-1}^{\rm reg}(\betat)}.
\label{eq:eosbas}
\end{equation}
For both QFT and kinetic theory, in figure \ref{fig:bas} our approximations can be seen to hold in a vicinity of the boundary.
In the kinetic theory case (right-hand plots in figure \ref{fig:bas}) the analytic expressions (\ref{eq:kinetic_bas_P}, \ref{eq:kinetic_bas_E}, \ref{eq:kinetic_bas_FC}) for $FC_{-1}(\betat)$, the energy density $E_{-1}(\betat)$ and pressure $P_{-1}(\betat)$ are good approximations to the exact results (\ref{eq:kinetic_m_E}, \ref{eq:kinetic_m_P}, \ref{eq:kineticFC}) everywhere in the space-time and for all values of the fermion mass shown, although the resulting approximation (\ref{eq:kinetic_bas_w}) for the equation of state $w_{-1}(\betat)$ is not very accurate away from the boundary when $k$ is small.
In the QFT case (left-hand plots in figure \ref{fig:bas}), the analytic expressions (\ref{eq:qft_bas_ppsi}--\ref{eq:qft_bas_P}) for $FC_{\beta }$, the energy density $E_{\beta }$ and pressure $P_{\beta }$ are good approximations to the exact results (\ref{eq:ppsi}, \ref{eq:qft_m_E}, \ref{eq:qft_m_P}) in only a small neighbourhood of the boundary.

The analysis presented in this subsection points to a fundamental
difference between the kinetic theory and QFT approaches. While in the kinetic theory
formulation, the equation of state on the space-time boundary can only have two values
(namely $1/3$ for massless particles and $0$ for any nonzero mass), the QFT results
allow $w$ to undertake a slow, temperature-independent transition from $1/3$ for massless
fermions down to $0$ as $k = m/\omega$ tends to large values.

\subsection{Behaviour at the origin}
\label{sec:comparison:origin}

We have seen from the profiles in figure \ref{fig:gen} that the FC, energy density and pressure all take their maximum values at the origin for fixed inverse temperature $\beta $ and fermion mass $m$.
To explore in more detail how the results from both QFT and kinetic theory depend on the fermion mass and temperature, in this section we consider the behaviour of all quantities at the space-time origin.
The maximal symmetry of adS is broken by fixing the inverse temperature at the point selected to be the origin, and the local inverse temperature (\ref{eq:tolman}) is then defined relative to the inverse temperature at the origin.

\begin{figure}
\begin{center}
\begin{tabular}{cc}
 \includegraphics[width=0.45\linewidth]{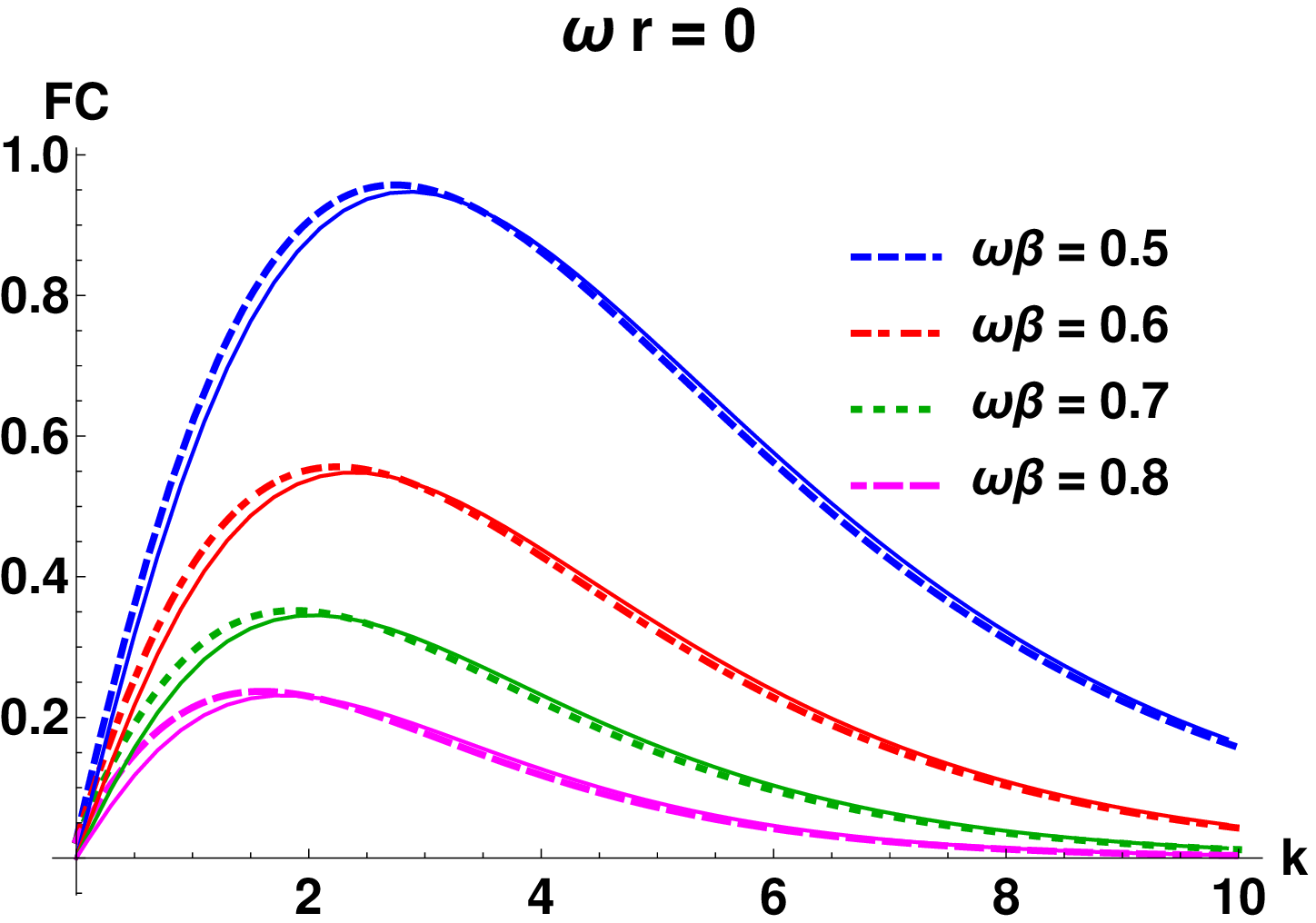} &
 \includegraphics[width=0.45\linewidth]{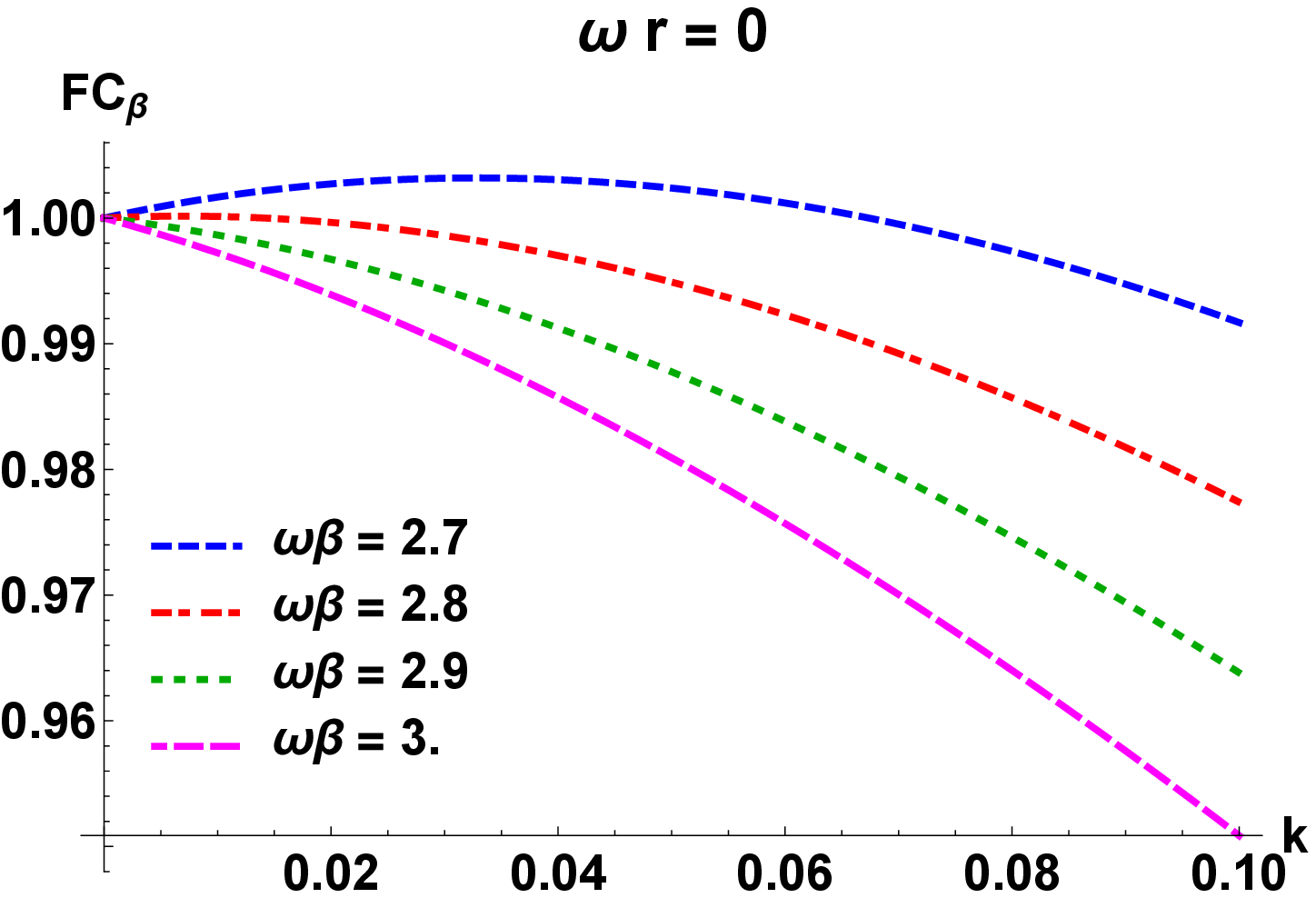} \\
 (a) & (b)
\end{tabular}
\end{center}
\caption{(a) $FC_{\beta}$ (\ref{eq:ppsi}) (dashed lines) and $FC_{-1}(\betat)$ (\ref{eq:kineticFC}) (continuous lines) evaluated at the origin $\omega r=0$  as functions of $k=m/\omega $ for various values of the inverse temperature $\beta $. (b) $FC_{\beta}$ (\ref{eq:ppsi}) for a massive fermion field at the origin divided by $FC_{\beta}$ for a massless fermion field at the origin, again as a function of $k$ for various values of $\beta \omega $.}
\label{fig:ppsiorigin}
\end{figure}

We begin by considering $FC_{\beta}$ (\ref{eq:ppsi}) and $FC_{-1}(\betat)$ (\ref{eq:kineticFC}) evaluated at the origin in figure \ref{fig:ppsiorigin}.
In the left-hand plot we show the values at the origin in both QFT and kinetic theory as a function of $k=m/\omega $ for a selection of values of the inverse temperature $\beta \omega $.
For fixed temperature, the QFT $FC_{\beta}$ at the origin is very small (but nonzero) when the fermion field is massless ($k=0$), whereas the kinetic theory $FC_{-1}(\betat)$ vanishes when $k=0$.
As $k$ increases with $\beta $ fixed for the values of $\beta \omega $ in figure \ref{fig:ppsiorigin}(a), both $FC_{\beta}$ and $FC_{-1}(\betat)$ at the origin increase until they reach a maximum at a certain value of the fermion mass.  They then decrease monotonically as $k$ increases for these values of $\beta $.

At the origin, the kinetic theory quantity $FC_{-1}(\betat)$ (\ref{eq:kineticFC}) has the same value as it does in Minkowski space-time (\ref{eq:minkowski}).
Therefore the qualitative shape of the profile of $FC_{-1}(\betat)$ as a function of $k=m/\omega $ does not change as $\beta $ varies and for all $\beta $ there is a maximum at some value of $k$.
In contrast, for the QFT quantity $FC_{\beta }$ considered at the origin as a function of $k$, the behaviour shown in figure \ref{fig:ppsiorigin}(a) at comparatively small values of $\beta \omega $ (corresponding to high temperatures) does not persist for larger values of $\beta \omega $.
This can be seen in figure \ref{fig:ppsiorigin}(b), where we show the $FC_{\beta }$ at the origin for nonzero $k$ divided by $FC_{\beta }$ at the origin when $k=0$ for a selection of larger values of $\beta \omega $.
We divide by the FC at the origin for $k=0$ in figure \ref{fig:ppsiorigin}(b) to make the behaviour easier to see.
When $\beta \omega =2.7$, we see that $FC_{\beta }$ at the origin initially increases for increasing $k$, reaches a maximum and then decreases as $k$ increases further.
However, for
$\beta\omega \ge 2.82857$,
we find that $FC_{\beta }$ is monotonically decreasing as $k$ increases from zero.
In order to find this value of $\beta\omega$ where $FC_{\beta}$ no longer initially increases with $k$, we compute the following derivative:
\begin{equation}
\fl \left.\frac{d(FC_\beta )}{dk}\right\rfloor_{k = 0, \omega r =0} = -\frac{\omega^3}{2\pi^2}
 \sum_{j = 1}^\infty (-1)^j \left(\frac{1}{\cosh\frac{\omega j \beta}{2}  \sinh^2\frac{\omega j \beta}{2}} -
 \frac{\ln \left[ \sinh^2\frac{\omega j \beta}{2} \right]}{\cosh^3\frac{\omega j \beta}{2}} \right).
\end{equation}
The value of $\beta\omega$ where the above derivative vanishes is found numerically to be $\sim 2.82857$, which is close, but not equal to $\sqrt{8} \simeq 2.82843$.

\begin{figure}
\begin{center}
\begin{tabular}{cc}
 \includegraphics[width=0.45\linewidth]{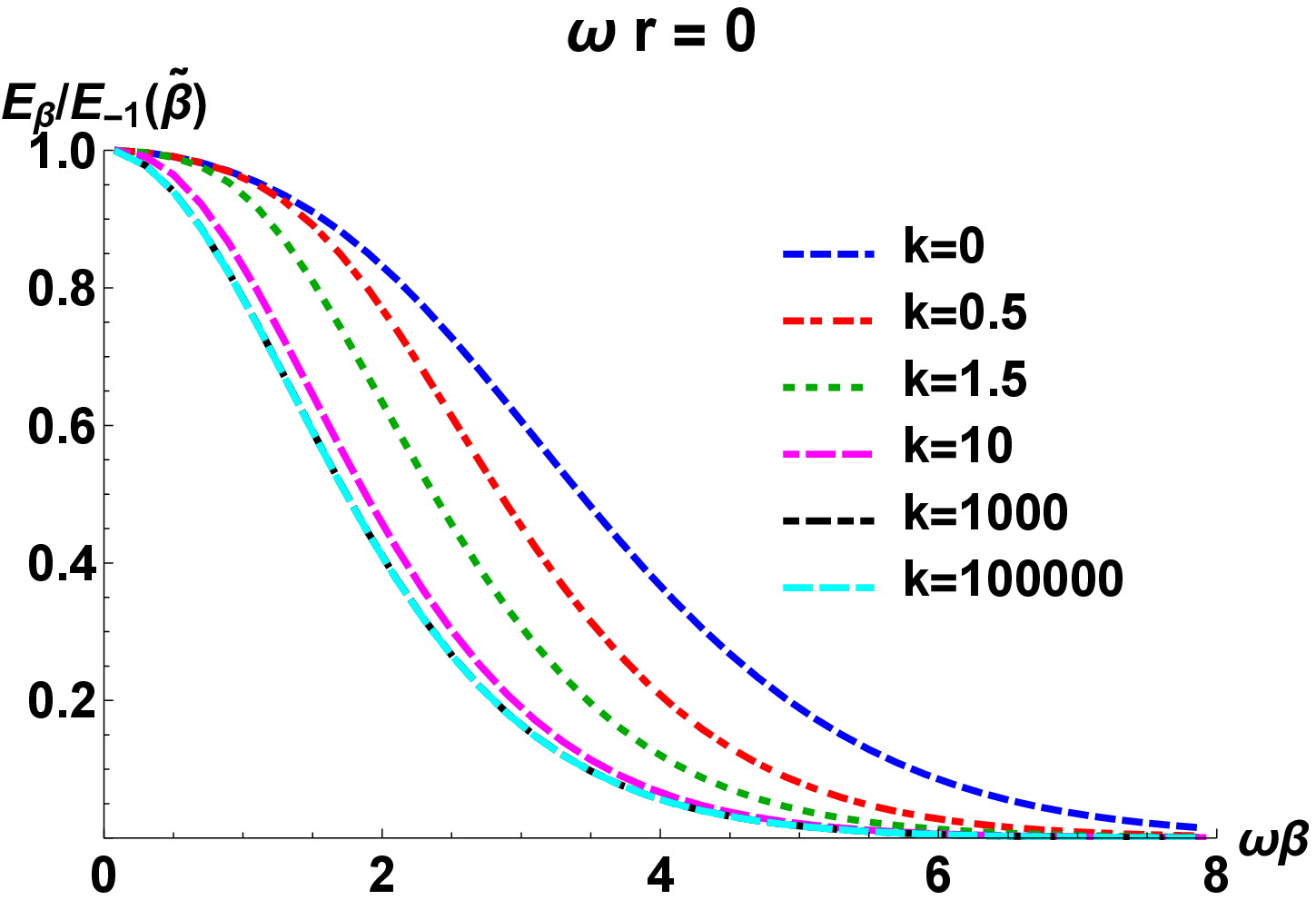} &
 \includegraphics[width=0.45\linewidth]{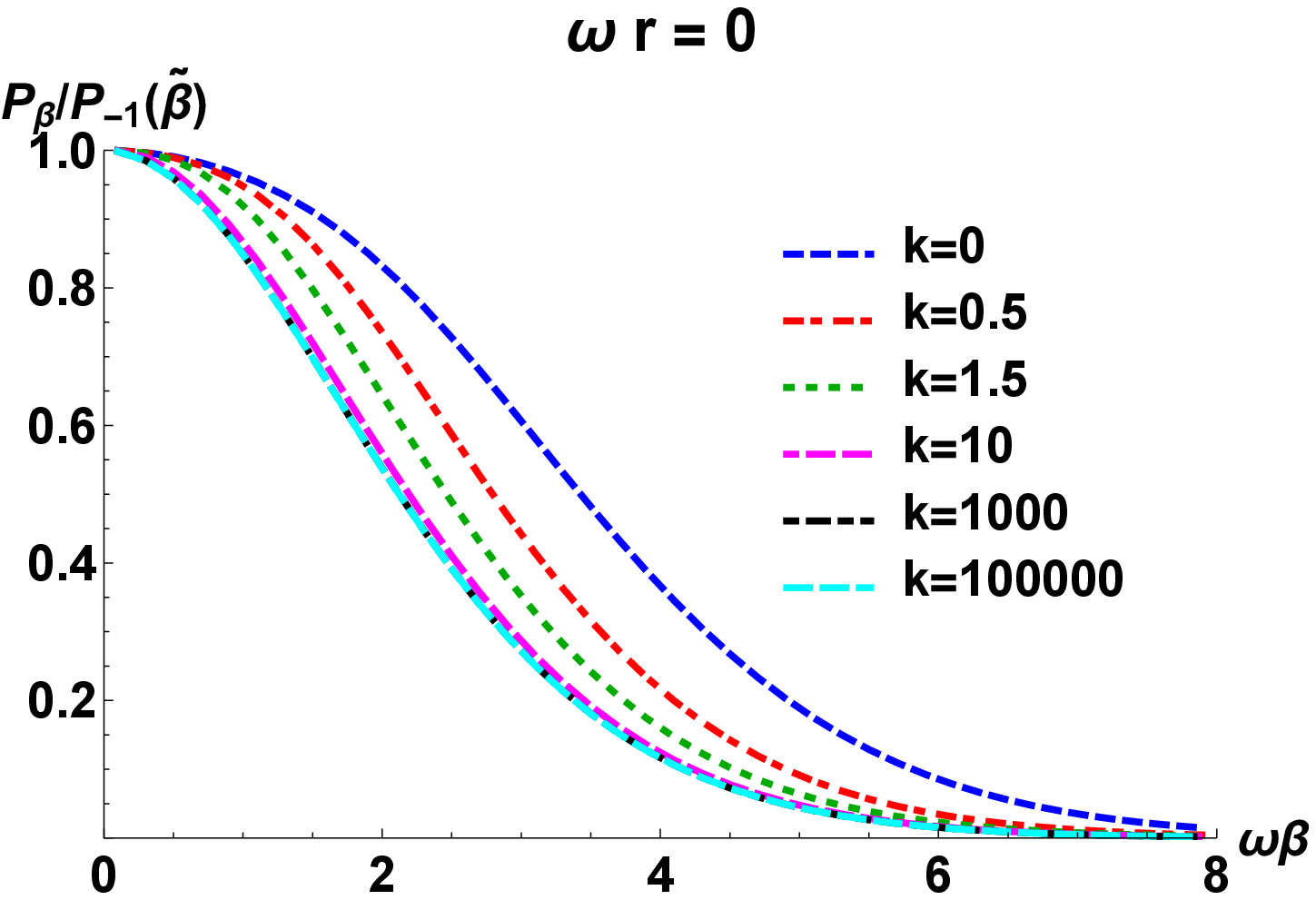} \\
 (a) & (b)
\end{tabular}
\begin{tabular}{c}
 \includegraphics[width=0.45\linewidth]{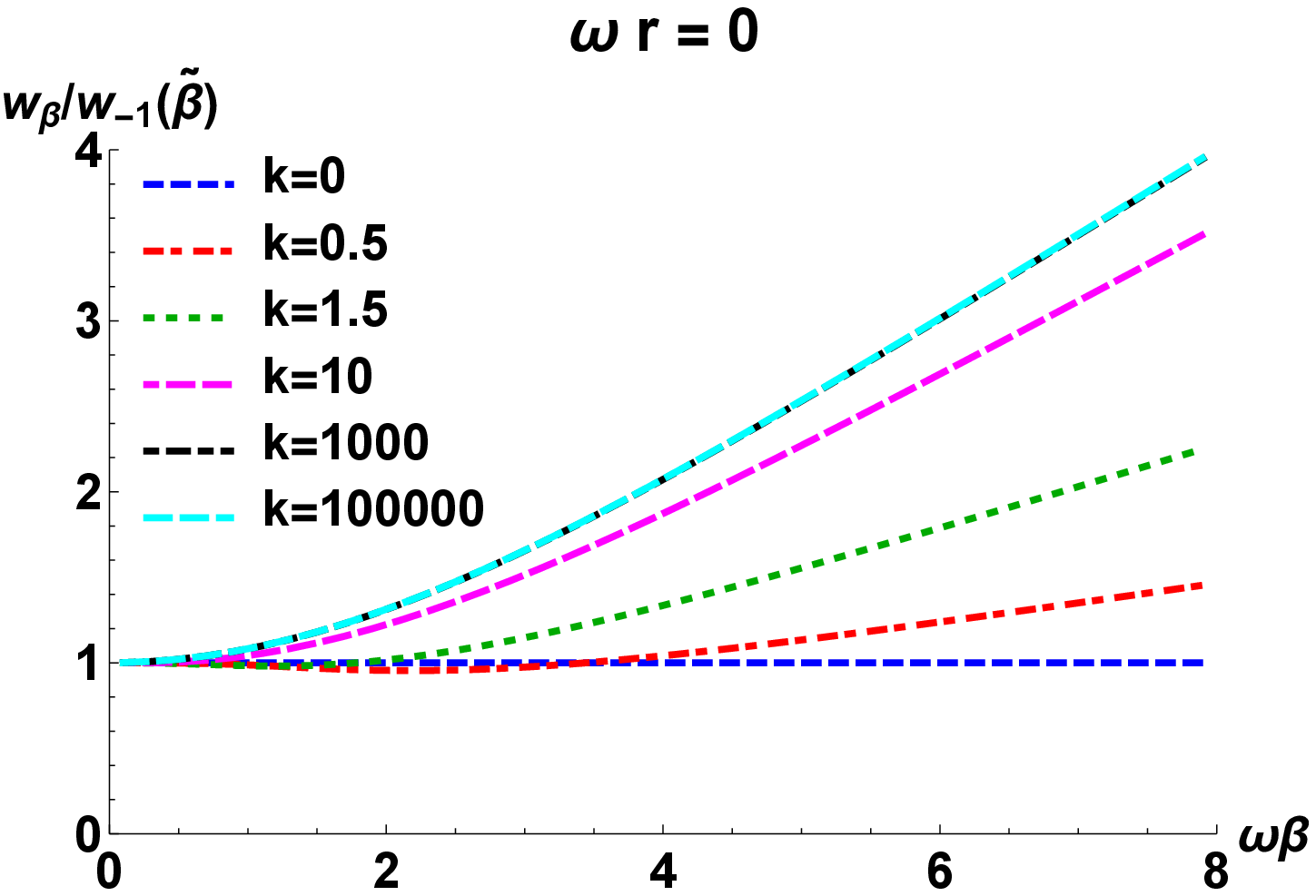}\\
 (c)
\end{tabular}
\end{center}
\caption{
(a) The ratio $E_\beta / E_{-1}(\betat)$ between
the energy densities obtained using QFT \eqref{eq:qft_m_E} and
kinetic theory \eqref{eq:kinetic_m_E},
(b) the ratio $P_\beta / P_{-1}(\betat)$ between the pressures obtained from QFT (\ref{eq:qft_m_P}) and kinetic theory (\ref{eq:kinetic_m_P}),
(c) the ratio $w_{\beta }/w_{-1}(\betat) $ between the equations of state corresponding to
QFT and kinetic theory.
All quantities are plotted as functions of $\beta \omega $ for a selection of fixed values of $k = m/\omega$.}
\label{fig:kr0}
\end{figure}

For the energy density and pressure, in both QFT and kinetic theory, we find a much simpler dependence on $k$ and $\beta \omega $ at the origin; as either $k$ increases or $\beta \omega $ increases, the values at the origin decrease.
We examine how the QFT and kinetic theory results compare in figure \ref{fig:kr0}.
At finite values of $k$, figure \ref{fig:kr0}(a) shows that
the ratio between the QFT result for the energy density \eqref{eq:qft_m_E} and the
kinetic theory energy density \eqref{eq:kinetic_m_E} decreases as $k$ increases for fixed values of
$\omega \beta$, approaching an asymptotic value as $k \rightarrow \infty$.
As $\beta \omega  \rightarrow 0$ and
the temperature increases to infinity, the ratio $E_\beta / E_{-1}(\betat)$
approaches unity for all values of $k$ examined.
This is to be expected since as the temperature increases, particles of
finite mass eventually behave as though they were massless.
Furthermore, at large values of the temperature, the quantum corrections to the energy density are negligible and the system becomes effectively classical.

Figure~\ref{fig:kr0}(b) presents a similar plot for the ratio between the QFT pressure \eqref{eq:qft_m_P}
and the kinetic theory pressure \eqref{eq:kinetic_m_P}.
It can be seen that the asymptotic behaviour
is approached at a smaller value of $k$, while
$P_\beta / P_{-1}(\betat)$ decreases more slowly with $\omega \beta$ than
the energy density ratio presented in figure \ref{fig:kr0}(a).

Finally, the equation of state $w_{\beta } = P_\beta / E_\beta$ is shown in figure \ref{fig:kr0}(c),
divided by the equation of state $w_{-1}(\betat) = P_{-1}(\betat)/ E_{-1}(\betat)$ predicted by kinetic theory.
It can be seen that, for large enough values of $\beta\omega$,
the ratio $w_{\beta} / w_{-1}(\betat)$ increases as $\beta\omega$ increases, approaching an asymptotic value
as $k \rightarrow \infty$.
These results indicate that the energy density is more strongly quenched by quantum corrections than the pressure.

\subsection{Massless limit}
\label{sec:comparison:m0}

We close this section by further examining the effect of quantum corrections on the energy density, restricting our attention to the massless limit $k=0$,
when the pressure is always one third of the energy density in both QFT and kinetic theory.
The expressions for the energy density obtained using QFT (\ref{eq:qft_m0}) and kinetic theory (\ref{eq:kinetic_m0}) simplify considerably for massless fermions, which aids our analysis. In both the QFT (\ref{eq:qft_m0}) and kinetic theory (\ref{eq:kinetic_m0}) energy densities, the coordinate dependence is contained only in an overall factor of $(\cos \omega r )^{4}$.
To see the effect of quantum corrections, it is instructive to consider the low- and high-temperature approximations to the QFT energy density (\ref{eq:qft_m0}).

To examine the high-temperature limit, we consider an expansion of \eqref{eq:qft_m0} in powers of $\beta\omega\ll 1$:
\begin{equation}
 E_\beta = \frac{7\pi^2}{60\beta^4} (\cos\omega r)^4
 \left[1 - \frac{5\beta^2\omega^2}{14\pi^2}
 - \frac{17 \beta^4 \omega^4}{112 \pi^4} + O([\beta  \omega ]^6)\right].
 \label{eq:qft_E_k0_small}
\end{equation}
The first term in (\ref{eq:qft_E_k0_small}) coincides with the
kinetic theory result in \eqref{eq:kinetic_m0}, while the second and third terms
represent quantum corrections which go to $0$ as $\beta\omega \rightarrow 0$.

The large $\beta \omega $  (low-temperature) behaviour of \eqref{eq:qft_m0} can be investigated by considering the following
series \cite{ambrus14phd}:
\begin{equation}
 \frac{\cosh \frac{j \omega \beta}{2}}{(\sinh \frac{j \omega \beta}{2})^4} =
 8 e^{-\frac{3}{2}j \omega \beta} \sum_{n = 0}^\infty
 \left(1 + \frac{13n}{6} + \frac{3n^2}{2} + \frac{n^3}{3}\right) e^{-n j \omega \beta}.
 \label{eq:phdseries}
\end{equation}
Substituting (\ref{eq:phdseries}) into \eqref{eq:qft_m0} gives an alternative formula for the QFT energy density:
\begin{equation}
\fl E_\beta = -\frac{6\omega^4}{\pi^2}
 \frac{(\cos \omega r)^4}{1 + e^{\frac{3}{2}\omega \beta}} \sum_{n = 0}^\infty
 e^{-n \omega \beta} \left(1 + \frac{13n}{6} + \frac{3n^2}{2} + \frac{n^3}{3}\right)
 \frac{1 + e^{-\frac{3}{2} \omega \beta}}{1 + e^{-(\frac{3}{2} + n) \omega \beta}}.
 \label{eq:qft_E_k0_large}
\end{equation}
It is clear that this tends to zero exponentially quickly as $\beta \omega \rightarrow \infty $, in contrast to the kinetic theory result (\ref{eq:kinetic_m0}) which tends to zero like $\left( \beta \omega \right) ^{-4}$ as $\beta \omega \rightarrow \infty $.
Thus, for low temperatures, the QFT energy density is considerably smaller than that arising from kinetic theory and quantum corrections are significant.

\begin{figure}
\begin{center}
\begin{tabular}{c}
 \includegraphics[width=0.75\linewidth]{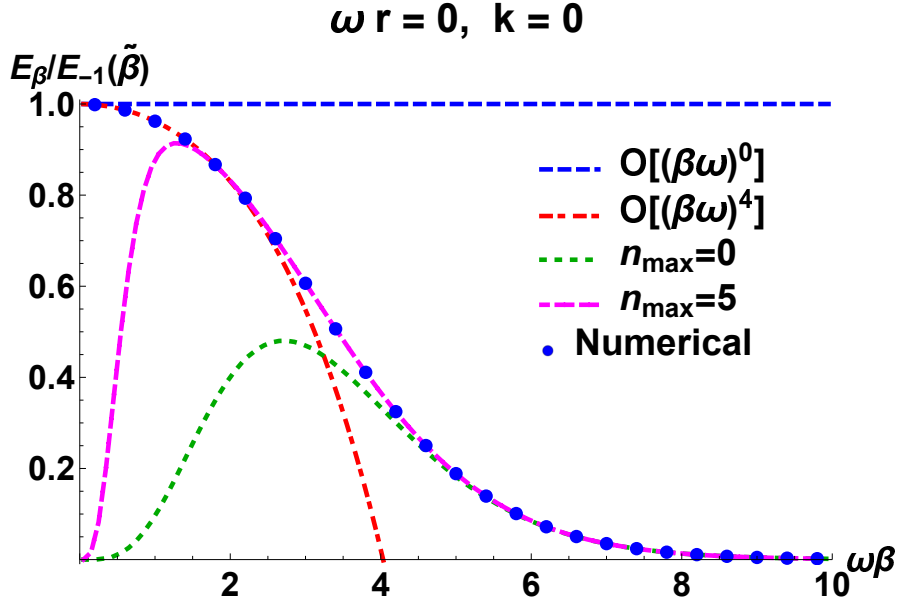}
\end{tabular}
\end{center}
\caption{
The effect of quantum corrections on the energy density. Blue dots show the QFT energy density $E_{\beta }$ computed numerically from \eqref{eq:qft_m0}
divided by the kinetic theory result $E_{-1}(\betat)$ (\ref{eq:kinetic_m0}), evaluated at the origin $\omega r =0$ as a function of $\beta \omega $ for massless fermions $k=0$.
We also plot the low- and high-temperature approximations to $E_{\beta }$ divided by $E_{-1}(\betat)$.
 For the blue line we use the leading order $O([\beta \omega ]^{0})$ term in the high-temperature approximation (\ref{eq:qft_E_k0_small}) to $E_{\beta }$ and for the red curve the high-temperature approximation (\ref{eq:qft_E_k0_small}) up to and including terms of $O([\beta \omega] ^{4})$.
The green and pink curves are, respectively, plotted using the low-temperature expression (\ref{eq:qft_E_k0_large}) for the QFT energy density $E_{\beta }$, with only the first ($n=0$) term in the series included for the green curve, while terms up to and including $n=5$ are included for the pink curve.}
\label{fig:k0r0}
\end{figure}

The effect of quantum corrections on the energy density can be seen in figure \ref{fig:k0r0}, where the blue dots show the QFT energy density $E_{\beta }$ (\ref{eq:qft_m0}) divided by the kinetic theory energy density $E_{-1}(\betat)$  (\ref{eq:kinetic_m0}).
As $\beta \omega \rightarrow 0$, this ratio tends to unity, as expected from (\ref{eq:qft_E_k0_small}), and quantum corrections are negligible.
For large values of $\beta \omega $, it can be seen that the quantum corrections are significant, with the QFT energy density being many times smaller than the kinetic theory result.

In figure \ref{fig:k0r0} we also show low- and high-temperature approximations to the QFT energy density $E_{\beta }$, again dividing by the kinetic energy density $E_{-1}(\betat)$.
As expected, the high-temperature approximation (\ref{eq:qft_E_k0_small}) to $E_{\beta }$ works well when $\beta \omega $ is small (especially the red curve, in which we retain terms up to and including $O([\beta \omega ]^{4})$ in (\ref{eq:qft_E_k0_small})), while the first few terms in the series in (\ref{eq:qft_E_k0_large}) are a good approximation for the low-temperature regime (particularly the pink curve for which we use terms up to and including $n=5$ in the series).
There is also a neighbourhood of $\beta \omega \sim 2$ where both the low-temperature series (\ref{eq:qft_E_k0_large}) including terms up to $n=5$ and the high-temperature approximation (\ref{eq:qft_E_k0_small}) with terms up to $O([\beta \omega ]^{4})$ are  good approximations to the QFT energy density.

\section{Conclusions}
\label{sec:conc}

In this paper we have studied t.e.v.s for a massive quantum fermion field propagating on four-dimensional adS space-time.
The maximal symmetry of the space-time enables us to derive a closed-form expression for the bispinor of parallel transport, which, when combined with the
imaginary time anti-periodicity property of the Feynman Green's function at finite temperature,
yields expressions for the differences between the t.e.v.s and the v.e.v.s as infinite sums involving hypergeometric functions.
We have also computed the corresponding quantities within relativistic kinetic theory and compared the results with the QFT expectation values.
In Minkowski space-time, relativistic kinetic theory results for fermions are exactly equal to the corresponding t.e.v.s computed in QFT. We have found that this is not the case in adS space-time, and quantum corrections to the kinetic theory results are significant.

De Sitter (dS) space-time is also maximally symmetric, but has positive rather than negative cosmological constant.
Due to the presence of a cosmological horizon, there is a preferred temperature for thermal states on dS space-time, namely the Gibbons-Hawking temperature \cite{Gibbons:1977mu}.
For a thermal state at the Gibbons-Hawking temperature, the t.e.v.~of the SET is proportional to the metric tensor and given exactly by the Brown-Ottewill-Page approximation \cite{page}.

In contrast, for adS space-time there is no such preferred temperature.
Any nonzero temperature breaks the maximal symmetry of the space-time by singling out an origin at which the local inverse temperature is $\beta $.
Thus, while for a massive quantum fermion field, the v.e.v.~of the FC is constant everywhere in adS, and that of the SET is proportional to the metric tensor \cite{art:ambrus15plb}, the t.e.v.~of the FC is not constant and that of the SET is not proportional to the metric tensor.
Similar behaviour has been found in \cite{Allen:1986ty} for thermal states for a massless, conformally coupled, scalar field on adS.
On the other hand, for all values of the fermion mass $m$ we find that the difference between the t.e.v.~and the v.e.v.~of the SET has the form of an ideal fluid throughout the space-time (\ref{eq:SETfluid}), whereas this is not the case for the massless, conformally coupled scalar field \cite{Allen:1986ty}.

In the massless case, the expressions (\ref{eq:qft_fc_mo}, \ref{eq:qft_m0}) for the t.e.v.s of the FC and energy density are rather simpler than the corresponding expressions in \cite{Allen:1986ty} for the t.e.v.s of the vacuum polarization and SET for a massless, conformally coupled, scalar field.
Applying the Brown-Ottewill-Page approximation \cite{page} to adS space-time, the difference between the t.e.v.~and the v.e.v.~of the SET has the perfect fluid form with energy density $E_{\beta }^{BOP}$:
\begin{equation}
E_{\beta }^{BOP} = \frac {7\pi ^{2}}{60\beta ^{4}}\left( \cos \omega r \right)^{4} \left[ 1 - \frac {5\omega ^{2}\beta ^{2}}{14\pi ^{2}}
- \frac {17\omega ^{4}\beta ^{4}}{112\pi ^{4}} \right] .
\label{eq:BOP}
\end{equation}
While the Brown-Ottewill-Page approximation (\ref{eq:BOP}) does not exactly reproduce (\ref{eq:qft_m0}) for all values of the inverse temperature $\beta $,
it does give the overall $\left( \cos \omega r\right) ^{4}$ dependence of (\ref{eq:qft_fc_mo}, \ref{eq:qft_m0}), as well as the perfect fluid form.
Furthermore, it agrees with the high-temperature limit (\ref{eq:qft_E_k0_small}) of our exact QFT results up to $O([\omega \beta ] ^{6})$.

For all values of the fermion mass, we find that the difference between the t.e.v.s and the v.e.v.s of all the quantities we have considered vanish on the space-time boundary, in both QFT and kinetic theory.  Similar behaviour has been found for a massless, conformally coupled, scalar field in both pure adS \cite{Allen:1986ty} and for the renormalized vacuum polarization on a Schwarzschild-adS black hole background \cite{Flachi:2008sr}.
Furthermore, in the black hole case the value of the renormalized vacuum polarization on the space-time boundary matches that of the v.e.v.~on pure adS space-time.
We see that thermal radiation (whether produced by a black hole or in pure adS) tends to ``clump'' away from the boundary.
This can be understood from the Tolman relation (\ref{eq:tolman}) for the local temperature, which vanishes as the boundary is approached.

While both the energy density $E$ and pressure $P$ vanish on the boundary, the equation of state $w=P/E$ can remain finite as the boundary is approached and has some interesting properties.
In kinetic theory, $w$ can take only two values on the boundary, either $1/3$ for massless fermions or $0$ for fermions of any nonzero mass.
In QFT, $w$ is again equal to $1/3$ for massless fermions. For massive fermions, the QFT $w$ on the boundary is independent of the inverse temperature $\beta $, and continuously decreases from $1/3$ down to $0$ as the fermion mass increases.

In this paper we have considered nonrotating thermal states on adS space-time.
In the past few years, there has been renewed interest in the properties of rotating thermal fermion states, both within kinetic theory \cite{ambrus16cota,ambrus15wut}
and in QFT on Minkowski space-time \cite{rot1,rot2,rotating}.
The properties of equilibrium states of gases undergoing rigid rotation in adS have been recently studied \cite{ambrus16cota}, focussing on the bulk viscosity.
It would be interesting to extend our QFT analysis in this paper to rotating states for fermions on adS and compare with the kinetic theory results.
On unbounded Minkowski space-time, it is known that rotating scalar and fermion fields have rather different properties \cite{rot1}.
In particular the only possible vacuum state for scalars is the nonrotating Minkowski vacuum, while for fermions a rotating vacuum can also be defined.
This means that rotating thermal states for fermions can be defined, whereas for scalar fields rotating thermal states are ill-defined everywhere in the space-time.
Rotating thermal states for fermions on unbounded Minkowski space-time are regular everywhere inside the speed-of-light (SOL) surface, where they diverge \cite{rot1}.
On adS, due to the time-like boundary, there may or may not be an SOL depending on the angular speed of rotation \cite{ambrus16cota}.
For a quantum scalar field, the only possible global vacuum state on adS is the nonrotating vacuum \cite{Kent:2014wda}.
We conjecture that the situation for a quantum fermion field is likely to be different and expect to be able to define a rotating vacuum.
If the angular speed is sufficiently small and there is no SOL,  one might expect that rotating t.e.v.s for the fermion field would have some properties similar
to those in bounded Minkowski space-time \cite{rot2}, while if there is an SOL, it seems likely that rotating t.e.v.s would diverge there, in analogy with the situation in unbounded Minkowksi space-time \cite{rot1}.
Testing these conjectures requires a full analysis of the QFT of rigidly-rotating fermions on adS, to which we plan to return in a forthcoming publication.

\ack
V.E.A.~was partially supported by a studentship from the
School of Mathematics and Statistics at the University of Sheffield,
as well as by a grant from the
Romanian National Authority for Scientific Research and Innovation,
CNCS-UEFISCDI, project number PN-II-RU-TE-2014-4-2910.
V.E.A.~thanks the Mathematics and Statistics Research Centre in the School of Mathematics and Statistics at the University of Sheffield for hospitality while this work was completed.
The work of E.W.~is supported by the Lancaster-Manchester-Sheffield Consortium for
Fundamental Physics under STFC grant ST/L000520/1.

\appendix

\section{Analytic expression for the bispinor of parallel transport}\label{app:bispinor}

The bispinor of parallel transport $\Lambda(x,x')$ is the solution of \eqref{eq:lambda_pt} subject to the initial
conditions \eqref{eq:lambda_ic}. On adS space-time, $\Lambda(x,x')$ also satisfies \eqref{eq:DLambda}.
For later convenience, we introduce an auxiliary bispinor $\lambda (x,x')$ as follows:
\begin{equation}
 \Lambda(x,x') = \frac{1}{\cos\frac{\omega s}{2} \sqrt{\cos \omega r \cos \omega r'}} \lambda(x,x').
 \label{eq:lambda_def}
\end{equation}
From~\eqref{eq:DLambda} we deduce the following equations for $\lambda (x,x')$:
\begin{eqnarray}
 D_\halpha \left[ \frac{\lambda}{\sqrt{\cos\omega r \cos\omega r'}} \right] = \frac{\omega}{2} \tan\left(\frac{\omega s}{2}\right)
 \gamma_\halpha  \frac{\slashed{n}\lambda}{\sqrt{\cos\omega r \cos \omega r'}},
\label{eq:Dlambda}\\
 D_{\halpha'}\left[ \frac{\lambda}{\sqrt{\cos\omega r \cos\omega r'}} \right] = \frac{\omega}{2} \tan\left(\frac{\omega s}{2}\right)
 \frac{\lambda}{\sqrt{\cos\omega r \cos \omega r'}} \slashed{n}' \gamma_{\halpha'},
 \label{eq:Dplambda}
\end{eqnarray}
and the initial conditions \eqref{eq:lambda_ic} become:
\begin{eqnarray}
 \lambda(x, x) = \cos\omega r, \qquad
 \overline{\lambda}(x, x') = \lambda(x', x),  \nonumber \\
 \lambda(x, x') \lambda(x', x) = \cos^2\frac{\omega s}{2} \cos\omega r \cos\omega r'.\label{eq:lambda_inv}
\end{eqnarray}

The construction of the solution of (\ref{eq:Dlambda}, \ref{eq:Dplambda}) will be performed in four steps:
(1) finding the time dependence of $\lambda$ (\ref{app:bispinor:time}); (2)
finding the radial dependence of $\lambda$ (\ref{app:bispinor:radial});
(3) simplifying the solution in terms of functions of the angular coordinates (\ref{app:bispinor:sol});
and finally (4) determining the angular dependence of the remaining functions of the angular coordinates
 (\ref{app:bispinor:eq}).

\subsection{Time dependence}\label{app:bispinor:time}

For the construction of $\lambda(x,x')$, both (\ref{eq:Dlambda}, \ref{eq:Dplambda}) need to be
considered. For brevity, the details  will be shown for \eqref{eq:Dlambda} (which involves derivatives with respect to the unprimed indices).
The analysis of (\ref{eq:Dplambda}) follows similarly.

The time dependence of $\lambda$ can be determined by setting $\halpha = \hatt$
in \eqref{eq:Dlambda}, yielding:
\begin{eqnarray}
 \fl (\cos\omega\Delta t + C_\gamma)\partial_{\omega t/2} \lambda + \sin\left(\omega \Delta t \right) \lambda \nonumber\\
 = -\tan\left(\frac{\omega r}{2}\right)(\cos\omega r' + C_\gamma) \gamma^\hatt \frac{\vx \cdot \vgamma}{r} \lambda
 - \sin\left(\omega r'\right) \gamma^\hatt \frac{\vx' \cdot \vgamma}{r'} \lambda,
 \label{eq:dtlambda_aux}
\end{eqnarray}
where we have defined a quantity $C_{\gamma }$ by
\begin{equation}
 C_\gamma = \cos\omega r \cos\omega r'(1 - \cos\gamma \tan\omega r \tan\omega r'),
 \label{eq:Cgamma_def}
\end{equation}
in terms of which the coefficient of $\partial_{\omega t / 2}$ was
obtained using the following relation:
\begin{equation}
 \cos\omega r \cos\omega r'(1 + \cos\omega s) = \cos\omega\Delta t + C_\gamma .
\end{equation}

To solve \eqref{eq:dtlambda_aux}, it is convenient to cast it in the form:
\begin{equation}
 \begin{pmatrix}
  A_t & B_t \\ B_t & A_t
 \end{pmatrix}
 \lambda(x,x') = 0,
 \label{eq:At_def}
\end{equation}
where $A_t$ and $B_t$ are $2\times 2$ matrices.
Using the result
\begin{equation}
 \begin{pmatrix}
  A & -B \\
  -B & A
 \end{pmatrix}
 \begin{pmatrix}
  A & B \\
  B & A
 \end{pmatrix}
 =
 \begin{pmatrix}
  A^2 - B^2 & [A, B]\\
  {[}A, B] & A^2 - B^2,
 \end{pmatrix}
\end{equation}
it can be seen that \eqref{eq:At_def} can be diagonalised if $A_t$ and $B_t$ commute. The choice
\begin{eqnarray}
 A_t = ( \cos\omega \Delta t + C_\gamma) \partial_{\omega t / 2} + \sin\omega \Delta t,\nonumber \\
 B_t = \tan\left(\frac{\omega r}{2}\right) (\cos\omega r' + C_\gamma) \frac{\vx \cdot \vsigma}{r} +
 \sin\omega r' \frac{\vx' \cdot \vsigma}{r'}, \label{eq:At_Bt}
\end{eqnarray}
satisfies $[A_t, B_t] = 0$, since $A_t$ is proportional to the $2\times 2$ identity matrix and $B_t$ has no time dependence.
Squaring $A_t$ and $B_t$ from \eqref{eq:At_Bt} gives:
\begin{eqnarray}
 A_t^2 = (\cos\omega \Delta t + C_\gamma)^2 \left[ \frac{\partial^2}{\partial(\omega t / 2)^2} + 1\right] +
 1 - C_\gamma^2,\nonumber\\
 B_t^2 = 1 - C_\gamma^2.\label{eq:AtBt_sq}
\end{eqnarray}
A similar analysis can be performed on \eqref{eq:Dplambda}. In this case a suitable choice of matrices $A_{t'}$ and $B_{t'}$ is
\begin{eqnarray}
 A_{t'} = \overleftarrow{\partial}_{\omega t' / 2} ( \cos\omega \Delta t + C_\gamma) - \sin\omega \Delta t,\nonumber \\
 B_{t'} = -\tan\left(\frac{\omega r'}{2}\right) (\cos\omega r + C_\gamma) \frac{\vx' \cdot \vsigma}{r'} -
 \sin\omega r \frac{\vx \cdot \vsigma}{r},
\end{eqnarray}
so that:
\begin{equation}
 \lambda(x,x')
 \begin{pmatrix}
  A_{t'} & B_{t'} \\ B_{t'} & A_{t'}
 \end{pmatrix} = 0.
 \label{eq:Atp_def}
\end{equation}
The squares of $A_{t'}$ and $B_{t'}$ can be obtained from \eqref{eq:AtBt_sq} by replacing $t\leftrightarrow t'$.

Therefore, $\lambda(x,x')$ obeys the following two differential equations:
\begin{equation}
 \frac{\partial \lambda}{\partial (\omega t / 2)^2} + \lambda = 0,\qquad
 \frac{\partial \lambda}{\partial (\omega t' / 2)^2} + \lambda = 0,
\end{equation}
so that $\lambda(x,x')$ is a harmonic function of $\omega t / 2$ and $\omega t' / 2$.
Since in \eqref{eq:lambda_def} the coordinates $t$ and $t'$ appear only through
the combination $\Delta t = t - t'$, it can be assumed that $\lambda$ depends on $t$ and $t'$ only through $\Delta t$.

For our later analysis, it is convenient to
introduce the $2 \times 2$ constituents $\lambda_{ij} \equiv \lambda_{ij}(x,x')$ ($i, j = 1, 2$) of $\lambda$ by:
\begin{equation}
 \lambda(x,x') =
 \begin{pmatrix}
  \lambda_{11}(x,x') & \lambda_{12}(x,x') \\
  \lambda_{21}(x,x') & \lambda_{22}(x,x')
 \end{pmatrix}.
 \label{eq:lambdaij_def}
\end{equation}
Since $\lambda(x,x')$ is a harmonic function of $\Delta t$, its constituents can be written as:
\begin{equation}
 \lambda_{ij}(x,x') = \mathcal{C}_{ij} \cos\frac{\omega \Delta t}{2} + \mathcal{S}_{ij} \sin\frac{\omega \Delta t}{2},
 \label{eq:csdef}
\end{equation}
where the coefficients $\mathcal{C}_{ij}$ and $\mathcal{S}_{ij}$ are $2\times 2$ matrices that depend only on $\vx$ and $\vx'$.
In the coincidence limit, the second term above vanishes, while, from (\ref{eq:lambda_inv}), $\mathcal{C}_{ij}$ should satisfy
\begin{equation}
 \mathcal{C}_{ij}\rfloor_{\vx' = \vx} = \delta_{ij} \cos\omega r.
\end{equation}

\subsection{Radial dependence}\label{app:bispinor:radial}

The equation giving the dependence of $\lambda$ on $r$ can be obtained by multiplying \eqref{eq:Dlambda}
by $x^j \omega_j^\halpha / r$. After some rearrangement, we find the following relation:
\begin{eqnarray}
 \fl \Bigg[(\cos\omega \Delta t + C_\gamma) \partial_{\omega r / 2} +
 \tan\frac{\omega r}{2} (\cos\omega r' + C_\gamma)\nonumber\\
 - \sin\omega r' \frac{\vx \cdot \vgamma}{r} \frac{\vx'\cdot \vgamma}{r'} -
 \sin\omega \Delta t \frac{\vx \cdot \vgamma}{r} \gamma^\hatt\Bigg] \lambda = 0.
\end{eqnarray}
The above equation can be brought into the form \eqref{eq:At_def} by multiplying by
\begin{equation}
i \varepsilon_{ijk} x^i \gamma^\hatj \gamma^\hatk / 2r =
\begin{pmatrix}
\vx \cdot \vsigma /r & 0 \\
0 & \vx \cdot \vsigma /r
\end{pmatrix} ,
\end{equation}
yielding:
\begin{eqnarray}
 \fl A_r = \frac{\vx\cdot\vsigma}{r} \left[(\cos\omega\Delta t + C_\gamma)\partial_{\omega r / 2} +
 \tan\left(\frac{\omega r}{2}\right)(\cos\omega r' + C_\gamma)\right]
 + \frac{\vx'\cdot \vsigma}{r'} \sin\omega r' ,
 \nonumber \\
 \fl B_r = \sin\omega \Delta t.
 \label{eq:Ar}
\end{eqnarray}
After some algebra, it can be shown that:
\begin{equation}
 A_r^2 = (\cos\omega\Delta t + C_\gamma)^2 \left[\frac{\partial^2}{\partial(\omega r / 2)^2} + 1\right]
 + \sin^2\omega \Delta t,
\end{equation}
while $B_r^2 = \sin^2 \omega \Delta t$. By applying the same methodology as in \ref{app:bispinor:time},
it can be shown that $\lambda$ is a harmonic function of $\omega r / 2$:
\begin{equation}
 \frac{\partial^2\lambda}{\partial(\omega r / 2)^2} + \lambda = 0.\label{eq:lambdar}
\end{equation}
Employing the same method for the $r'$ dependence, it can be shown that $\lambda$ also satisfies:
\begin{equation}
 \frac{\partial^2\lambda}{\partial(\omega r' / 2)^2} + \lambda = 0.\label{eq:lambdarp}
\end{equation}
Equations (\ref{eq:lambdar}--\ref{eq:lambdarp}) imply that $\lambda$ is a harmonic function of both $\omega r/2$ and $\omega r'/2$.
In particular, the $2\times 2$ components $\lambda_{ij}$, as well as the matrices
${\mathcal {C}}_{ij}$ and ${\mathcal {S}}_{ij}$ (\ref{eq:csdef}) must also be harmonic functions of $\omega r/2$ and $\omega r'/2$.
We therefore write ${\mathcal {C}}_{ij}$ and ${\mathcal {S}}_{ij}$ in the following form:
\begin{eqnarray}
\fl \mathcal{C}_{ij} = c^{cc}_{ij} \cos\frac{\omega r}{2} \cos\frac{\omega r'}{2} +
 c^{cs}_{ij} \cos\frac{\omega r}{2} \sin\frac{\omega r'}{2}
 +c^{sc}_{ij} \sin\frac{\omega r}{2} \cos\frac{\omega r'}{2} +
 c^{ss}_{ij} \sin\frac{\omega r}{2} \sin\frac{\omega r'}{2},\nonumber\\
\fl \mathcal{S}_{ij} = s^{cc}_{ij} \cos\frac{\omega r}{2} \cos\frac{\omega r'}{2} +
 s_{ij}^{cs} \cos\frac{\omega r}{2} \sin\frac{\omega r'}{2}
+ s_{ij}^{sc} \sin\frac{\omega r}{2} \cos\frac{\omega r'}{2} +
 s_{ij}^{ss} \sin\frac{\omega r}{2} \sin\frac{\omega r'}{2},
 \nonumber \\
 \label{eq:lambda_coeffs_def}
\end{eqnarray}
where the coefficients $c_{ij}^{ab}$ and $s_{ij}^{ab}$ depend only on the orientations $\vn = \vx / r$ and
$\vn' = \vx' / r'$ of $\vx$ and $\vx'$.
In other words, $c_{ij}^{ab}$ and $s_{ij}^{ab}$ depend only on the angular coordinates $\theta $, $\theta '$, $\varphi $ and $\varphi '$.

\subsection{Simplifying the solution}\label{app:bispinor:sol}

We now simplify the expressions (\ref{eq:lambdaij_def}, \ref{eq:csdef},  \ref{eq:lambda_coeffs_def}) for the auxiliary bispinor $\lambda (x,x')$
by finding relations between the coefficients $c_{ij}^{ab}$ and $s_{ij}^{ab}$ which will enable us to reduce the number of independent functions of the angular coordinates.

Some relations involving the terms in $\mathcal{C}_{11}$ and $\mathcal{S}_{11}$ can be obtained
by considering the expression for the off-diagonal component $\lambda_{21}$,
which can be obtained via \eqref{eq:At_def}:
\begin{equation}
 \lambda_{21} = -B_t^{-1} A_t \lambda_{11}.
\end{equation}
The operator $A_t$, defined in \eqref{eq:At_Bt}, has the following action on the harmonic
functions of argument $\omega \Delta t / 2$:
\begin{equation}
\fl A_t \cos\frac{\omega \Delta t}{2} = \sin\left(\frac{\omega \Delta t}{2}\right) (1 - C_\gamma), \qquad
 A_t \sin\frac{\omega \Delta t}{2} = \cos\left(\frac{\omega \Delta t}{2}\right) (1 + C_\gamma),
 \label{eq:At_harm}
\end{equation}
where $C_\gamma$ is defined in \eqref{eq:Cgamma_def}. Furthermore, the inverse of the matrix
$B_t$ \eqref{eq:At_Bt} can be written as:
\begin{equation}
B_t^{-1} = \frac{1}{1-C_\gamma^2} B_t.
\end{equation}
Thus, $\lambda_{21}$ takes the form:
\begin{eqnarray}
\fl \lambda_{21} &= \mathcal{C}_{21} \cos\frac{\omega \Delta t}{2} + \mathcal{S}_{21} \sin\frac{\omega \Delta t}{2}\nonumber\\
\fl &= -\left[\tan\left(\frac{\omega r}{2}\right)
(\cos\omega r' + C_\gamma)
\frac{\vx \cdot \vsigma}{r} +
 \sin\omega r' \frac{\vx' \cdot \vsigma}{r'}\right]
 \left(\frac{\mathcal{C}_{11} \sin\frac{\omega \Delta t}{2}}{1 + C_\gamma} +
 \frac{\mathcal{S}_{11} \cos\frac{\omega \Delta t}{2}}{1 - C_\gamma}\right).
 \nonumber \\ \fl &
 \label{eq:lambda21_aux}
\end{eqnarray}
When $r' = 0$,  the expression \eqref{eq:lambda21_aux} reduces to:
\begin{eqnarray}
\fl \lambda_{21}\rfloor_{r' = 0} = -\frac{\vx \cdot \vsigma}{r} \left[
 \sin\frac{\omega \Delta t}{2} \tan\frac{\omega r}{2}
 \left(c_{11}^{cc} \cos\frac{\omega r}{2} + c_{11}^{sc} \sin\frac{\omega r}{2}\right)\right.\nonumber\\
 \left. + \cos\frac{\omega \Delta t}{2} \cot\frac{\omega r}{2}
 \left(s_{11}^{cc} \cos\frac{\omega r}{2} + s_{11}^{sc} \sin\frac{\omega r}{2}\right)\right].
\end{eqnarray}
Since $\lambda_{21}\rfloor_{r' = 0}$ must be a harmonic function of $\omega r / 2$,
it follows that:
\begin{equation}
 c_{11}^{sc} = s_{11}^{cc} = 0,
\end{equation}
which holds for all $\vn$ and $\vn'$.
In the case when $r = 0$, the expression \eqref{eq:lambda21_aux} simplifies to:
\begin{eqnarray}
\fl \lambda_{21}\rfloor_{r = 0} = -\frac{\vx' \cdot \vsigma}{r'} \left[
 \sin\frac{\omega \Delta t}{2} \tan\frac{\omega r'}{2}
 \left(c_{11}^{cc} \cos\frac{\omega r'}{2} + c_{11}^{cs} \sin\frac{\omega r'}{2}\right)\right.\nonumber\\
 \left.+ \cos\frac{\omega \Delta t}{2} \cot\frac{\omega r'}{2}
 s^{cs}_{11} \sin\frac{\omega r'}{2}\right],
\end{eqnarray}
implying that $c_{11}^{cs} = 0$.

We now consider the construction of $\lambda_{21}$ using the radial matrices $A_r$ and $B_r$:
\begin{equation}
 \lambda_{21} = -B_r^{-1} A_r \lambda_{11},
 \label{eq:lambda21const}
\end{equation}
where $B_r^{-1} = 1 / \sin\omega \Delta t$. The action of $A_r$ \eqref{eq:Ar} on harmonic functions
of $\omega r / 2$ is given by:
\begin{eqnarray}
\fl A_r\cos\frac{\omega r}{2} = \sin\left(\frac{\omega r}{2}\right) (-\cos\omega \Delta t + \cos\omega r') \frac{\vx\cdot\vsigma}{r}
 + \cos\left(\frac{\omega r}{2}\right) \sin\omega r' \frac{\vx'\cdot\vsigma}{r'},\nonumber\\
\fl A_r\sin\frac{\omega r}{2} = \cos\left(\frac{\omega r}{2}\right) (\cos\omega \Delta t + \cos\omega r') \frac{\vx\cdot\vsigma}{r}
 - \sin\left(\frac{\omega r}{2}\right) \sin\omega r' \frac{\vx\cdot\vsigma}{r} \frac{\vx'\cdot\vsigma}{r'} \frac{\vx\cdot\vsigma}{r}.
 \nonumber \\
\end{eqnarray}
To make use of the above properties, it is useful to write:
\begin{equation}
 \lambda_{11} = C^{r}_{11} \cos\frac{\omega r}{2} + S^r_{11} \sin\frac{\omega r}{2},
\end{equation}
where
\begin{eqnarray}
 C^r_{11} = c_{11}^{cc} \cos\frac{\omega \Delta t}{2} \cos\frac{\omega r'}{2} +
 s_{11}^{cs} \sin\frac{\omega \Delta t}{2} \sin\frac{\omega r'}{2}, \nonumber\\
 S^r_{11} = c_{11}^{ss} \cos\frac{\omega \Delta t}{2} \sin\frac{\omega r'}{2} +
 \sin\frac{\omega \Delta t}{2} \left(s_{11}^{sc} \cos\frac{\omega r'}{2} +
 s_{11}^{ss} \sin\frac{\omega r'}{2}\right).
\end{eqnarray}
Thus, $\lambda_{21}$ can be written as:
\begin{eqnarray}
\fl \lambda_{21} = -\frac{1}{\sin\omega \Delta t} \left\{
 \left[\sin\left(\frac{\omega r}{2}\right)(-\cos\omega \Delta t + \cos\omega r') \frac{\vx\cdot\vsigma}{r} +
 \cos\left(\frac{\omega r}{2}\right) \sin\omega r'
 \frac{\vx' \cdot \vsigma}{r'}\right]C^r_{11} \right.\nonumber\\
\fl \left. + \left[\cos\left(\frac{\omega r}{2}\right)(\cos\omega\Delta t + \cos\omega r') \frac{\vx\cdot\vsigma}{r} -
 \sin\left(\frac{\omega r}{2}\right)\sin\omega r' \frac{\vx\cdot\vsigma}{r} \frac{\vx'\cdot\vsigma}{r'} \frac{\vx\cdot\vsigma}{r}\right]S^r_{11}\right\}.
 \label{eq:lambda21_r}
\end{eqnarray}
In the limit $r = 0$, the above equation reduces to:
\begin{eqnarray}
\fl \lambda_{21}\rfloor_{r = 0} =
\sin \left(\frac{\omega\Delta t}{2}\right) \sin\left(\frac{\omega r'}{2} \right) \frac{\vx\cdot\vsigma}{r} c_{11}^{ss}
\nonumber \\ -
 \cos\left(\frac{\omega\Delta t}{2}\right) \frac{\vx\cdot \vsigma}{r} \left[\cos\left(\frac{\omega r'}{2}\right) s_{11}^{sc} + \sin\left(\frac{\omega r'}{2} \right) s_{11}^{ss}\right]\nonumber\\
 -\frac{1}{\sin\omega\Delta t}\left[\sin\omega r' \cos\left(\frac{\omega r'}{2}\right) \cos\left(\frac{\omega \Delta t}{2}\right)
 \left(\frac{\vx'\cdot \vsigma}{r'} c_{11}^{cc} + \frac{\vx \cdot \vsigma}{r} c_{11}^{ss}\right) \right.\nonumber\\
 \left. + \sin\frac{\omega \Delta t}{2} \sin\frac{\omega r'}{2} \sin\omega r' \left(\frac{\vx'\cdot\vsigma}{r'}s_{11}^{cs} -
 \frac{\vx\cdot\vsigma}{r} s_{11}^{sc} - \tan\left(\frac{\omega r'}{2}\right) \frac{\vx\cdot\vsigma}{r} s_{11}^{ss}\right)\right].
 \nonumber \\
\end{eqnarray}
Since $\lambda_{21}$ must be a harmonic function of $\omega\Delta t/2$, the coefficient of $1/\sin\omega \Delta t$ must
vanish. Inside the square brackets, the coefficients of
$\cos\frac{\omega \Delta t}{2}$ and $\sin\frac{\omega \Delta t}{2}$ must vanish individually,
since they are linearly independent functions. Furthermore,
$\sin\frac{\omega r'}{2} \sin\omega r' = (\cos\frac{\omega r'}{2} - \cos\frac{3\omega r'}{2})/2$ and
$\sin\frac{\omega r'}{2} \sin\omega r' \tan\frac{\omega r'}{2} = (3\sin\frac{\omega r'}{2} - \sin\frac{3\omega r'}{2})/2$
are also linearly independent, thus requiring that their coefficients vanish separately. Altogether, the following
relations are obtained:
\begin{equation}
 s_{11}^{ss} = 0, \qquad
 c_{11}^{ss} = -\frac{\vx\cdot\vsigma}{r} \frac{\vx'\cdot\vsigma}{r'} c_{11}^{cc}, \qquad
 s_{11}^{sc} = \frac{\vx\cdot\vsigma}{r} \frac{\vx'\cdot\vsigma}{r'} s_{11}^{cs}.
\end{equation}
Using the above information, \eqref{eq:lambda21_r} simplifies to:
\begin{eqnarray}
\fl \lambda_{21} = \cos\left( \frac{\omega\Delta t}{2}\right) \left[\sin\left( \frac{\omega r}{2}  \right) \sin\left( \frac{\omega r'}{2}\right) \frac{\vx\cdot\vsigma}{r} -
 \cos\left( \frac{\omega r}{2}\right) \cos\left( \frac{\omega r'}{2}\right) \frac{\vx'\cdot\vsigma}{r'} \right]s_{11}^{cs} \nonumber\\
\!\!\!\!\!\!\!\!\! - \sin\left(\frac{\omega\Delta t}{2} \right) \left[\sin\left(\frac{\omega r}{2}\right) \cos\left(\frac{\omega r'}{2}\right) \frac{\vx\cdot\vsigma}{r} +
 \cos\left(\frac{\omega r}{2}\right) \sin\left(\frac{\omega r'}{2}\right) \frac{\vx'\cdot\vsigma}{r'} \right] c_{11}^{cc}.
 \nonumber \\
 \label{eq:lambda_21_coeff}
\end{eqnarray}

From the above analysis, we conclude that $\lambda_{11}$ can be written using two unknown coefficients, namely
$c_{11}^{cc}$ and $s_{11}^{cs}$:
\begin{eqnarray}
\fl \lambda_{11} = \cos\left( \frac{\omega\Delta t}{2} \right) \left[\cos\left(\frac{\omega r}{2}\right) \cos\left(\frac{\omega r'}{2}\right) -
 \sin\left(\frac{\omega r}{2}\right) \sin\left(\frac{\omega r'}{2}\right)\frac{\vx\cdot\vsigma}{r} \frac{\vx'\cdot\vsigma}{r'} \right]c_{11}^{cc} \nonumber\\
\!\!\!\!\!\!\!\!\! + \sin\left(\frac{\omega\Delta t}{2}\right) \left[\cos\left( \frac{\omega r}{2}\right) \sin\left(\frac{\omega r'}{2}\right)
 + \sin\left(\frac{\omega r}{2}\right) \cos\left(\frac{\omega r'}{2}\right) \frac{\vx\cdot\vsigma}{r} \frac{\vx'\cdot\vsigma}{r'} \right]s_{11}^{cs}.
 \nonumber \\
 \label{eq:lambda_11_coeff}
\end{eqnarray}
In (\ref{eq:lambda_21_coeff}, \ref{eq:lambda_11_coeff}) we have expressions
for $\lambda_{21}$ and $\lambda_{11}$ in terms of two $2\times 2$ matrices
$s_{11}^{cs}$ and $c_{11}^{cc}$, whose entries can only depend
on $\bm{n}$ and $\bm{n}'$. In the following subsection, the
angular dependence of these matrices will be determined by solving
\eqref{eq:Dlambda} directly.

\subsection{Determining the functions of the angular coordinates }\label{app:bispinor:eq}

Having derived expressions for  $\lambda_{11}$ and $\lambda_{21}$ in terms of $s_{11}^{cs}$ and $c_{11}^{cc}$, we can now solve
\eqref{eq:Dlambda} for the left half of the matrix $\lambda$ \eqref{eq:lambdaij_def}.
However, before we can solve (\ref{eq:Dlambda}), we need to find $\slashed{n} \lambda$ and therefore we require the equivalent of \eqref{eq:DLambda} for
$\slashed{n} \Lambda$.
Using the following property:
\begin{equation}
 D_\halpha \slashed{n} = - A(n_\halpha + \gamma_\halpha \slashed{n}) \slashed{n},
\end{equation}
it can be shown that $\slashed{n} \Lambda$ satisfies the equation:
\begin{equation}
 D_\halpha (\slashed{n} \Lambda) = - \frac{A - C}{2} (n_\halpha + \gamma_\halpha \slashed{n})
 (\slashed{n} \Lambda).\label{eq:DnLambda_aux}
\end{equation}
Substituting \eqref{eq:AC} for $A$ and $C$  into \eqref{eq:DnLambda_aux}
gives the equation
\begin{equation}
 D_\halpha (\slashed{n} \Lambda) = - \frac{\omega}{2} \cot\left(\frac{\omega s}{2}\right)
 (n_\halpha + \gamma_\halpha \slashed{n}) (\slashed{n} \Lambda).
 \label{eq:DnLambda}
\end{equation}
Using \eqref{eq:lambda_def} to write (\ref{eq:DnLambda}) in terms of $\lambda$
and taking advantage of the factor of $\cot\frac{\omega s}{2}$ on the right hand side,
the following equation is obtained:
\begin{equation}
\fl  D_\halpha \left[\cot\left(\frac{\omega s}{2}\right) \frac{\slashed{n} \lambda}{\sqrt{\cos\omega r \cos\omega r'}}\right] =
 -\frac{\omega}{2} \gamma_\halpha \slashed{n}
 \left[\cot\left(\frac{\omega s}{2}\right) \frac{\slashed{n} \lambda}{\sqrt{\cos\omega r \cos\omega r'}}\right].
 \label{eq:Dnlambda}
\end{equation}
The similarity between \eqref{eq:Dlambda} and \eqref{eq:Dnlambda} enables the construction
of $\slashed{n} \lambda$ directly.
Indeed, using \eqref{eq:ntetrad} for $n_{\halpha}$
and (\ref{eq:lambda_21_coeff}, \ref{eq:lambda_11_coeff}) for
$\lambda_{21}$ and $\lambda_{11}$, a tedious but otherwise straightforward calculation
allows $(\slashed{n}\lambda)_{11}$ to be written as:
\begin{eqnarray}
\fl \frac{(\slashed{n} \lambda)_{11}}{\cot(\omega s/2)} =
 \sin\left(\frac{\omega\Delta t}{2}\right) \left[\cos\left(\frac{\omega r}{2}\right) \cos\left(\frac{\omega r'}{2}\right) +
 \sin\left(\frac{\omega r}{2}\right) \sin\left(\frac{\omega r'}{2}\right) \frac{\vx\cdot\vsigma}{r} \frac{\vx'\cdot\vsigma}{r'}
 \right]c_{11}^{cc} \nonumber\\
\!\!\!\!\!\!\!\!\! - \cos\left(\frac{\omega\Delta t}{2}\right) \left[\cos\left(\frac{\omega r}{2}\right) \sin\left(\frac{\omega r'}{2}\right) -
 \sin\left(\frac{\omega r}{2}\right) \cos\left(\frac{\omega r'}{2}\right)\frac{\vx\cdot\vsigma}{r}\frac{\vx'\cdot\vsigma}{r'}
 \right]s_{11}^{cs},
 \nonumber \\
 \label{eq:nlambda_11_coeff}
\end{eqnarray}
while $(\slashed{n}\lambda)_{21}$ reduces to:
\begin{eqnarray}
\fl \frac{(\slashed{n} \lambda)_{21}}{\cot(\omega s/2)} =
 \cos\left(\frac{\omega\Delta t}{2}\right) \left[\sin\left(\frac{\omega r}{2}\right) \cos\left(\frac{\omega r'}{2}\right) \frac{\vx\cdot\vsigma}{r} -
 \cos\left(\frac{\omega r}{2}\right) \sin\left(\frac{\omega r'}{2}\right) \frac{\vx'\cdot\vsigma}{r'} \right]c_{11}^{cc} \nonumber\\
\!\!\!\!\!\!\!\!\! + \sin\left(\frac{\omega\Delta t}{2}\right) \left[\sin\left(\frac{\omega r}{2}\right) \sin\left(\frac{\omega r'}{2}\right) \frac{\vx\cdot\vsigma}{r}
 + \cos\left(\frac{\omega r}{2}\right) \cos\left(\frac{\omega r'}{2}\right) \frac{\vx'\cdot\vsigma}{r'} \right]s_{11}^{cs}.
 \nonumber \\
 \label{eq:nlambda_21_coeff}
\end{eqnarray}

In order to solve \eqref{eq:Dlambda}, it is convenient to cast it in the following form:
\begin{eqnarray}
 \fl \frac{2}{\omega} \frac{\partial \lambda}{\partial x^i} +
 \frac{x^i}{r} \left\{ \tan\left(\frac{\omega r}{2}\right) \lambda +
 \frac{1}{\cos\omega r}\left(1 - \frac{\sin\omega r}{\omega r}\right)\left[\tan\left(\frac{\omega r}{2}\right) \lambda -
 \frac{\vx \cdot \vgamma}{r} \frac{\slashed{n} \lambda}{\cot (\omega s/2)}\right]\right\} \nonumber\\
 - \frac{\tan \omega r}{\omega r} \gamma^{\hat{i}} \frac{\vx \cdot \vgamma}{r}
 \left[\tan\left(\frac{\omega r}{2}\right) \lambda -
 \frac{\vx \cdot \vgamma}{r} \frac{\slashed{n} \lambda}{\cot (\omega s/2)}\right] = 0.
 \label{eq:appa4intermediate}
\end{eqnarray}
It is sufficient to consider the $(11)$ component of \eqref{eq:appa4intermediate}. Considering those terms proportional
to $\cos\frac{\omega \Delta t}{2}$ and $\sin\frac{\omega \Delta t}{2}$ separately, we find:
\begin{equation}
 \partial_i c^{cc}_{11}=0, \qquad
 \partial_i s^{cs}_{11} = 0,
\end{equation}
thus showing that $c^{cc}_{11}$ and $s^{cs}_{11}$ are constants.
In the coincidence limit, $\lambda_{11}$ and $\lambda_{21}$ reduce to:
\begin{equation}
 \left.\lambda_{11}\right\rfloor_{x = x'} = (\cos\omega r) c^{cc}_{11}, \qquad
 \left.\lambda_{21}\right\rfloor_{x = x'} = -(\cos\omega r) \left(\frac{\vx \cdot \vsigma}{r}\right) s^{cs}_{11}.
\end{equation}
According to \eqref{eq:lambda_inv}, in the coincidence limit
$\lim_{x' \rightarrow x} \lambda(x,x') = \cos\omega r$, from which we determine the constants $c^{cc}_{11}$
and $s^{cs}_{11}$ to be:
\begin{equation}
 c^{cc}_{11} = 1, \qquad s^{cs}_{11} = 0.
\end{equation}
This result allows us to write (\ref{eq:lambda_21_coeff}, \ref{eq:lambda_11_coeff}) as:
\begin{eqnarray}
\fl \lambda_{11} = \cos\left(\frac{\omega\Delta t}{2}\right) \left[\cos\left(\frac{\omega r}{2}\right) \cos\left(\frac{\omega r'}{2}\right) -
 \sin\left(\frac{\omega r}{2}\right) \sin\left(\frac{\omega r'}{2}\right)\frac{\vx\cdot\vsigma}{r} \frac{\vx'\cdot\vsigma}{r'} \right] ,
 \label{eq:lambda_11}\\
\fl \lambda_{21} = -\sin\left(\frac{\omega\Delta t}{2}\right) \left[
 \sin\left(\frac{\omega r}{2}\right) \cos\left(\frac{\omega r'}{2}\right) \frac{\vx\cdot\vsigma}{r} +
 \cos\left(\frac{\omega r}{2}\right) \sin\left(\frac{\omega r'}{2}\right) \frac{\vx'\cdot\vsigma}{r'} \right].
 \label{eq:lambda_21}
\end{eqnarray}
The $2\times 2$ component $\lambda_{12}$ can be found by using the second property \eqref{eq:lambda_inv}:
\begin{equation}
 \lambda_{12}(x,x') = -\lambda^\dagger_{21}(x',x) = \lambda_{21}(x,x').\label{eq:lambda_12}
\end{equation}
Finally, $\lambda _{22}$ can be found from $\lambda_{12}$ via (\ref{eq:At_def}):
\begin{equation}
\lambda_{22}(x,x') = -B_t^{-1} A_t \lambda_{12},
\end{equation}
and hence we have
\begin{equation}
 \lambda_{22}(x,x') = \lambda_{11}(x,x').
 \label{eq:lambda_22}
\end{equation}
It can be checked that introducing (\ref{eq:lambda_11}--\ref{eq:lambda_22}) into \eqref{eq:lambdaij_def} and using \eqref{eq:lambda_def}
to find $\Lambda(x,x')$ recovers \eqref{eq:Lambda}.

Finally, the $(12)$ and $(22)$ components of $\slashed{n}\lambda$ can be found starting from
(\ref{eq:lambda_12}, \ref{eq:lambda_22}):
\begin{equation}
 (\slashed{n} \lambda)_{12} =
 -(\slashed{n} \lambda)_{21},
 \qquad
 (\slashed{n} \lambda)_{22} = -(\slashed{n} \lambda)_{11},
\end{equation}
thus establishing \eqref{eq:Lambdan}.


\section*{References}

\begin{thebibliography}{99}

\bibitem{Maldacena:1997re}
Gubser S S, Klebanov I R and Polyakov A M 1998 \PL B {\bf 428} 105--14

\nonum
Maldacena J M 1999 {\it {Int.~J.~Theor.~Phys.}}~{\bf 38} 1113--33

\nonum
Witten E 1998 {\it {Adv.~Theor.~Math.~Phys.}}~{\bf 2} 253--91

\bibitem{art:aharony00}
Aharony O, Gubser S S, Maldacena J M, Ooguri H and Oz Y 2000 {\it {Phys.~Rept.}}~{\bf {323}} 183--386

\bibitem{Avis:1977yn}
  Avis S J, Isham C J and Storey D 1978 \PR D {\bf 18} 3565--76

\bibitem{Burgess:1984ti}
  Burgess C P and L\"utken C A 1985 \PL B {\bf 153} 137--41

\nonum
Cot\u{a}escu I I 1999 \PR D {\bf 60} 107504

\bibitem{Kent:2014nya}
  Kent C and Winstanley E 2015 \PR D {\bf 91}  044044

\bibitem{Camporesi:1992wn}
  Camporesi R and Higuchi A 1992 \PR D {\bf 45} 3591--603

\bibitem{Caldarelli:1998wk}
 Caldarelli M M  1999 \NP B {\bf 549} 499--515

\bibitem{Allen:1986ty}
  Allen B, Folacci A and Gibbons G W 1987 \PL B {\bf 189} 304--10

\bibitem{art:allen86a}
Allen B and Jacobson T 1986 {\it {Commun.~Math.~Phys.}}~{\bf {103}} 669--92

\bibitem{art:belo}
  Belokogne A, Folacci A and Queva J 2016 \PR D {\bf 94} 105028

\bibitem{art:allen86b}
Allen B and L\"utken C A 1986 {\it {Commun.~Math.~Phys.}}~{\bf {106}} 201--10

\bibitem{art:muck00}
M\"{u}ck W 2000 \JPA {\bf {33}} 3021--6

\bibitem{art:cotaescu07}
Cot\u{a}escu I I 2007 {\it {Rom.~J.~Phys.}}~{\bf {52}} 895--940

\bibitem{art:ambrus15plb}
Ambru\cb{s} V E and Winstanley E 2015 \PL B {\bf 749} 597--602

\bibitem{Buhl-Mortensen:2016jqo}
 Buhl-Mortensen I, de Leeuw M, Ipsen A C, Kristjansen C and Wilhelm M 2017
  \JHEP {\bf 1701} 098

\bibitem{tolman30}
Tolman R C  1930 \PR {\bf 35} 904--24

\nonum
Tolman R C and Ehrenfest P  1930 \PR {\bf 36} 1791--8


\bibitem{art:decanini08}
D\'ecanini Y and Folacci A 2008 \PR D {\bf {78}}  044025

\bibitem{ambrus16cota}
Ambru\cb{s} V E and Cot\u{a}escu I I 2016 \PR D {\bf 94} 085022

\bibitem{ambrus15wut}
Ambru\cb{s} V E and Blaga R 2015 {\it {Annals of West University of Timi\cb{s}oara - Physics}} {\bf 58} 89--108

\bibitem{florkowski15}
Florkowski W and Maksymiuk E 2015 \jpg {\bf 42} 045106

\bibitem{Brill:1957fx}
  Brill D R and Wheeler J A 1957 \RMP {\bf 29} 465--79

\nonum
Weldon H A 2001 \PR D {\bf 63} 104010

\nonum
Gies H and Lippoldt S 2014 \PR D {\bf 89} 064040

\bibitem{Christensen:1976vb}
Christensen S M 1976 \PR D {\bf 14} 2490--501

\bibitem{poisson}
Poisson E 2004 \CQG {\bf {21}} R153--232

\bibitem{art:christensen78}
Christensen S M 1978 \PR D {\bf {17}} 946--963

\bibitem{art:groves02}
Groves P B, Anderson P R and Carlson E D 2002 \PR D {\bf {66}} 124017

\bibitem{birrell82}
 Birrell N D and Davies P C W 1982
 {\em Quantum fields in curved space} (Cambridge: Cambridge University Press).

\bibitem{rezzolla13}
Rezzolla L and Zanotti O 2013 {\em Relativistic hydrodynamics} (Oxford: Oxford University Press)

\bibitem{mezzacappa13}
Cardall C Y, Endeve E and Mezzacappa A 2013 \PR D {\bf 88} 023011

\bibitem{cercignani02}
Cercignani C and Kremer G M 2002 \emph{The relativistic Boltzmann equation: theory and applications} (Basel: Birkh\"auser)

\bibitem{ambrus14phd}
Ambru\cb{s} V E 2014 {\em Dirac fermions on rotating space-times} (PhD thesis, University of Sheffield) [http://etheses.whiterose.ac.uk/id/eprint/7527]

\bibitem{book:olver10}
Olver F W J, Lozier D W, Boisvert R F and Clark C W 2010
{\emph{NIST Handbook of Mathematical Functions}}
(New York: Cambridge University Press)

\bibitem{Gibbons:1977mu}
Gibbons G W and Hawking S W 1977 \PR D {\bf 15} 2738--51

\bibitem{page}
Brown M R, Ottewill A C and Page D N 1986 \PR D {\bf 33} 2840--50

\bibitem{Flachi:2008sr}
  Flachi A and Tanaka T 2008 \PR D {\bf 78} 064011

\bibitem{rot1}
Ambru\cb{s} V E and Winstanley E 2014 \PL B {\bf 734} 296--301

\bibitem{rot2}
Ambru\cb{s} V E and Winstanley E 2016 \PR D {\bf 93} 104014

\bibitem{rotating}
Manning A 2015 Fermions in rotating reference frames {\tt {arXiv:1512.00579 [hep-th]}}

\nonum
Ebihara S, Fukushima K and Mameda K 2017 \PL B {\bf 764} 94--9

\nonum
Chernodub M N and Gongyo S 2017 \JHEP {\bf 1701} 136

\nonum
Chernodub M N  and Gongyo S 2017 \PR D {\bf 95} 096006

\bibitem{Kent:2014wda}
  Kent C and Winstanley E 2015 \PL B {\bf 740} 188--91

\end{thebibliography}
\end{document}